\documentclass[12pt,reqno]{article}

\usepackage{amsmath,amsfonts,amsthm,subfloat,algorithm,url,float,epsf,psfrag,bm}

\usepackage[pdftex]{color,graphicx}
\usepackage[colorlinks,citecolor=blue,urlcolor=blue]{hyperref}

\usepackage[noabbrev,capitalize]{cleveref}
\usepackage{algorithm, algorithmic, verbatim, float}
\usepackage{titlesec}
\titlespacing*{\section}{0pt}{1ex plus 1ex minus .2ex}{1ex plus .2ex}
\titlespacing*{\subsection}{0pt}{1ex plus 1ex minus .2ex}{1ex plus .2ex}
\usepackage[pagewise]{lineno}

\usepackage[numbers]{natbib}
\usepackage{setspace,caption}
\usepackage{blindtext}
\usepackage{subcaption}
\usepackage[utf8]{inputenc}

\usepackage{mathrsfs}
\usepackage{amsfonts}
\usepackage{enumerate}
\usepackage{latexsym}
\usepackage{multirow}

\floatname{algorithm}{Algorithm}

\newtheorem{theorem}{Theorem}[section]

\newtheorem{lemma}[theorem]{Lemma}
\newtheorem{defn}[theorem]{Definition}
\newtheorem{corollary}[theorem]{Corollary}
\newtheorem{remark}{Remark}[section]
\newtheorem{example}[theorem]{Example}

\newtheorem{proofpart}{Proof of}

\newcommand{\bigO}{{\mathcal{O}}}

\newcommand{\E}{{\mathbb E}}

\newcommand{\R}{{\mathbb R}}
\renewcommand{\P}{{\mathbb P}}
\newcommand{\ac}{{\mathcal{A}}}
\newcommand{\Di}{{\mathfrak{D}}}
\newcommand{\A}{{\mathcal{A}}}

\newcommand{\Ps}{{\mathcal{P}}}

\newcommand{\F}{{\mathcal F}}
\newcommand{\f}{{\mathfrak F}}
\newcommand{\samp}{{\mathcal{X}}}
\newcommand{\sampy}{{\mathcal{Y}}}
\newcommand{\X}{{\mathcal{X}}}

\newcommand{\N}{{\mathcal{N}}}
\newcommand{\pilo}{\check \pi^{(\underline \alpha)}}
\newcommand{\pihi}{\check \pi^{(\overline \alpha)}}
\newcommand{\fonelo}{\check f_1^{(\underline \alpha)}}
\newcommand{\fonehi}{\check f_1^{(\overline \alpha)}}
\newcommand{\alo}{\underline \alpha}
\newcommand{\ahi}{\overline \alpha}
\newcommand{\vertiii}[1]{{\left\vert\kern-0.25ex\left\vert\kern-0.25ex\left\vert #1 
		\right\vert\kern-0.25ex\right\vert\kern-0.25ex\right\vert}}
\newcommand{\Fga}{{\mathcal{F}_{\text{Gauss}}}} 
\newcommand{\Fde}{{\mathcal{F}_{\downarrow}}}
\newcommand{\dga}{{\mathfrak{F}_{\text{Gauss}}}} 
\newcommand{\dde}{{\mathfrak{F}_{\downarrow}}} 
\newcommand{\logi}{{\Pi_{\text{logit}}}} 
\newcommand{\probit}{{\Pi_{\text{probit}}}}
\newcommand{\Pide}{{\Pi_{\uparrow}}} 
\newcommand{\Picon}{{\Pi_{\equiv}}}
\newcommand{\Pig}{{\Pi_{\text{g}}}}
\newcommand{\lfdr}{{\mathfrak{l}}}
\newcommand{\Rs}{{\mathcal{R}}}

\newcommand{\keywords}[1]{\textbf{\textit{Keywords:}} #1}
\newcommand\norm[1]{\left\lVert#1\right\rVert}
\newcommand*{\defeq}{\mathrel{\vcenter{\baselineskip0.5ex \lineskiplimit0pt
			\hbox{\scriptsize.}\hbox{\scriptsize.}}}%
	=}
\numberwithin{algorithm}{section}
\allowdisplaybreaks

\newcounter{rcnt}[section]

\def\qt#1{\;\text{#1}}

\def\argmin{\mathop{\rm argmin}}
\def\argmax{\mathop{\rm argmax}}
\makeatletter
\@addtoreset{proofpart}{theorem}
\makeatother

\usepackage{graphicx}
\usepackage{amsmath}
\makeatletter
\newcommand{\distas}[1]{\mathbin{\overset{#1}{\kern\z@\sim}}}%
\newsavebox{\mybox}\newsavebox{\mysim}
\newcommand{\distras}[1]{%
	\savebox{\mybox}{\hbox{\kern3pt$\scriptstyle#1$\kern3pt}}%
	\savebox{\mysim}{\hbox{$\sim$}}%
	\mathbin{\overset{#1}{\kern\z@\resizebox{\wd\mybox}{\ht\mysim}{$\sim$}}}%
}
\makeatother

\usepackage{xr}
\usepackage{enumerate}

\newcommand{\blind}{1}

\begin{document}
	


\if1\blind
{
  \title{\LARGE \bf Two-component Mixture Model in the Presence of Covariates}
  \author{Nabarun Deb\thanks{E-mail: nd2560@columbia.edu}\hspace{.2cm}\\
    \small{Department of Statistics, Columbia University}\\
  Sujayam Saha\thanks{E-mail: sujayam@berkeley.edu} \hspace{.2cm}\\
    \small{Data Scientist at Google} \\
    Adityanand Guntuboyina\thanks{Supported by NSF CAREER Grant DMS-16-54589; e-mail: aditya@stat.berkeley.edu} \hspace{.2cm}\\
    \small{Department of Statistics, University of California at Berkeley} \\
    Bodhisattva Sen\thanks{Supported by NSF Grants DMS-17-12822 and AST-16-14743; e-mail: bodhi@stat.columbia.edu} \hspace{.2cm}\\
    \small{Department of Statistics, Columbia University}     
    }
  \maketitle
} \fi

\if0\blind
{
  \bigskip
  \bigskip
  \bigskip
  \begin{center}
    {\LARGE\bf Two-component Mixture Model in the Presence of Covariates}
\end{center}
  \medskip
} \fi
	\renewcommand{\baselinestretch}{1.4} 
\bigskip
\begin{abstract}
	In this paper, we study a generalization of the two-groups model in the presence of covariates --- a problem that has recently received much attention in the statistical literature due to its applicability in multiple hypotheses testing problems. The model we consider allows for infinite dimensional parameters and offers flexibility in modeling the dependence of the response on the covariates. We discuss the identifiability issues arising in this model and systematically study several estimation strategies. We propose a tuning parameter-free nonparametric maximum likelihood method, implementable via the EM algorithm, to estimate the unknown parameters. Further, we derive the rate of convergence of the proposed estimators --- in particular, we show that the finite sample Hellinger risk for every `approximate' nonparametric maximum likelihood estimator achieves a near-parametric rate (up to logarithmic multiplicative factors). In addition, we propose and theoretically study two `marginal' methods that are more scalable and easily implementable. We demonstrate the efficacy of our procedures through extensive simulation studies and relevant data analyses --- one arising from neuroscience and the other from astronomy. We also outline the application of our methods to multiple testing. The companion R package \texttt{NPMLEmix} implements all the procedures proposed in this paper. 
\end{abstract}


\section{Introduction}\label{sec:Intro}

Consider i.i.d.~observations $Y_1,\ldots, Y_n$ from the following two-component mixture model:
\begin{equation}\label{eq:TwoComp}
\qquad \qquad Y_i \sim \bar \pi F_1^* + (1-\bar \pi) F_0, \qquad \mbox{ for } i = 1,\ldots, n,
\end{equation}
where $F_0$ is assumed to be a completely {\it known} distribution function (DF) while $F_1^*$, along with $\bar \pi$, are the unknown quantities of interest. We will call $F_0$ the {\it noise} distribution, $F_1^*$ the {\it signal} distribution and $\overline{\pi}$ the {\it signal} proportion. Such a model has received a lot of attention in the statistical literature,  particularly in the context of multiple testing problems (microarray analysis, neuroimaging, etc.) where it is usually referred to as the {\it two-groups} model; see e.g.,~\cite{Efron08},~\cite[Chapter 2]{Efron10},~\cite{storey2003},~\cite{Storey02} and~\cite{cai10}. In the multiple testing problem, the obtained $p$-values or $z$-values ($Y_i$'s as per \eqref{eq:TwoComp}), from the numerous (independent) hypotheses tests, have a Uniform$(0,1)$ or $N(0,1)$ distribution (under the null hypothesis), which we call $F_0$, while their distribution (i.e., $F_1^*$) under the alternative is {\it unknown}; here $\bar \pi$ is the proportion of non-null hypotheses. The two-groups model has also been used in contamination problems, where the (unknown) distribution $F_1^*$ may be contaminated by the known distribution $F_0$, yielding a sample drawn from $F$ as in~\eqref{eq:TwoComp}; see e.g.,~\cite{McPeel00},~\cite{walker2009clean},~\cite{dai07} and~\cite{lemdani99}.


However, quite often in real applications, additional information is available on each observation in the form of {\it covariates} which is ignored by the two-groups model. The following two examples describe two such applications and illustrate the need to model the observed covariates.

\begin{example}[Neuroscience example]\label{ex:Neuro} 
	\begin{figure}
		\centering
		\includegraphics[width = 0.45\textwidth]{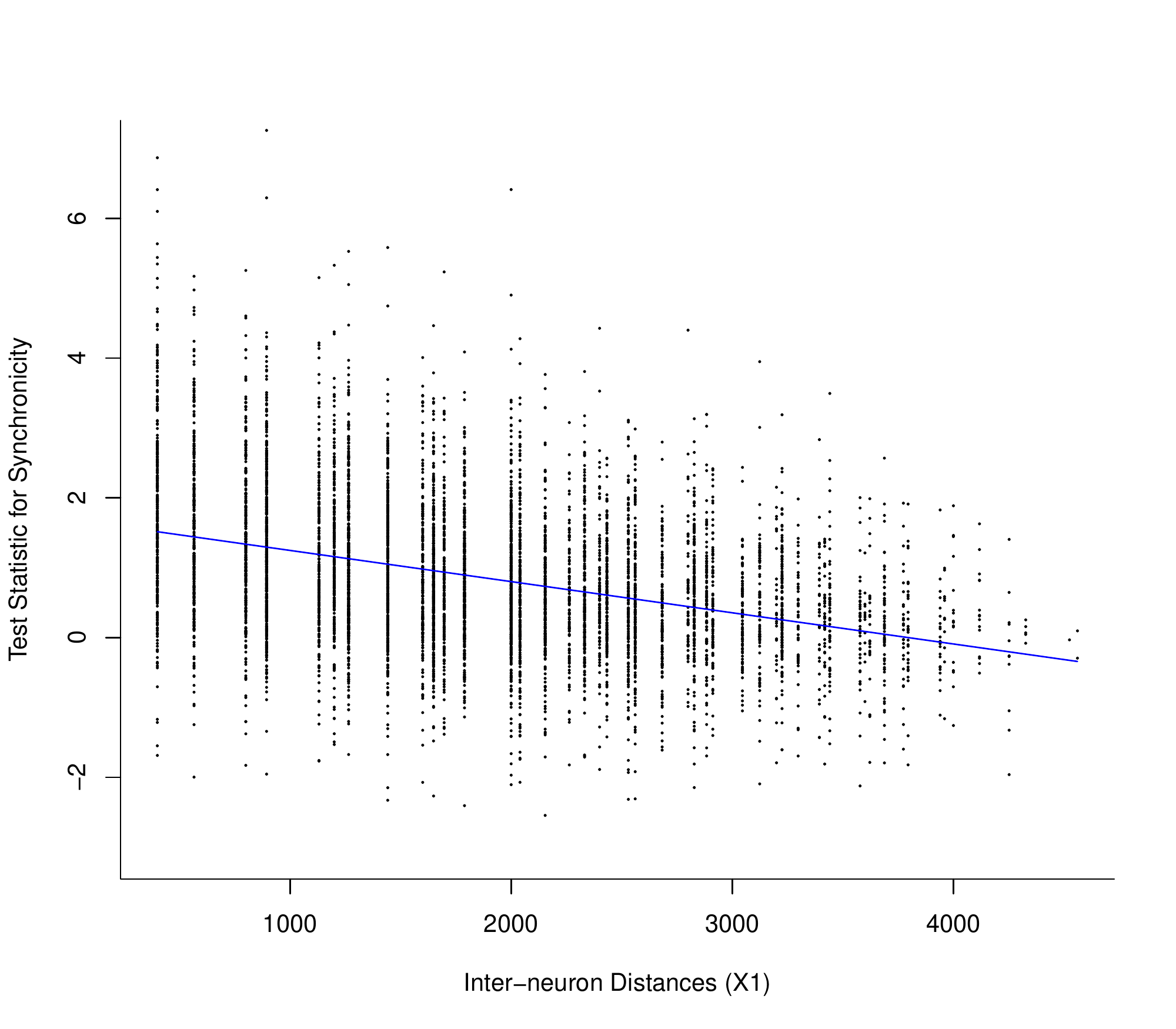}
		\includegraphics[width = 0.45\textwidth]{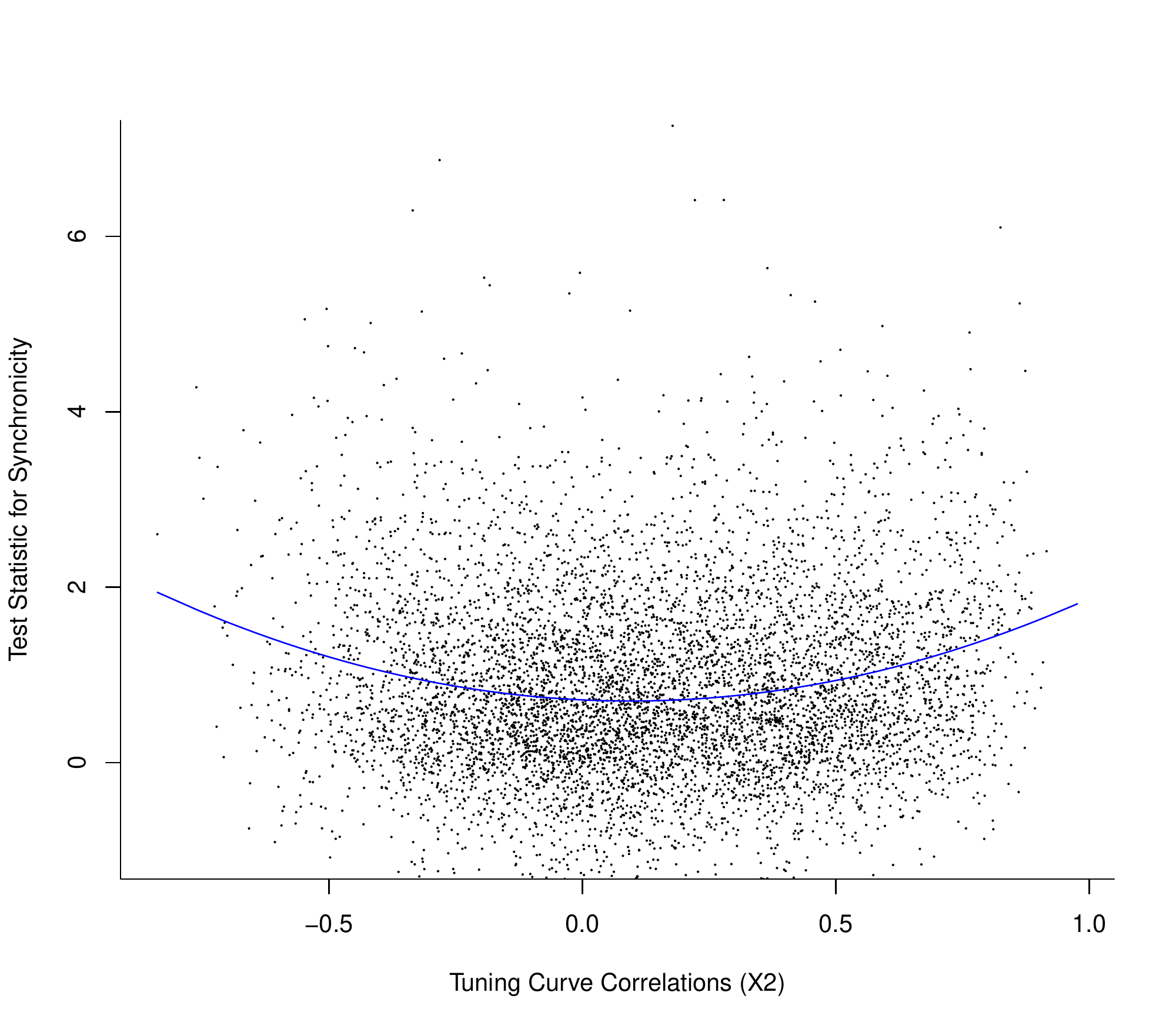}
		\caption{Scatter plots of test statistics computed on each pair of units (higher is more significant), plotted against covariates: distance between units (left), tuning curve correlation between units (right). The loess fit is overlaid upon each scatter plot, visually indicating that the test statistics are dependent upon covariate information.}\label{fig:scott_scatterplots}	
	\end{figure}
	\citet{ScottEtAl15} analyzed data arising from a
	multi-unit recording experiment consisting of
	measurements from $128$ units (either neurons or
	multi-unit groups) from the primary visual cortex of a
	rhesus macaque monkey in response to visual stimuli
	(see~\cite{Kelly07} for details). The goal of the
	experiment was to detect fine-time-scale neural 
	interactions (``synchrony''). The data consisted of thousands of test
	statistics $Y_i$'s, each one arising from testing the null hypothesis
	of no interaction between a pair of units. Let $F_0$ be the null
	distribution of $Y_i$ (assumed to be known) and $F_1^*$ the unknown
	signal distribution. A natural approach for modeling the distribution of $Y_i$'s is via the two-groups model, see e.g.,~\cite{ScottEtAl15}. However the data set also included two interesting covariates: (a) physical distance between units, and
	(b) tuning curve correlation between
	units. Figure~\ref{fig:scott_scatterplots} illustrates the
	relationship between the observed test statistics and the two
	covariates. It clearly shows that the covariates are related to the
	$Y_i$'s. However, as was also observed by~\cite{ScottEtAl15}, the two-groups model~\eqref{eq:TwoComp}
	inappropriately ignores the known spatial and functional relationships
	among the neurons. This motivates the need to develop and study models
	that generalize~\eqref{eq:TwoComp} to include covariates. We discuss
	this data and its analysis in more detail in \cref{realdata:neuro}.  
\end{example}

\begin{example}[Astronomy example]\label{ex:Astro} \citet{walker2009clean} analyzed data on individual stars obtained from nearby dwarf spheroidal (dSph) galaxies. The data contain measurements on line-of-sight velocity (denoted by $Y$), projected distance from the center of the dSph galaxy (denoted by $X$), and other variables (e.g., metallicity) for around 1000-2500 stars per dSph, including some fraction of contamination from foreground Milky Way stars (in the field of view of the dSph galaxy); see e.g.,~\cite{walker09stellar}. The primary goal is to identify the dSph galaxy stars in the sample and recover their line-of-sight velocity distribution. Due to foreground contamination, $Y$ is distributed marginally as in the two-component mixture model~\eqref{eq:TwoComp}; see the right panel of Figure~\ref{fig:Carina}. Here we plot the estimated density (obtained using kernel density estimation) of the observed $Y_i$'s (for the Carina dSph) along with (scaled) $f_0$ --- the density of $F_0$ --- which is known from the Besancon Milky Way model (see~\cite{RobinEtAl03}). However, the left panel of Figure~\ref{fig:Carina}, which shows the scatter plot of $X$ and $Y$, reveals that $Y$ indeed depends on $X$ which the {\it two-groups} model fails to capture. In this paper we develop a methodology that incorporates this covariate information to yield: (a) better estimation of $F_1^*$, the distribution of the line-of-sight velocity for stars in the dSph; and (b) more reliable ``posterior'' probability estimates of each star (in the sample) being a dSph member; see~\cref{realdata:astro} in the appendix for details. 
	
	\begin{figure}
		\centering
		\includegraphics[width=3in, height=2.5in]{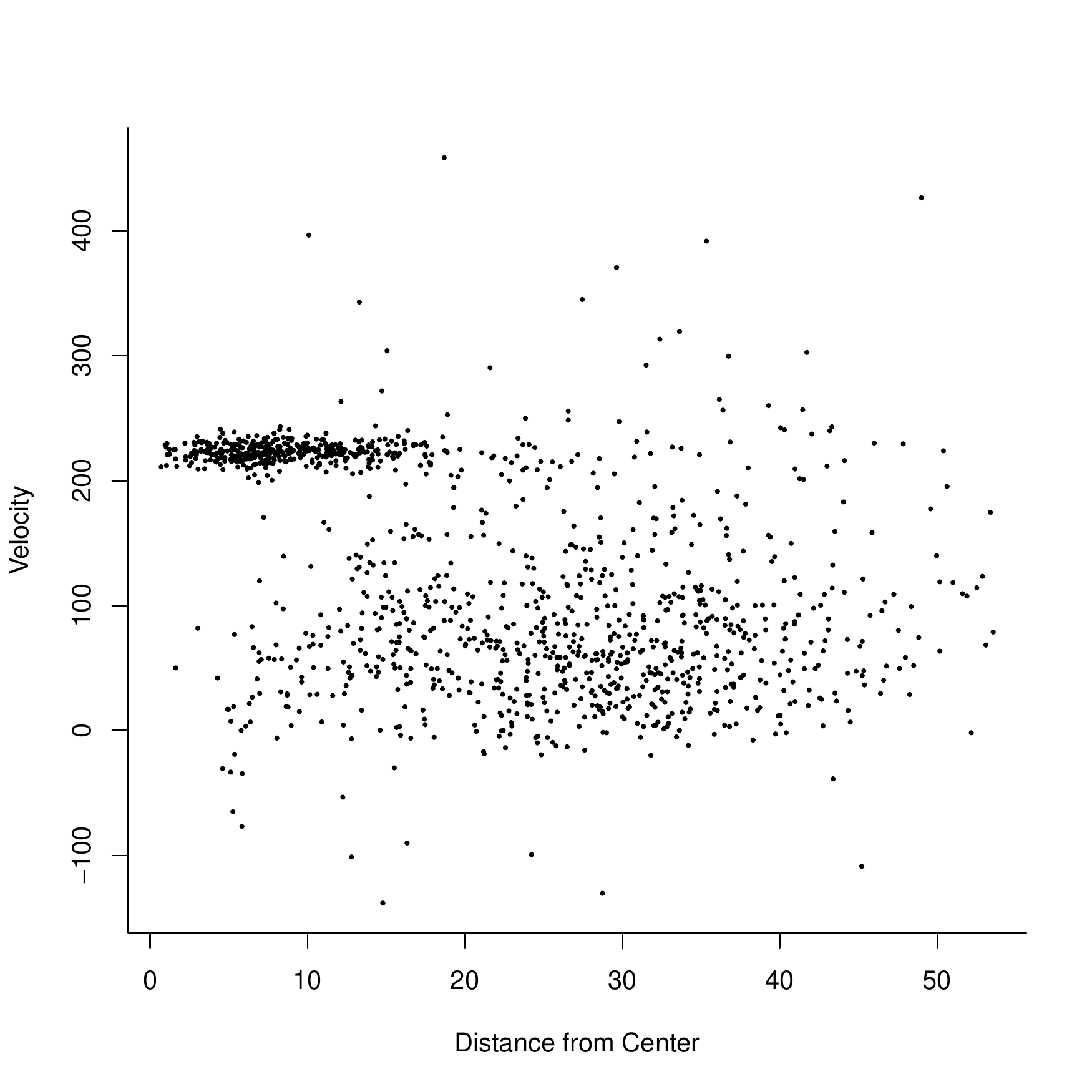}
		\includegraphics[width=3in, height=2.5in]{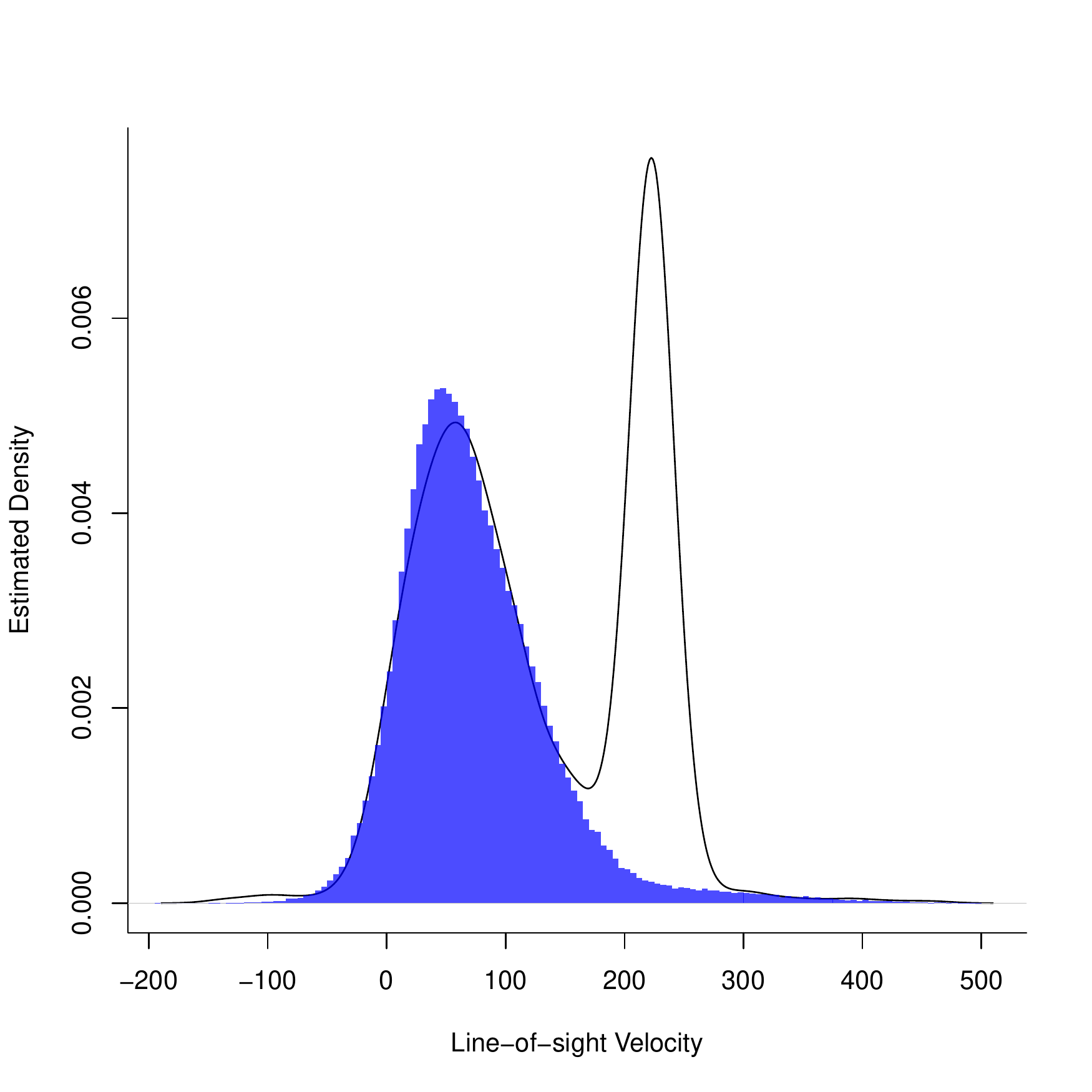}
		
		\caption{The left hand figure shows the joint scatter plot of $Y$ vs. $X$. The right hand plot shows the standard kernel density estimate of $Y$ in this dataset with a scaled $f_0$ overlaid in blue. This indicates that the tight point cloud to the top left on left figure comprises mostly of stars from Carina galaxy, while the sparser point cloud to bottom center comprises mostly of stars from Milky Way. From the scatter plot, Carina stars are clearly more frequent at lower values of $X$ (distance from center). Thus, a classification procedure which utilizes $X$ should be more accurate.}\label{fig:Carina}
	\end{figure}
\end{example}

Applications such as Examples \ref{ex:Neuro} and \ref{ex:Astro} motivate the need to generalize \eqref{eq:TwoComp} to incorporate covariate information; also see~\cite{schildknecht2016more} and~\cite{li2015accumulation} for two more relevant applications in neural imaging and genetics data respectively. Towards this direction, suppose that $(Y_1, X_1), \ldots, (Y_n, X_n)$ are i.i.d.~having a distribution on $\R \times \R^p$ ($p \ge 1$). As studied in~\citet{ScottEtAl15} and~\citet{walker2009clean}, a natural way to model the joint distribution of $(Y,X)$ that generalizes \eqref{eq:TwoComp} would be to consider
\begin{equation}\label{mdl}
Y | X = x \sim \pi^*(x) F_1^* +  (1 -  \pi^*(x)) F_0 ~~~ \text{  and  } ~~~ X \sim m  
\end{equation}
where: 
\begin{enumerate}
	\item  $m$ is a fixed probability measure supported on some space $\samp \subseteq \R^p$.  
	\item The random variable $Y$ takes values in a subset $\sampy$ of
	$\R$ (e.g., $\sampy = [0, 1]$ or $\sampy = \R$) and $F_0 \ne F_1^*$ are two DFs on $\R$. We assume that $F_0$ is {\it known} and $F_1^*$ is {\it unknown} and belongs to a parametric or nonparametric class $\F$. Note that model \eqref{mdl} assumes that $F_0$ and $F_1^*$ do not depend on the covariates.   
	\item $\pi^*: \samp \rightarrow [0, 1]$ is an {\it unknown} function
	belonging to a parametric or nonparametric class of functions $\Pi$.   
\end{enumerate}
The crucial difference between models~\eqref{mdl} and~\eqref{eq:TwoComp} is that~\eqref{mdl} allows the prior probability of an observation coming from the signal distribution to depend on the covariates. In fact, model~\eqref{mdl} is indeed a generalization of the two-groups model (which is obtained by taking $\pi^*(\cdot)$ to be the constant function). It is worth mentioning that~\eqref{mdl} can be treated as a regression model with a special structure: Suppose that $Z$ is the unobservable latent variable corresponding to $Y$ that decides which of the two populations ($F_0$ or $F_1^*$) $Y$ is drawn from; i.e., $Y|Z = 0 \sim F_0$ and $Y|Z = 1 \sim F_1^*$. Then, under model~\eqref{mdl}, $Y$ is {\it conditionally independent} of $X$ given $Z$; of course, $Y$ is dependent on $X$ unconditionally. This observation can be interpreted in the following way: Model~\eqref{mdl} implies that $X$ provides some information about $Y$, but $X$ does not provide any additional information about $Y$ if we knew the value of $Z$.

To motivate model~\eqref{mdl} further, we mention a few special cases of~\eqref{mdl} that are of significant interest in the multiple testing  problem. Let us start with two natural examples of $\F$ and $F_0$. 

\noindent {\bf Decreasing densities}: In this case, $\F$ denotes the class of all DFs having a nonincreasing density on $[0, 1]$ and $F_0$ is the uniform distribution on $[0, 1]$. This situation naturally arises in multiple testing
problems where $Y$ denotes the $p$-value corresponding to a hypothesis test and we assume that under $H_0$, the $p$-values have the uniform distribution on $(0,1)$; see e.g.,~\cite{GW04},~\cite{Efron10}. Further, in this problem it is quite natural to assume that, under the alternative, the $p$-values will tend to be stochastically smaller (or they will have a nonincreasing density on $[0,1]$); see e.g.,~\cite{LangaasEtAl05},~\cite{schweder1982plots}. Let us denote the class of all distributions with nonincreasing densities on $[0, 1]$ by $\Fde$.  

\noindent {\bf Gaussian location mixtures}: In this case, $\F\equiv \Fga$ denotes the class of all Gaussian location mixtures, i.e., any $F_1^* \in \Fga$ has the form $$\qquad \qquad \qquad F_1^*(x) \defeq \int \Phi(x - \theta) d G(\theta), \qquad \mbox{ for } x \in \R,$$ where $G$ is some unknown probability measure on $\R$ and $\Phi$ is the standard normal DF.  Moreover, we take $F_0 \defeq \Phi$; see e.g.,~\cite{ScottEtAl15},~\cite{cai10}. In the above display $G$ models the {\it effect size} distribution (see \cref{subsec:esize} in the appendix for the details) and naturally arises when dealing with $z$-scores (as opposed to $p$-values). Note that $\Fga$ contains all finite Gaussian location mixtures (with unit variance).


Next we consider some natural models for the class $\Pi$. 

\noindent {\bf Constant functions:} Let us first consider the case when $\Pi$ consists of all constant functions. This reduces model~\eqref{mdl} to the well-known  two-groups model (see~\eqref{eq:TwoComp}). We   shall denote this class by $\Picon$.  

\noindent {\bf Nondecreasing functions:} Assume $p = 1$ and $\samp$ is a subinterval of $\R$. Quite often when testing a set of (ordered) hypotheses, the practitioner may have reason to believe that the test statistics earlier in the set are less likely to be signals; see e.g.,~\cite{li2016},~\cite{li2015accumulation}. In such a situation, it is natural to consider $\Pi$ to be the class of all nondecreasing functions on $\samp$. We shall denote this class by $\Pide$.

\noindent {\bf Generalized linear model:} In the absence of strong prior information on the class $\Pi$, a general modeling strategy would be to consider the following class of functions:	\begin{equation}\label{eq:parametric_link}
\qquad \qquad \pi(x) \defeq g(\beta_0+\beta^\top x) \qquad \mbox{as $(\beta_0,\beta)$ varies over $\mathbb{R}\times \R^p$.}
\end{equation}
Here $g: \R \rightarrow [0, 1]$ is a fixed and known link function. We shall denote this class of functions by $\Pig$. When $g(z) \defeq (1 + \exp(-z))^{-1}$ (logistic link), we shall denote $\Pig$ by $\logi$. This is a special case of the model considered in~\cite{ScottEtAl15}. When $g(z)\defeq \Phi(z)$, we denote $\Pig$ by $\probit$. We will study these classes in detail in this paper.


\subsection{Our contributions}\label{contribute}
In this paper, we propose and study likelihood based methods for estimating the functions $\pi^*(\cdot)$ and $F_1^*$ (and its density $f_1^*$) as described in model~\eqref{mdl}. We conduct a systematic study of the statistical and computational properties of our proposed methods, which very naturally yield a multiple testing procedure; see~\cref{sec:testing} in the appendix. 
We summarize our contributions below:

\noindent {\bf Identifiability:} Model~\eqref{mdl}, as posited, need
not be identifiable. In~\cref{secIdenti}, we study identifiability of
model~\eqref{mdl} and give easily verifiable necessary and sufficient
conditions in a rather general setting; see~\cref{lem:bid}. In
addition, we demonstrate how to use~\cref{lem:bid} to prove
identifiability for a wide range of choices of $\Pi$ and $\F$,
including the ones used in~\cite{ScottEtAl15} (see~\cref{ide}). In
fact, our results can accommodate general descriptions of the
covariate space $\X$ and the accompanying measure $m$ (on
$\X$).~\cref{revide} (in the appendix) provides a sufficient condition for
identifiability in~\eqref{mdl} when $\X$ has both discrete and
continuous components. To the best of our knowledge, the issue of
identifiability in \eqref{mdl} has not been properly addressed
before. Note that the two-groups model, as posited in~\eqref{eq:TwoComp}, is not identifiable; see e.g.,~\cite{GW04},~\cite{ patra2016estimation}. However, it is interesting to note that the presence of covariates can make model~\eqref{mdl} identifiable.  

\noindent {\bf Joint maximum likelihood:} In Sections~\ref{NPMLE}
and~\ref{sec:EM-Gauss} we develop a general (nonparametric) maximum
likelihood based procedure to estimate $\pi^*$ and $f_1^*$ from
i.i.d.~observations drawn according to model~\eqref{mdl}. We propose
iterative procedures based on the expectation-maximization (EM)
algorithm to compute the maximum likelihood estimates (MLEs). Our
procedure can handle both parametric and nonparametric specifications
for $\F$ and $\Pi$ and, in particular, covers the important scenarios
discussed above. The resulting estimates of $\pi^*$ and $f_1^*$ yield
accurate estimators of the conditional density of $Y$ given $X$. We
show in \cref{sr1} that when we maximize the likelihood over the
nonparametric class of all Gaussian location mixtures ($\Fga$), the
resulting estimator of this conditional density has a {\it parametric}
rate of convergence, up to logarithmic factors (see
\cref{sec:GaussMix}). In fact,~\cref{sr1} holds for a much larger
class of estimators (we call these {\it approximate MLEs}) which
includes the MLE as a special case. This generalization is important for analyzing 
the statistical properties of our estimators as we are dealing with a
non-convex optimization problem where exact maximizers
are computationally difficult to obtain. 

We also propose specialized algorithms for solving the M-step in the EM algorithm for estimating $\pi^*$ and $f_1^*$, depending on the choices of $\Pi$ and $\F$. For $\Pi=\Pide$ and $\F=\Fde$, we are able to relate the underlying optimization problem to relevant variants of weighted isotonic regression for which exact and efficient algorithms exist. 
When $\F=\Fga$, we observe the corresponding connections to the
Kiefer-Wolfowitz MLE (see~\cite{ghosal2001entropies}). We discretize
the resulting infinite dimensional optimization problem that can now
be expressed in the separable convex optimization framework (as
discussed in~\cite{koenker2014convex}) and can be solved efficiently
using the optimization suite \texttt{Rmosek}. 


\noindent {\bf Marginal methods:} We propose two other 
methods for estimating $\pi^*(\cdot)$ and $f_1^*$ that are based on
appropriately marginalizing the joint distribution of $(X,Y)$; see
\cref{sec:MargMeth} for the details. These marginal methods bypass the
joint maximization of the likelihood (which is a non-convex problem in general) and are easily implementable. These marginal methods can also be successfully used to properly initialize the EM algorithm to compute the joint MLE. We establish a finite sample risk bound of our estimator of $f_1^*$ (see~\cref{marg-1}) and derive the asymptotic distribution of the coefficient vector for certain parametrically specified link functions $\pi^*(\cdot)$ (see~\cref{cor:Mar-2}). 

Even though we can handle nonparametric classes $\F$ (and $\Pi$), both our proposed methods --- namely, the joint maximization and marginal procedures --- are tuning parameter-free, and are thus completely automated.



\noindent {\bf Simulations and real data examples:} We conduct extensive simulation studies (see \cref{sec:sim}) that point to the superior performance of the proposed estimators, when compared to its competitors. A direct consequence of our proposed methodology is a comprehensive procedure that addresses the multiple testing problem (see~\cref{sec:testing} in the appendix). We demonstrate the accuracy of the estimated local false discovery rate (lFDR) through extensive simulations. Further,  we analyze the two real data examples introduced above (see Sections~\ref{realdata:neuro} and Appendix~\ref{realdata:astro}). These illustrate the applicability of our methods.
Both marginal methods and the joint maximum likelihood method have been implemented in the companion R package \texttt{NPMLEmix} which has been made available in the authors' \texttt{GitHub} page - \url{https://github.com/NabarunD/NPMLEmix}. It also includes relevant codes for all our simulations and data analyses.

Before considering estimation in the framework of~\eqref{mdl}, as we have done above, it is natural to ask: ``Do the covariates indeed have any effect in the multiple testing problem?''. In~\cref{sec:hypotest}, we show that the above question reduces to testing for statistical independence between $X$ and $Y$, and we advocate the use of distance covariance (see~\cite{Szekely07}) to address this issue.

The accompanying appendix contains proofs of our main results, detailed discussions on some of the algorithms we propose in the paper and additional computational studies.
\subsection{Literature review}\label{sec:Litrev}
The two-groups mixture model (without covariates) has been studied and applied extensively; see e.g., \cite{efron2004large},~\cite{Efron08},~\cite[Chapter 2]{Efron10},~\cite{scott2006exploration},~\cite{johnstone2004needles},~\cite{muller2006fdr},~\cite{robin07},~\cite{MeinRice06},~\cite{mclachlan2010use,mclachlan2002mixture},~\cite{Storey02,storey2003},~\cite{walker2009clean},~\cite{patra2016estimation},~\cite{GW04} and the references therein. 

However, in a variety of multiple testing applications, arising from neural imaging, genetics, finance, etc., there is often additional information available on the individual test statistics (e.g., $p$-values or $z$-scores)  --- for example, the $p$-values may be naturally ordered, grouped, contain inherent clusters, etc. One of the first papers that incorporated such auxiliary information in the multiple testing procedure was~\cite{geno06}, where the authors suggest using such prior knowledge to choose weights $w_i$ (corresponding to $p$-values $p_i$) non-adaptively (without observing the $p$-values) for the $i^{th}$ hypothesis and then applying the Benjamini-Hochberg procedure on the reweighed $p$-values $p_i/w_i$. Since then, several authors have proposed various methods that use the idea of reweighing $p$-values utilizing known group or hierarchical structure, e.g.,~\cite{dobriban2016general},~\cite{ben14},~\cite{hu10}, etc. The use of weighted $p$-values can be useful when there is strong prior information. However, when this is not the case, we believe that modeling the weights is itself a difficult problem and there is no generally accepted strategy. These limitations have prompted some recent advances in this area; e.g.,~\cite{ignatiadis2016data},~\cite{li2016},~\cite{lei2016adapt},~\cite{ScottEtAl15}, etc. We next compare and contrast our proposed approach with some of this more recent work.

\citet{ignatiadis2016data} propose grouping the hypotheses and choosing weights for each group so as to maximize the number of rejections after a usual reweighing procedure. In~\cite{ignatiadis2018}, using a slightly modified censoring $p$-value based approach, the authors are able to guarantee finite sample FDR control. To study their procedure they consider a further generalization of model~\eqref{mdl} where the distribution of non-null $p$-values are allowed to depend on the covariates. In contrast, we take a more direct approach by proposing natural models (see e.g., $\Pig$, $\Pide$, $\Fde$, $\Fga$) for $p$-values or $z$-scores and focus on accurate estimation of the unknown quantities. 
Moreover, our approach avoids grouping hypotheses based on covariates (which may be difficult if the covariate space is complex) and does not need the choice of any tuning parameters.

The papers~\cite{li2016} and~\cite{lei2016adapt} study the problem of multiple testing in the presence of covariate information under minimal assumptions (and allow the distribution of the non-null $p$-values to depend on the covariates) and develop methods with guaranteed finite sample FDR control. In particular,~\cite{lei2016adapt} considers a generalization of~\eqref{mdl} but indexed by finitely many parameters. We, on the other hand, are able to accommodate natural nonparametric classes. Both the approaches above (\cite{li2016} and~\cite{lei2016adapt}) rely on masking $p$-values lower than a certain threshold. Note that it is likely that most signals will have low $p$-values and so, masking them may not be desirable if the analyst is also interested in estimating the distribution of the non-null $p$-values. This is the case, for example, in the contamination problem mentioned in \cref{ex:Astro}. All the theoretical results proved in~\cite{li2016} and~\cite{lei2016adapt} are in terms of FDR control (which was their primary object of interest). Whereas, we are able to provide theoretical guarantees on the finite sample risk behavior of our estimators.

The paper~\cite{ScottEtAl15} is perhaps the closest to our work. The authors use $\Pi=\logi$, $F_0=\Phi$ and $\F=\Fga$ and illustrate the
superiority of such a model over the traditional two-groups model~\eqref{eq:TwoComp} in terms of signal detection through extensive
simulations and by analyzing the neural synchrony data (\cref{ex:Neuro}). The big difference between our paper and~\cite{ScottEtAl15} is that our main recommended procedure is based on (nonparametric) MLE while their recommended procedure (which they call FDRreg) is more like one of our marginal ones (see~\cref{rem:sim} for the details). Note that~\cite{ScottEtAl15} also proposes a full Bayes procedure and an empirical Bayes procedure. We, however, resort to a frequentist approach and obtain estimators by maximizing the likelihood function. Moreover,~\cite{ScottEtAl15} does not provide any theoretical guarantees for their estimators (as we do). In \cref{sec:sim}, we
argue through extensive simulations that our method yields more accurate estimates of $\pi^*(\cdot)$, $f_1^*$ and lFDRs (particularly when the signal varies significantly with the covariates) as we make better use of the available covariate information. 

\begin{remark}
	Note that some of the references above focus on finite sample control of FDR (e.g.,~\cite{lei2016adapt},~\cite{li2016}). However in our problem setting the (oracle) optimal testing procedure should reject hypotheses with low lFDRs; see~\cite{basu2018weighted} where the authors view the multiple testing problem from a decision theoretic perspective and prove such an optimality result. Thus, we focus on accurate estimation of lFDRs (and the associated model parameters). This is a crucial point of difference between our approach and some of those listed above. 
	Although our proposed method does not  guarantee finite sample FDR control, our multiple testing setup yields asymptotic control (see Section~\ref{sec:sim} for such a detailed study).  We believe that asymptotic control generally yields much higher power compared to guaranteed finite sample FDR control which can sometimes be quite conservative. 
\end{remark}

\section{Identifiability in model~\eqref{mdl}}\label{secIdenti} 
Identifiability issues arise naturally in the study of mixture models (see e.g.,~\cite{Teicher61},~\citet[Section 3.1]{Titt85}) and  model~\eqref{mdl} is no exception. We detail these issues in this section before proceeding to estimate $\pi^*(\cdot) \in \Pi$ and $F_1^* \in \F$ from model \eqref{mdl}. 

Recall that $X \sim m$ having support $\X \subset \R^p$. For a fixed $\pi(\cdot) \in \Pi$ and $F_1 \in \F$, let $P_{\pi, F_1}$ denote the joint distribution of $(X, Y)$ defined in \eqref{mdl}. Also
let $\Ps\defeq \Ps(\Pi, \F)$ denote the class $\{P_{\pi, F_1} : \pi \in 
\Pi, F_1 \in \F\}$.  The main issue with identifiability arises from
the fact that, in general, it is possible to represent a given $P \in
\Ps$ as $P_{\pi,   F_1}$ for two (or more) different choices of $\pi
\in \Pi$ and $F_1 \in \F$. 
\begin{defn}[Identifiability]\label{def:Iden}
	We say that $P_{(\pi^*,F_1^*)} \in \Ps(\Pi,\F)$ is {\it identifiable}
	if for every function $(\pi,F_1) \in \Pi \times \F$, the condition
	$P_{(\pi^*,F_1^*)} = P_{(\pi,F_1)}$ implies $\pi(x) = \pi^*(x)$ for
	$m$-almost everywhere (a.e.)~$x$, and $F_1(y) = F_1^*(y)$ for all $y \in \R$. 
\end{defn}

Although model~\eqref{mdl} has been considered before by~\citet{ScottEtAl15} there has not been a rigorous study of the associated identifiability issues. The following lemma characterizes identifiability in the setting of~\eqref{mdl}.  

\begin{lemma}\label{lem:bid}
	Let $\pi, \pi'$ be two functions from $\samp$ to $[0, 1]$ and let $F_1, F_1'$ be two DFs on $\R$. Consider the following two statements: 
	\begin{enumerate}
		\item[(a)] The probability distributions $P_{\pi, F_1}$ and $P_{\pi',
			F_1'}$ are identical. 
		\item[(b)] There exists a real number $c \ne 1$ such that 
		\begin{equation}\label{bid.eq1}
		\qquad\qquad\pi'(x) = \pi(x)/(1 - c) \qquad \qt{for $m$-a.e.~x, and} 
		\end{equation}
		\begin{equation}\label{bid.eq2}
		\qquad\qquad F_1'(y) = c F_0(y) + (1 - c) F_1(y)\qquad \qt{for every $y \in \R$}. 
		\end{equation}
	\end{enumerate}
	Then, 
	\begin{enumerate}
		\item The second statement (b) always implies the first one (a). 
		\item If we have the conditions $F_0 \neq F_1$ and
		$\pi(x) >0 $ with positive probability under $m$ (or $F_0 \neq
		F_1'$ and $\pi'(x) >  0$ with positive probability under $m$), then
		the first statement (a) implies the second statement (b).   
	\end{enumerate}
\end{lemma}


\begin{remark}[Non-identifiability under two-groups model without covariates]
	When $\Pi\defeq \Picon$ and $\F$ denotes any of the classes $\Fde$ or $\Fga$, then the model $P_{(\pi^*,F_1^*)} \in \Ps(\Pi,\F)$, where $\pi^*\in (0,1)$, is not identifiable. This is an immediate consequence of Lemma~\ref{lem:bid} and has also been observed in~\cite{GW04},~\cite{ patra2016estimation} among others. Thus, for many nonparametric classes $\F$, the absence of covariate information always leads to a non-identifiable model and it is not possible to recover $\bar \pi$. However, there is indeed a way of defining an identifiable mixing proportion in these problems; see e.g.,~\cite{GW04},~\cite{ patra2016estimation}.

\end{remark}

\begin{remark}[Non-identifiability under two-groups model with covariates] Quite often there is a natural ordering among the hypotheses to be tested; see e.g.,~\cite{li2016}. In this scenario, a natural choice for the parameters in model~\eqref{mdl} are $\Pi\defeq \Pide$, $F_0\sim\mathrm{Uniform}(0,1)$ and $\F\defeq\Fde$. In this setting~\cref{lem:bid} immediately yields that the model $P_{(\pi^*,F_1^*)} \in \Ps(\Pi,\F)$, where $\pi^*(x)<\delta<1$ for $m$-a.e.$~x\in\mathcal{X}$ (for some $\delta$), is not identifiable. As a result, for the multiple testing problem when we have $p$-values for each test, the natural model $F_0\sim\mathrm{Uniform}(0,1)$ and $\F\defeq\Fde$ is non-identifiable if we model the non-null proportion as a nondecreasing function of the covariates. 
\end{remark}
\begin{remark}[Presence of covariates can restore identifiability] Let $\pi^* \in \Pi$ and $F^*_1 \in \F$. Lemma~\ref{lem:bid} shows that if $c\pi^*(\cdot) $ does not belong to $\Pi$, for any $c\in (0,1)$, then $P_{(\pi^*,F_1^*)}$ is  identifiable. This shows that for many reasonable model classes $\Pi$ and $\F$, the presence of covariates (if we can model the observed data correctly) can lead to identifiability. Some examples of such model classes are provided below.
\end{remark}
Let us recall the definitions of $\logi$, $\probit$, $\Fde$ and $\Fga$ from~\cref{sec:Intro}. In the following discussion we will use~\cref{lem:bid} to investigate the issue of identifiability (in the sense of~\cref{def:Iden}) in model~\eqref{mdl} when $\Pi = \logi$ or $\probit$ and $\F =\Fga$ or $\Fde$. The following result states that under some assumptions on $\mathcal{X}$ and $m$, the probability measure $P_{(\pi^*,F_1^*)}$, where $\pi^*\in \logi$ or $\probit$, $F_1^*\in \Fga$ or $\Fde$, is identifiable as long as $\pi^*(\cdot)$ is not a constant function for $m$-a.e.~$x$ and $F_1^*\neq F_0$. 

\begin{lemma}\label{ide}
	Consider the class of distributions $\Ps(\Pi, \F)$, with $\Pi\defeq \Pig$ where $g(x)=(1+\exp(-x))^{-1}$ or $g(x)=\Phi(x)$, and $\mathcal{F}\defeq \Fga$ or $\Fde$. Suppose that the set $\samp$ contains a non-empty open subset $\samp'$ of $\R^p$ such that the probability measure $m$ assigns strictly positive probability to every open ball contained in $\samp'$. Assume that $F_0\neq F_1^*$ and $\pi^* \in  \Pig$ is given by  $\pi^*(x)\defeq g(\beta_0^*+(\beta^*)^\top x)\mbox{ for } x \in \samp$ and some $(\beta_0^*,\beta^*) \in \R\times\R^p$. Then $P_{(\pi^*, F_1^*)}$ is identifiable if $\beta^* \ne 0$.  
\end{lemma}

It is worth noting that the assumption on $m$ in~\cref{ide} --- namely, there exists an open set $\samp'$ such that $m$ assigns positive probability to every non-empty open subset of
$\samp'$, is not very stringent as any absolutely continuous (with respect to the Lebesgue measure) distribution satisfies this. The other key assumption in Lemma \ref{ide} is that $\beta^* \ne 0$. This means that if the covariates are relevant (i.e., $\beta^* \ne 0$), then identifiability is restored; compare this with the two-groups model (which corresponds to $\beta^* = 0$) in which case we already know that~\eqref{eq:TwoComp} is not identifiable.

However, the way Lemma~\ref{ide} has been stated, it may
not accommodate all discrete covariates alongside the test
statistics ($p$-values or $z$-scores).~\cref{revide} (see~\cref{sec:idenmore} in the appendix) is aimed at addressing this issue. In the appendix, see~\cref{logitnoniden}, we present a simple example which shows that in the presence of discrete covariates, without certain additional assumptions, model~\eqref{mdl} may fail to be identifiable.



\section{(Nonparametric) Maximum likelihood estimation}\label{NPMLE}

In this section we propose and discuss our main estimation strategy --- maximum likelihood --- for estimating the unknown parameters in model~\eqref{mdl}, and state our main theoretical result on the estimation accuracy of our proposed estimators. We will assume in this section that every $F \in \F$ admits a probability density on $\R$ and will denote the class of probability densities corresponding to DFs in $\F$ by $\mathfrak{F}$. Our main examples for $\F$ will be $\Fga$ and $\Fde$; we have already seen that these classes arise naturally in multiple testing problems. Our examples for $\Pi$ will be $\Picon$, $\Pig$ and $\Pide$. Further, we will denote by $\dga$ and $\dde$ the classes of densities corresponding to $\Fga$ and $\Fde$ respectively. As we will show, the nonparametric classes $\dga$ and $\dde$ lend themselves to tuning parameter-free estimation through the method of maximum likelihood. Further, for estimation in the class $\dga$ we establish an almost parametric rate of convergence of the MLE (see~\cref{sr1}).

\subsection{Maximum likelihood estimation}
Let us denote by $f_1^*$ the unknown density of $F_1^*$. This reduces model~\eqref{mdl} to
\begin{equation}\label{denmodel}
Y | X = x \sim \pi^*(x) f^*_1 + (1 - \pi^*(x)) f_0 ~~~~ \text{ and } ~~ X \sim m,
\end{equation}
where $f_0$ is a known density (corresponding to the DF $F_0$), and $\pi^*(\cdot) \in \Pi$ and $f_1^* \in \mathfrak{F}$ are the unknown parameters of interest. Here, we discuss estimation of $(\pi^*,f_1^*)$ based on the principle of maximum likelihood.  For any $\pi \in \Pi, f_1 \in \mathfrak{F}$, let us denote the normalized log-likelihood at $(\pi, f_1)$, up to a constant not depending on the parameters, by
\begin{equation}\label{fulik}
\ell(\pi, f_1) \defeq \frac{1}{n} \sum_{i=1}^n \log \Big(\pi(X_i) f_1(Y_i) + (1-\pi(X_i))f_0(Y_i)\Big)
\end{equation}
and consider the MLE
\begin{equation} \label{eq:full_MLE}
\big( \hat{\pi}, \hat{f}_1 \big) \defeq \argmax_{\pi \in \Pi, f_1 \in
	\mathfrak{F}} \ell(\pi, f_1).
\end{equation}
As $\mathfrak{F}$ and $\Pi$ can be nonparametric classes of functions, the estimator $(\hat \pi, \hat f_1)$ can be thought of as the nonparametric (NP) MLE in model~\eqref{denmodel}. 
However, the optimization problem in \eqref{eq:full_MLE} is often non-convex which makes it difficult to guarantee the convergence of algorithms to global maximizers. To bypass this issue, we define another class of estimators: call any estimator $(\hat{\pi}^A, \hat{f}_1^A)$ satisfying
\begin{equation}\label{appMLE}
\prod\limits_{i=1}^n\frac{\left(1-\hat{\pi}^A(X_i)\right)f_0(Y_i)+\hat{\pi}^A(X_i)\hat{f}_1^A(Y_i)}{(1-\pi^*(X_i))f_0(Y_i)+\pi^*(X_i)f^*_1(Y_i)}\geq 1.
\end{equation}
an {\it approximate NPMLE} (AMLE). In other words, $(\hat{\pi}^A(\cdot),\hat{f}_1^A)$ is an AMLE if it yields a higher likelihood (as in~\eqref{fulik}) compared to the true model parameters $(\pi^*(\cdot),f_1^*)$. 

\subsection{Gaussian location mixtures}\label{sec:GaussMix}
Let us specialize to the case where $f_0$ is standard normal, $f_1^* \in \dga$ and $\pi^* \in \Pi$
for some class of functions $\Pi$. Note that this setting has received a lot of attention in the multiple testing literature (see~\cite{ScottEtAl15}). In the following discussion, we quantify the Hellinger accuracy of any AMLE in estimating $(\pi^*, f_1^*)$. As is common in regression problems, we state our results conditional on the
covariates $X_1, \dots, X_n$. For each $i = 1, \dots, n$, and any  $\tilde{\pi}\in\Pi$, $\tilde{f}_1\in \dga$, define $h_i^2\big((\tilde{\pi}, \tilde{f}_1), (\pi^*, f_1^*) \big)$ as
\begin{equation*}
\int\left(\sqrt{(1 - \tilde{\pi}(X_i))f_0(y) + \tilde{\pi}(X_i) \tilde{f}_1(y)
} - \sqrt{(1 - \pi^*(X_i))f_0(y) + \pi^*(X_i) {f}^*_1(y)
} \right)^2 dy. 
\end{equation*}
Thus, $h_i^2\big((\hat{\pi},\hat{f}_1),(\pi^*,f_1^*)\big)$ denotes the squared Hellinger distance between the true and estimated conditional density of $Y_i$ given $X_i$. Our loss function will be the average of $h_i^2$, for $i = 1,
\dots, n$: 
\begin{equation}\label{avhel}
\Di^2 \left((\tilde{\pi}, \tilde{f}_1) , (\pi^*, f_1^*) \right) \defeq
\frac{1}{n} \sum_{i=1}^n h_i^2 \left((\tilde{\pi}, \tilde{f}_1), (\pi^*,
f_1^*) \right). 
\end{equation}
Our main result below gives a nonasymptotic finite sample upper
bound on \newline $\Di ((\tilde{\pi}, \tilde{f}_1) , (\pi^*, f_1^*))$ conditional on the covariates $X_1, \dots, X_n$. The bound will involve the complexity of the class $\Pi$ as measured through covering numbers and metric entropy; see \citep[Chapter 2, pp.~83-86]{van1996weak} for the definitions. 

\begin{theorem}\label{sr1}
	Suppose that the data $(X_1, Y_1), \dots, (X_n, Y_n)$ are drawn from
	model \eqref{denmodel} for some $\pi^* \in \Pi$ and $f_1^* \in \dga$
	which can be written as 
	$f_1^*(x) = \int \phi(x - u) dG^*(u),  ~ x \in \R$,
	for some probability measure $G^*$ that is supported on $[-M, M]$ for
	some $M > 0$, where $\phi(\cdot)$ denotes the standard normal density. Also let $M^* \defeq \max(M, \sqrt{\log n})$. 
	Define the sequence $\{\epsilon_n\}$ as
	\begin{equation*}
	\epsilon^2_n \defeq n^{-1}\max \left(M^* (\log n)^{3/2},
	\inf_{\gamma > 0} \left\{n \sqrt{\gamma} M^* + H(\gamma, \Pi_n,
	L^{\infty}) \right\} \right)
	\end{equation*}
	where $H(\gamma,\Pi_n,L^{\infty})$ is the $\gamma$-metric entropy of the class of $n$-dimensional vectors \newline $\Pi_n\defeq \{(\pi(X_1),\ldots ,\pi(X_n)):\pi\in\Pi\}$ with respect to the uniform metric. Then, given an AMLE $(\hat{\pi}^A,\hat{f}_1^A)$ for estimating $(\pi^*,f_1^*)$, there exists a universal positive constant $K$ such that for every $t \geq 1$ and $n \geq 2$, we have 
	\begin{equation}\label{sr1.eq}
	\P  \left\{\Di \left((\hat{\pi}^A, \hat{f}_1^A), (\pi^*, f_1^*) \right) \geq
	t K \epsilon_n \biggr \vert X_1, \dots, X_n \right\} \leq 2 n^{-t^2}.  \vspace{-0.1in}
	\end{equation}
	Moreover, there exists a universal positive constant $C$ such that for
	every $n \geq 2$, we have \vspace{-0.1in}
	\begin{equation}\label{expe.eq}
	\E \left[\Di^2 \left((\hat{\pi}^A, \hat{f}_1^A), (\pi^*, f_1^*)
	\right) \biggr \vert X_1, \dots, X_n \right] \leq 
	C \epsilon_n^2. 
	\end{equation}
\end{theorem}

\begin{remark}
	Note that $(\hat{\pi},\hat{f}_1)$ as defined in \eqref{eq:full_MLE} clearly satisfies \eqref{appMLE} and thus \cref{sr1} implies that \eqref{sr1.eq} and \eqref{expe.eq} are true with $(\hat{\pi}^A,\hat{f}_1^A)$ replaced by $(\hat{\pi},\hat{f}_1)$.
\end{remark}

\begin{remark}\label{nonconv}
	The optimization problem in \eqref{eq:full_MLE} is non-convex and thus there may be multiple local maxima. Consequently, our proposed algorithms (see~\cref{sec:EM-Gauss}) do not guarantee convergence to a global maximizer. Therefore, \cref{sr1} is of particular importance (more generally useful in estimation involving non-convex optimization problems) as it establishes finite sample risk bounds for any AMLE. Moreover, our simulations in \cref{sec:sim} and~\cref{appMLEsim} (in the appendix) illustrate that our proposed algorithms almost always yield estimates that are AMLEs.
\end{remark}
The above theorem might look a bit abstract at first glance. Let us consider a typical function class $\Pi$ to demonstrate the conclusions of \cref{sr1}. Let $\Pi$ be given by a generalized linear model, i.e.,
each function $\pi \in \Pi$ is of the form $x \mapsto g(x^\top  \beta)$ for some $\beta \in \R^p$ and known link function $g(\cdot)$. Then~\cref{sr1} gives a parametric rate of convergence $p/n$, up to a logarithmic
factor of $n$, in the average Hellinger metric (see~\eqref{avhel}), for all standard choices of $g(\cdot)$. This is illustrated in the subsequent corollary and remarks.

\begin{corollary}\label{sr.li}
	Suppose $g : \R \rightarrow [0, 1]$ is a fixed link function that is Lipschitz with some constant $L > 0$, i.e., $|g(z_1) - g(z_2)| \leq L |z_1 - z_2|$, for all $z_1,z_2 \in \R.$ Suppose that the covariate space $\samp$ is contained in a
	$p$-dimensional Euclidean ball of radius $T$ and that the function
	class $\Pi$ is given by $\{\pi_{\beta}: \beta \in \R^p, \|\beta\| \leq
	R\}$ for some $R > 0$ where $\pi_{\beta}(x) \defeq g(x^\top \beta)$ for $x \in \samp$. Then, under the same assumptions on $f_1^*$ as in Theorem \ref{sr1}, inequalities \eqref{sr1.eq}  and \eqref{expe.eq} both hold with $$\epsilon_n^2=\frac{1}{n} \max \left(M^* (\log n)^{3/2},M^* + p \log \left(1
	+ 2 L T R n^2 \right) \right).$$ The quantities $L, M, R$ and $T$ can be taken to be either fixed or changing with $n$.
\end{corollary}

\begin{remark}\label{fcor}
	The most common example of the link function $g$ in Theorem \ref{sr1}
	is the logistic link given by $g(z) \defeq (1 + e^{-z})^{-1}$, for $z \in \R$. This function $g$ is clearly Lipschitz with constant $L = 1$  because $|g'(z)| =  e^z (1 + e^z)^{-2} \leq 1$ for every $z \in \R$. Another example of the link function $g(\cdot)$ in Theorem~\ref{sr1} is the probit link given by
	$g(z) \defeq \Phi(z)$ for $z \in \R$. This function $g$ is also Lipschitz with constant $L = (2\pi)^{-1/2}$ because
	$|g'(z)| = \frac{1}{\sqrt{2\pi}}\exp (-{z^2}/{2}) \leq (2\pi)^{-1/2}$, for every $z \in \R$.  Both the logit and probit links arise from symmetric (about $0$) densities which may sometimes be undesirable, specially in some survival models. As a result, often the complementary log-log link is recommended in survival models; e.g.,~\cite{jenkins1995easy}. In this case $g(z) \defeq 1-\exp\left(-\exp(z)\right)$, $z \in \R$. Observe that $|g'(z)| = \exp(-\exp(z))\exp(z)\leq 1$. 
	Therefore Corollary \ref{sr.li} applies to all the three link functions above. 
\end{remark}

\begin{remark}
	If $L, M, R$ and $T$ are all constant, then the rate $\epsilon_n$
	given by Corollary \ref{sr.li} is parametric up to logarithmic factors
	in $n$. 
\end{remark}
In the following section, we describe an iterative approach based on the expectation-maximization (EM) algorithm (\citet{dempster1977maximum}, also see~\cite{Lange16},~\cite{McPeel00}) to compute the MLE described in~\eqref{eq:full_MLE}. We had also looked into an  alternative maximization based approach for solving~\eqref{eq:full_MLE}. Our simulations revealed that the EM algorithm significantly and consistently outperformed the alternative maximization scheme. Hence we only describe the details of the EM based algorithm. 
\section{EM algorithm for joint likelihood maximization}\label{sec:EM-Gauss}
Let us first recall a familiar setting from~\cref{sec:Intro}. Consider $n$ independent but unobserved (latent) Bernoulli random variables $Z_1,Z_2,\ldots ,Z_n$ such that $\mathbb{P}(Z_i=1|X_i)=\pi^*(X_i)$ for some $\pi^*(\cdot)\in\Pi$ and suppose that the conditional densities of $(Y_i|Z_i=1,X_i)$ and $(Y_i|Z_i=0,X_i)$ are $f_1^*$ and $f_0$ respectively. The EM algorithm then, proceeds as follows. We first write down the ``complete data'' likelihood which involves the joint density of our observed data $\{(Y_i, X_i)\}_{i=1}^n$ and the latent variables $Z_1, \dots, Z_n$. Observe that the joint (complete) average log-likelihood of $(X_i, Y_i, Z_i)$, for $i = 1, \dots, n$, equals  
\begin{equation*}
\frac{1}{n} \sum_{i=1}^n \Big\{ Z_i \log
\left[\pi(X_i) f_1(Y_i) \right]  + (1 - Z_i) \log \left[(1 -
\pi(X_i)) f_0(Y_i) \right] \Big\}, 
\end{equation*}
where we have ignored some terms that do not depend on the parameters of interest. Observe that the conditional expectation of $Z_i$ given the data can be expressed as
\begin{equation}\label{conex}
\E_{\pi^*,f_1^*}[Z_i | Y_i = y, X_i = x] = \frac{\pi^*(x) f_1^*(y)}{\pi^*(x) f_1^*(y) + (1-\pi^*(x)) f_0(y)}, \quad \mbox{for } i = 1,\ldots, n. 
\end{equation}
As the random variables $Z_i$'s are unobserved, we replace them in the log-likelihood in the $E$-step of the algorithm by their conditional expectations evaluated as in~\eqref{conex} with $\pi^*(\cdot)$ and $f_1^*$ replaced by their estimates from the previous iteration; see~\cref{em} for details. The obtained expected log-likelihood function is then maximized in the M-step of the algorithm with respect to both the parameters $\pi \in \Pi$ and $f_1 \in \mathfrak{F}$. We provide the corresponding pseudo-code for the EM algorithm below.
\begin{algorithm}\label{EMimp}
	\caption{EM implementation of \eqref{eq:full_MLE}}
	\label{em}
	\begin{algorithmic}
		\STATE Input $\{(Y_i,X_i)\}_{i = 1}^n$ and initial estimates $\pi^{(0)}, f^{(0)}_1$
		\STATE $k \gets 1$ 
		\REPEAT \STATE \vspace{-0.3in}
		\begin{align*}
		\textbf{E-step:} ~ &w_i^{(k)} \gets \frac{\pi^{(k-1)}(X_i)f_1^{(k-1)}(Y_i)}{\pi^{(k-1)}(X_i)f_1^{(k-1)}(Y_i)  + \left(1 - \pi^{(k-1)}(X_i) \right)f_0(Y_i)},\;\;i=1,2,\ldots ,n.\\
		\textbf{M-step:} ~ &\pi^{(k)} \gets \hat \pi_{\text{EM}}(\mathbf{w}^{(k)}, \Pi) \qquad \mbox{and} \qquad 
		f_1^{(k)} \gets \hat f_{\text{EM}}(\mathbf{w}^{(k)}, \mathfrak{F})\\
		k \gets k+1&
		\end{align*}
		\vspace{-0.3in}
		\UNTIL convergence of $\mathbf{w}^{(k)}=(w_1^{(k)},\ldots ,w_n^{(k)}).$
	\end{algorithmic}
\end{algorithm}

In~\cref{em}, for any $\mathbf{w} = (w_1, \dots, w_n) \in [0,1]^n$,
\begin{equation} \label{piem} 
\hat \pi_{\text{EM}}(\mathbf{w}, \Pi) \defeq \argmax_{\pi \in \Pi} \frac{1}{n} \sum_{i=1}^n \left[w_i \log \pi(X_i) + (1 - w_i) \log \left(1 - \pi(X_i) \right) \right],\quad \mbox{ and }
\end{equation}
\begin{equation} \label{fem}
\hat f_{\text{EM}}(\mathbf{w}, \mathfrak{F}) \defeq \argmax_{f_1 \in \mathfrak{F}} \frac{1}{n}  \sum_{i=1}^n \left[w_i \log f_1(Y_i) + (1 - w_i) \log f_0(Y_i) \right]. 
\end{equation}
In order to implement the EM algorithm, it is necessary to solve the optimization problems \eqref{piem} and \eqref{fem}. In general, both of these problems are more tractable than \eqref{eq:full_MLE} (as explained in the next two subsections). Indeed, when the classes $\Pi$ and $\mathfrak{F}$ are convex (e.g., $\Pi\defeq \Pide$,  $\f\defeq\dga$ or $\dde$), the optimization problems \eqref{piem} and \eqref{fem} are also convex in $\pi$ and $f_1$, respectively. Further, due to the particular form of the expected log-likelihood, this joint maximization breaks into two isolated maximization problems, i.e., problems \eqref{piem} and \eqref{fem} are decoupled. Hence, solving \eqref{piem} (or \eqref{fem}) requires no knowledge of $\mathfrak{F}$ (or $\Pi$). However, as~\eqref{eq:full_MLE} is a non-convex problem we cannot guarantee the convergence of our EM algorithm to the global maximizer. Moreover, we need proper initial estimates of $(\pi^*,f_1^*)$ to start the iterative scheme in the EM algorithm (see~\cref{EMimp}). In Sections~\ref{sec:MargMeth-I} and~\ref{sec:MargMeth-II} we describe two easily implementable procedures that can be used as starting points for the EM algorithm. We now provide more specific details on the implementations of \eqref{piem} and \eqref{fem}.

\subsection{Implementation strategies for the optimization problem~\eqref{piem}} \label{imple_piem}
In the following we discuss the maximization of the expected log-likelihood function with respect to $\pi \in \Pi$. We focus on two types of $\Pi$'s: (1) when $\Pi = \Pig$ (see~\eqref{eq:parametric_link}) is parametrized by a finite-dimensional parameter, and (2) when $\Pi$ is infinite-dimensional, e.g.,~$\Pi = \Pide$, see~\cref{pinonpara} in the appendix.

\subsubsection{Parametric link function}\label{piparam} Suppose that we want to optimize~\eqref{piem} when $\Pi = \Pig$ and the known link function $g(\cdot)$ is assumed to be smooth so that $\pi_{\beta}(x)=g(x^\top\beta)$ is once differentiable with respect to $\beta$ (at every $\beta$), for every $x$. In this case we can employ various first-order iterative optimization algorithms to solve~\eqref{piem}, e.g., gradient descent or steepest descent method; see~\citet[Chapter 2]{NW06}. 
However, in this paper, we recommend using the Broyden-Fletcher-Goldfarb-Shanno (BFGS) algorithm; see e.g.,~\citet{broyden1969new},~\citet{fletcher1970new},~\citet{goldfarb1970family},~\citet{shanno1970conditioning}. This method is implemented in the \texttt{stats} package in the R language (using the command \texttt{optim}). Note that the BFGS algorithm is a quasi-Newton method; see~\citet[Chapters 3 and 8]{NW06}. It requires computing the gradients of the objective of~\eqref{piem} but instead of using the actual Hessian (like in Newton's method; see e.g.,~\citet[Chapter 9]{boyd04}), the  BFGS algorithm replaces it by an approximation. In our simulations, we found that this method was computationally faster (being a quasi-Newton method it generally requires much fewer iterations to converge) as compared to gradient descent methods. 




\subsection{Implementation of \eqref{fem}}

Let us now discuss optimization problem \eqref{fem} involving $f_1 \in \mathfrak{F}$. It is important to note that the objective function in
\eqref{fem} is concave in $f_1$. Therefore, \eqref{fem} is a convex optimization problem as long as $\mathfrak{F}$ is a convex class of densities.

Our two main examples of $\mathfrak{F}$ are the cases $\mathfrak{F} = \dga$, the class of all normal location mixtures (with unit variance), and $\mathfrak{F} = \dde$, the class of all nonincreasing densities on $[0, 1]$ (see~\cref{deninc} in the appendix). Both of these are
convex classes of densities so that \eqref{fem} becomes a convex
optimization problem. Below, we provide the details of solving
\eqref{fem} when $\mathfrak{F}=\dga$.

\subsubsection{When $\mathfrak{F} = \dga$ in~\eqref{fem}}\label{dengauss}
When $\mathfrak{F} \defeq \dga$,  optimization problem~\eqref{fem} is 
reminiscent of the NPMLE over normal
location mixtures (also referred to in the statistical literature as the Kiefer-Wolfowitz Maximum Likelihood Estimator (KWMLE); see the book-length treatments~\citet{lindsay1995mixture},~\cite{bohning2000computer} and~\citet{schlattmann2009medical}). However,~\eqref{fem} is an infinite-dimensional problem (as is the computation of the KWMLE). To find a finite-dimensional approximation to this problem we employ the following standard strategy (see~\citet{koenker2014convex}): we choose a large set $\ac \defeq \{a_1, \dots,
a_m\}$, $ m \ge 1$, and restrict our attention to densities of the form
$f(x) = \int \phi(x - u) dG(u) \in \dga$ 
where the DF $G(\cdot)$ is supported on $\ac$. Before we proceed to give the details of our algorithm for solving the above finite-dimensional problem, we first argue that such an approximation is valid (and useful). While the exact properties of this finite-dimensional approximation are not known, some indications of its accuracy are available in the related KWMLE problem. In the case of the KWMLE, when $\ac$ is chosen to be a set of equidistant grid points spanning the range of observations with at least $\lfloor \sqrt{n}\rfloor$ elements, \citet[Theorem 1]{dicker2014nonparametric} have proved that the approximate solution is within $O_p(\log n/\sqrt{n})$ of the true density in the Hellinger metric. As noted therein, this is both close to the parametric rate $O_p(1/\sqrt{n})$, up to a logarithmic factor, and further matches the rates established for the KWMLE (without approximation) in \citet{ghosal2001entropies} and \citet{zhang2009generalized}. We believe that a similar correspondence remains valid in the case of problem \eqref{fem}. 

Once the approximating set $\{a_1,\ldots ,a_m\}$ is chosen, problem~\eqref{fem} can now be solved via the following finite-dimensional convex optimization problem:
\begin{equation}\label{fdem}
\hat{\mathbf{p}}(\mathbf{w}) \defeq \argmax_{\mathbf{p} \in \mathcal{P}_m} \frac{1}{n} \sum_{i=1}^n w_i \log \Big[\sum_{j=1}^m p_j \phi(Y_i - a_j) \Big]
\end{equation}
where
\begin{equation*}
\mathcal{P}_m \defeq \Big\{ (p_1, \dots, p_m) :~ p_j \ge 0 ~~ \text{for all } j, \sum_{j=1}^m p_j = 1 \Big\}
\end{equation*}
denotes the $m$-dimensional probability simplex. Once we have computed a solution $\hat{\mathbf{p}}(\mathbf{w})$ to the above problem, we take 
$\hat f_{\text{EM}}(\mathbf{w}, \mathfrak{\Fga}) \defeq \sum_{j=1}^m \hat{p}_j(\mathbf{w}) \phi(x - a_j)$.
In practice we choose the atoms $\{a_j\}_{j=1}^m$ along a regular grid in the range of the data $\{Y_i\}_{i=1}^n$. We find the choice $m = \max\{100,  \sqrt{n} \}$ to be satisfactory in our numerical experiments. In the following we mention the optimization algorithm to solve \eqref{fdem}. We generalize the procedure laid down in \citet{koenker2014convex} using the optimization suite \citet{mosek2010mosek} via the R package \texttt{Rmosek} \cite{rmosekpackage}. More specifically, we solve:
\begin{align*}
\hat{\mathbf{p}}(\mathbf{w}) &\defeq \argmax_{\mathbf{v}, \mathbf{p}} \frac{1}{n} \sum_{i=1}^n w_i \log v_i \\
\text{subject to } ~~~~~~ v_i &= \sum_{j=1}^m p_j \phi(Y_i - a_j) ~~~~~~ i = 1, \dots, n; \qquad 
\mathbf{p}  \in \mathcal{P}_m, \; \mathbf{v} \in \R^n,
\end{align*}
using the \texttt{Separable Convex Optimization} interface available in \citet{mosek2010mosek}. This procedure has been implemented in the accompanying R package NPMLEmix. 
As the criterion function in~\eqref{fdem} is finite-dimensional, convex and smooth in $\mathbf{p}$, we can also employ the projected gradient descent algorithm (see~\cite{CP11}), a first order method, to solve this problem; see~\cref{sec:A-PGD} in the appendix for the details.

\section{Marginal methods}\label{sec:MargMeth}
Maximizing the joint likelihood (of $(X,Y)$; see~\eqref{fulik}) can be computationally expensive, especially when dealing with nonparametric classes for $\Pi$ or $\f$. Further, the EM algorithm proposed in~\cref{sec:EM-Gauss} to find the MLEs is iterative in nature and can get stuck at a local maxima, different from the global maximizer (as the underlying optimization problem is non-convex). In this subsection we propose two novel {\it marginal} methods that bypass the joint estimation of $\pi^*(\cdot)$ and $f_1^*$. As the name suggests, these methods do not deal with a joint maximization problem; instead they use properties of model~\eqref{denmodel} to isolate each of the parameters and estimate them separately. Both the proposed methods are conceptually simple and easy to implement. They also provide good estimates for the true parameters in model~\eqref{denmodel}; in~\cref{sec:sim}, we compare their performance to FDR regression (see~\cite{ScottEtAl15}). Our marginal methods can also be used to obtain preliminary estimators of $\pi^*(\cdot)$ and $f_1^*$ which can then be chosen as starting points for the EM algorithm outlined in Section~\ref{sec:EM-Gauss} (see~\cref{em}).

\subsection{Marginal method -- I}\label{sec:MargMeth-I}
To motivate this decoupled approach, first observe that the marginal distribution of $Y$ in model~\eqref{mdl} has the form~\eqref{eq:TwoComp} where $ \bar \pi \defeq \E_{X \sim m} [\pi(X)]$, which is the standard two-groups model with unknown $F_1^*$ and $\bar \pi$. The above observation can be used to directly estimate $f_1^*$ (the density of $F_1^*$), bypassing the estimation of $\pi^*(\cdot)$. Observe that, if $\bar \pi \equiv \alpha$ were known (assume $\overline{\pi}>0$), estimation of $f_1^* \in \mathfrak{F}$ could be accomplished by maximizing the marginal likelihood of the $Y_i$'s, i.e.,  
\begin{equation}\label{eq:Mar_f_1}
\hat f_1^{(\alpha)} \defeq  \argmax_{f_1 \in \mathfrak{F}} \frac{1}{n}
\sum_{i=1}^n \log \Big(\alpha f_1(Y_i) + (1-\alpha) f_0(Y_i)\Big). 
\end{equation}
The above optimization problem is indeed computationally more tractable --- note that for function classes $\mathfrak{F}$ that are convex (e.g., $\dga$ and $\dde$)~\eqref{eq:Mar_f_1} is a convex program and can be solved efficiently. For instance, we may directly use the convex optimization technique outlined in Section~\ref{dengauss} to solve~\eqref{eq:Mar_f_1} if $\f=\dga$. 

Once we obtain an estimator $\hat f_1^{(\alpha)}$ of $f_1^*$, we can maximize the joint log-likelihood just as a function of $\pi(\cdot) \in \Pi$ to obtain
\begin{equation}\label{eq:Mar-I-pi}
\hat \pi^{(\alpha)} \defeq \argmax_{\pi \in \Pi} \frac{1}{n}
\sum_{i=1}^n \log \left(\pi(X_i) \hat f_1^{(\alpha)}(Y_i) + (1- \pi(X_i)) f_0(Y_i) \right).  
\end{equation}
Problem~\eqref{eq:Mar-I-pi} is also tractable for a variety of choices of $\Pi$. In particular, if $\Pi\defeq\Pide$, one can once again use the convex optimization strategy discussed in Section~\ref{pinonpara} in the appendix, whereas if $\Pi\defeq\logi$, we can use the BFGS method discussed in Section~\ref{piparam}. Based on the above discussion, we end up with one-step estimators $\hat \pi^{(\alpha)}$ and $\hat{f}_1^{(\alpha)}$ of $\pi^*$ and $f_1^*$ (respectively), if we knew the value of $\bar \pi \equiv \alpha$. 

In practice $\bar \pi$ may not be known, in which case we will need to estimate $\bar \pi$ from the data to estimate $f_1^*$ using~\eqref{eq:Mar_f_1}. As we are now in the well-known two-groups model, there are many estimators available for $\bar \pi$; see e.g.,~\cite{Storey02},~\cite{tang2005bayes},~\cite{Efron10},~\cite{ patra2016estimation},~\cite{LangaasEtAl05},~\cite{zhang1990fourier}. However, the estimation of $\bar \pi$ is a difficult problem when $\mathfrak{F}$ is nonparametric (e.g., when $\mathfrak{F} = \dga$ or $\dde$) and there is no known $\sqrt{n}$-consistent estimator of $\bar \pi$ with finite variance; see e.g.,~\cite{NM14}. Note that, when $f_0 \in \mathfrak{F}$ and $\mathfrak{F}$ is convex (e.g., $f_0(\cdot)=\phi(\cdot)$, $\f=\dga$), we cannot obtain a consistent estimator of $\bar \pi$ by maximizing~\eqref{eq:Mar_f_1} jointly with respect to $f_1$ and $\alpha$ (as the likelihood in such a case will always be maximized at $\alpha =1$). In fact, $\bar \pi$ is a parameter for which a lower (honest) confidence bound can be provided easily (see e.g.,~\cite{GW04},~\cite{MeinRice06},~\cite{patra2016estimation}) but an upper confidence bound is difficult to obtain; see e.g.,~\cite{Donoho88} for a unified treatment of such `one-sided' parameters. 

The methods for estimating $\bar \pi$ cited above do not use the covariate information available in our model. Based on extensive simulation studies (see \cref{sec:sim}), we believe that incorporating covariate information in the estimation of $\overline{\pi}$ can lead to a better estimator. In the following display, we propose a possible strategy to estimate $\bar \pi$ that uses the joint likelihood of the available data. 
Note that as defined, both~\eqref{eq:Mar_f_1} and~\eqref{eq:Mar-I-pi}, depend on $\alpha \in (0,1]$. We can now consider the ``profiled'' one-dimensional MLE of $\bar \pi$:  
\begin{equation}\label{eq:MLE_pi_0}
\hat {\bar \pi} = \arg \max_{\alpha \in (0,1]} \frac{1}{n} \sum_{i=1}^n \log \Big(\hat \pi^{(\alpha)}(X_i) \hat f_1^{(\alpha)}(Y_i)  + (1- \hat \pi^{(\alpha)}(X_i)) f_0(Y_i)\Big),
\end{equation}
where $\hat f_1^{(\alpha)}(\cdot)$ is defined in~\eqref{eq:Mar_f_1}, and $\hat \pi^{(\alpha)}(\cdot)$ is defined in~\eqref{eq:Mar-I-pi}. To solve problem~\eqref{eq:MLE_pi_0}, we recommend a grid search over the unit interval $(0,1]$. One may also start with a standard estimator of $\overline{\pi}$ (using any of the methods from the references cited above), and restrict the grid search to a suitably small neighborhood of the initial estimate.

Below we state a theoretical result which gives finite sample risk bounds for the estimated marginal density of $Y$. In fact, the following can be interpreted as an estimation accuracy result in the two-groups model (without covariates), i.e., model~\eqref{eq:TwoComp} when an upper bound for the signal proportion ($\bar \pi$) is known.

\begin{theorem}\label{marg-1}
	Suppose that the data $(X_1, Y_1), \dots, (X_n, Y_n)$ are drawn from model \eqref{denmodel} for some $\pi^* \in \Pi$ and $f_1^* \in \mathfrak{F}=\dga$ which can be written as $f_1^*(x) = \int \phi(x - u) dG^*(u)$, for $x \in \R$, and for some probability measure $G^*$ supported on $[-M, M]$ (for some $M>0$). If $\overline{\pi}\leq \alpha\leq 1$, we have
	$$\mathbb{E}\left[h^2\left(\left(\alpha,\hat{f}_1^{(\alpha)}\right),\left(\overline{\pi},f_1^*\right)\right)\right]\leq \frac{C}{n}\max{\left(M^*(\log{n})^{3/2},M^*+\log{n}\right)}$$
	where $C$ is a universal constant, $M^*=\max{(M,\sqrt{\log{n}})}$ and 
	\begin{small}
		\begin{equation*}			h^2\left(\left(\alpha,\hat{f}_1^{(\alpha)}\right),\left(\overline{\pi},f_1^*\right)\right) \defeq \int
		\left(\sqrt{(1 - \alpha)f_0(y) + \alpha \hat{f}_1^{(\alpha)}(y)
		} - \sqrt{(1 - \overline{\pi})f_0(y) + \overline{\pi} {f}^*_1(y)
		} \right)^2 dy. 
		\end{equation*}
	\end{small}		
\end{theorem}

\begin{remark}
	If $M$ does not change with $n$ in \cref{marg-1}, then, for $n\geq 3$, we have $\mathbb{E}\left[h^2\left(\left(\alpha,\hat{f}_1^{(\alpha)}\right),\left(\overline{\pi},f_1^*\right)\right)\right]\leq C'(\log{n})^2/n$, where $C'$ is a constant free of $n$ but depending on $M$. In particular $C'$ can be taken as $C(M+1)$.
\end{remark}

\begin{remark}\label{rem:sim}
	The method recommended in~\cite{ScottEtAl15} (FDRreg) and Marginal - I may seem somewhat similar. In~\cite{ScottEtAl15}, the authors use the predictive recursion algorithm to estimate $\bar\pi$ and $f_1^*$ from the two-groups model~\eqref{eq:TwoComp}; see~\cite[Appendix A]{ScottEtAl15}. This step does not utilize covariate information. The subsequent estimation of $\pi^*(\cdot) \in\logi$ is done via the EM algorithm. For our Marginal - I, obtaining $\hat{f}_1^{(\alpha)}$ (for a fixed $\alpha$) also does not utilize covariate information (see~\eqref{eq:Mar_f_1}). This, in spirit, is similar to the estimation of $f_1^*$ in FDRreg, if $\alpha$ were equal to (or close to) $\bar\pi$. However, it must be noted that our proposal for obtaining $\hat{f}_1^{(\alpha)}$ is not based on predictive recursion (see~\cref{dengauss} for our recommended algorithm). A more important difference between FDRreg and Marginal - I is that the latter estimates $\bar\pi$ by solving the one-dimensional ``profiled" likelihood maximization problem as stated in~\eqref{eq:MLE_pi_0} and hence utilizes information provided by the covariates. This ensures that our final estimator of $f_1^*$ indeed takes account of the covariates yielding more accurate estimators, see~\cref{sec:estim} for a comparison with FDRreg using synthetic data.
\end{remark}

\subsection{Marginal method -- II}\label{sec:MargMeth-II}
In the previous marginal procedure we isolated the effect of the unknown density $f_1^*$ and used the marginal distribution of $Y$ to estimate $f_1^*$. In this subsection we describe a procedure that targets the estimation of $\pi^*(\cdot)$ first. Observe that the regression function of $Y$ on $X$ is
\begin{equation}\label{eq:E_YX}
\E (Y|X = x) = (1 - \pi^*(x)) \mu_0 + \pi^*(x) \mu^*,
\end{equation} 
where $\mu_0 \defeq \E_{Y \sim F_0} [Y]$ and  $\mu^*\defeq \E_{Y \sim F_1^*} [Y]$. Here $\mu_0$ is known (as $F_0$ is known) but $\mu^*$ is unknown. Thus, the regression function isolates the effect of $\pi^*(\cdot)$, modulo the estimation of $\mu^*$. If $\mu^* \ne \mu_0$ and $\pi^*(\cdot)$ is not a constant function,~\eqref{eq:E_YX} poses a nonlinear regression problem and we can use the method of least squares to estimate $(\pi^*, \mu^*)$:
\begin{equation}\label{eq:E_YX-LSE}
(\hat \pi,\hat \mu) \defeq \argmin_{\pi \in \Pi, \mu \in \R} \sum_{i = 1}^n \Big( Y_i - \mu_0 - \pi(X_i)(\mu  - \mu_0)  \Big)^2.
\end{equation} 
Once $\hat \pi(\cdot)$ is estimated, we can use the joint likelihood of $(X,Y)$ to estimate $f_1^*$ (plugging in the value of $\hat \pi(\cdot)$): 
\begin{equation}\label{eq:E_YX-f_s}
\hat f_1  \defeq \argmax_{f_1 \in \mathfrak{F}} \sum_{i = 1}^n \log \Big[ \hat \pi(X_i) f_1 (Y_i) + (1 - \hat \pi(X_i)) f_0(Y_i)\Big].
\end{equation} 
Note that the above display yields a convex program which can be solved easily to estimate $\hat f_1$.

Problems~\eqref{eq:E_YX-LSE} and~\eqref{eq:E_YX-f_s} can be solved based on the methods discussed in~\cref{imple_piem}. As the least squares problem in~\eqref{eq:E_YX-LSE} can be non-convex, we recommend fixing $\mu$ and optimizing over $\pi(\cdot)$ followed by a grid search in the space of $\mu$. In the following we discuss in detail the estimation of $\pi^*$. 
\subsubsection{Estimating a parametric $\pi^*(\cdot)$}
Suppose that $\pi^*(\cdot)\in\Pi$ is a parametric class of functions indexed by a subset of $\R^p$ (with $p < n$ fixed), e.g., $\Pi = \{\pi: \pi_{\beta}(x) =  (1 + e^{-\beta^\top x})^{-1}$, $\beta \in \R^p\}$. Thus the unknown parameters now are $\beta^*\in\R^p$ (as $\Pi$ is now parametrized by $\beta$) and $f_1^*\in\f$.
Without loss of generality, in this subsection we assume that $f_0$ has known mean $0$ (as otherwise we can just subtract the known constant $\mu_0$ from $Y$ and work with this ``centered" $Y$). Consider the following least squares estimator (LSE) of $\theta^* \defeq (\beta^*,\mu^*)$ (cf.~\eqref{eq:E_YX-LSE}):
\begin{equation}\label{eq:Beta-LSE}
\hat \theta_n \equiv (\hat \beta_n,\hat \mu_n) \defeq \argmin_{\theta = (\beta, \mu) \in \Theta \subset \R^{p+1}} \sum_{i = 1}^n (Y_i - \mu \pi_{\beta}(X_i))^2.
\end{equation}
An application of~\cite[Theorem 5.23]{van2000asymptotic} then yields the following result. For the sake of completeness, we present a proof of the above result in~\cref{cor:Mar-2pf} (see the appendix).

\begin{theorem}\label{cor:Mar-2}
	Suppose that $(X,Y)$ has a joint distribution described by~\eqref{denmodel} where $\pi^*(\cdot)\in \Pig$, i.e., $\pi^*\equiv \pi^*_{\beta^*}(x)=g(x^\top \beta^*)$. Also assume that $Y$ and each component of $X$ has a finite fourth moment. Let $g(\cdot)$ be thrice differentiable and the $i^{th}$ derivative of $g(\cdot)$ satisfy $\sup_{\lambda\in\mathbb{R}}|g^{(i)}(\lambda)|\leq c_i$ for some constants $c_i$, $i=0,1,2,3$ (Note that $c_0$ can be chosen as $1$). Further assume $\Theta\subset\mathbb{R}^{p+1}$ (which appears in~\eqref{eq:Beta-LSE}) is a fixed compact set and $\theta^* \equiv (\beta^*,\mu^*)\in \Theta$ is identifiable from~\eqref{eq:E_YX} in the sense that $\theta \ne \theta^*$ implies that $\mu g(X^\top\beta) \ne \mu^* g(X^\top\beta^*)$ with positive probability under the measure $m$. Then, the LSE $\hat \theta_n$, defined in~\eqref{eq:Beta-LSE}, is $\sqrt{n}$-consistent, if $\theta^*$ belongs to the interior of $\Theta$, and has an asymptotically normal limit given by
	$\sqrt{n} (\hat \theta_n - \theta^*) \overset{d}{\to} \N(0, V_{\theta^*}^{-1} (\E[\dot m_{\theta^*} \dot m_{\theta^*}^\top]) V_{\theta^*}^{-1} )$ as $n \to \infty$. Here $m_{\theta}(X,Y)\defeq-(Y-\mu\pi^*_{\beta}(X))^2$, $\dot m_{\theta}=\nabla_{\theta} m_{\theta}$ and $V_{\theta}\defeq \mathbb{E}[\nabla_{\theta}^2 m_{\theta}(X,Y)]$ is assumed to be invertible at $\theta^*$.
\end{theorem}

\begin{corollary}
	Recall the choices for $g(\cdot)$ in~\cref{fcor}: $g(z)=(1+\exp(-z))^{-1}$, $g(z)=\Phi(z)$ and $g(z)=1-\exp(-\exp(z))$. It is straight-forward to check that all these $3$ functions satisfy the assumptions on $g(\cdot)$ in~\cref{cor:Mar-2}. As a result, the asymptotic normality of the LSE $\hat{\theta}_n$ (stated in~\cref{cor:Mar-2}) holds for these $3$ choices of  $g(\cdot)$.
\end{corollary}

\section{Are the covariates at all important?} \label{sec:hypotest}
Till now we have focused on the estimation of parameters assuming that model~\eqref{denmodel} holds. A basic and important question that we have not yet addressed is: ``Do the covariates provide any information at all on the distribution of $Y$"? Put another way, we must first check that model~\eqref{eq:TwoComp} is inadequate. Only then does it make sense to model the dependence between $Y$ and $X$, as in~\eqref{denmodel}. In statistical parlance, this reduces to testing for {\it independence} between $X$ and $Y$. Observe that under the independence assumption (of $X$ and $Y$) model~\eqref{denmodel} reduces to model~\eqref{eq:TwoComp}; it necessarily implies that the function $\pi^*(\cdot)$ is a constant function.

There are several ways to test the hypothesis of statistical independence between $X$ and $Y$, given a random sample from their joint distribution; see e.g.,~\citet{BlumEtAl61},~\citet{Taskinen05},~\citet{Szekely07} and~\citet{gretton2007kernel}. The one that we prefer, mostly because of its simplicity and applicability, is the notion of ``distance covariance" introduced by~\citet{Szekely07}; also see~\citet{Szekely09}. We discuss this idea in further details in the appendix,~\cref{sec:hypotestcont}.
\section{Simulations}\label{sec:sim}
In this section we describe results from extensive simulation studies that illustrate the usefulness of~\eqref{denmodel} over the two-groups model and the utility of our proposed methodology. Further, we discuss the implementation of our methods and compare their performances with the closest existing method in the literature, namely FDRreg in~\cite{ScottEtAl15}. In our simulations, we confine ourselves to $\pi^*(\cdot)\in\logi$ and $f_1^*\in\dga$ (as in~\cite{ScottEtAl15}). In fact, most of our simulation settings are borrowed from~\cite{ScottEtAl15}.


\subsection{Estimation of \texorpdfstring{$f_1^*$} ~:~role of covariates}\label{sec:rolecovinit}
Recall models~\eqref{eq:TwoComp} and~\eqref{denmodel}. The estimation of $f_1^*$ in FDRreg by~\citet[Appendix A]{ScottEtAl15} is performed via the method of predictive recursion (see~\cite{newton2002nonparametric}) based on the marginal distribution of $Y_i$'s, as in model~\eqref{eq:TwoComp}, as opposed to the joint distribution of $(X_i,Y_i)$'s, as in model~\eqref{denmodel}. While the approach of~\cite{ScottEtAl15} may seem simpler, it is natural to ask if the two-groups model is enough to accurately estimate $f_1^*$? Or do we incur a loss of information by restricting ourselves to the two-groups model? In order to address this question, we performed extensive simulations where we generate data from model~\eqref{denmodel} and compute the MLE of $f_1^*$ using: (i) the marginal distribution of $Y_i$'s assuming $\bar \pi$ is known, and (ii) the joint distribution of $(X_i,Y_i)$'s assuming $\pi^*(\cdot)$ is known. These simulations demonstrate that using the information present in the covariates leads to significantly more accurate (in terms of Hellinger distance) estimates of $f_1^*$. For a detailed description of these simulations, the algorithms used to compute the MLEs, and relevant plots, see~\cref{sec:rolecov} in the appendix. 

\subsection{Estimation of parameters and multiple hypotheses testing}\label{sec:estim}
We now document an extensive set of simulations investigating the performance of all our proposed methods: (i) the first marginal method based on profile likelihood maximization (Marginal - I), (ii) the second marginal method based on nonlinear regression (Marginal - II), and (iii) the full MLE (fMLE) implemented via the EM algorithm (see the end of this section for a discussion on the initialization scheme). We also compare our methods to FDRreg, proposed in \cite{ScottEtAl15}. In order to evaluate the performance of these methods we compute six different metrics, described below. We use $(\check \pi, \check f_1)$ to denote any generic estimator of $(\pi^*,f_1^*)$. We also use $\check \lfdr_i$ to denote any generic estimator of $\lfdr^*_i$, where $\lfdr^*_i$, the lFDR of the $i^{th}$ observation, is defined as one minus the right hand side in~\eqref{conex} (we discuss the importance of the vector $(\lfdr_1^*,\ldots ,\lfdr^*_n)$ in greater detail in~\cref{sec:testing} of our appendix). The first three metrics below are directly aimed at understanding the accuracy in the estimation of $\pi^*$, $f_1^*$ and the lFDR's respectively.
\begin{itemize}
	\item[(a)] Root mean squared error (RMSE) in estimating the vector $(\pi^*(X_1), \dots, \pi^*(X_n))$:\newline $
	\left[ \frac{1}{n} \sum_{i = 1}^n \E (\check \pi(X_i) - \pi^*(X_i))^2 \right]^{1/2}.$
	
	\item[(b)] RMSE in estimating the vector $(f_1^*(Y_1), \dots, f_1^*(Y_n))$: $
	\left[ \frac{1}{n} \sum_{i = 1}^n \E (\check f_1(Y_i) - f_1^*(Y_i))^2 \right]^{1/2}.$
	
	\item[(c)] RMSE in estimating the vector $(\lfdr^*_1, \dots, \lfdr^*_n)$: $
	\left[\frac{1}{n} \sum_{i = 1}^n \E (\check \lfdr_i - \lfdr^*_i)^2\right]^{1/2}.$ Here $\check \lfdr_i$'s are evaluated as one minus the right hand side of~\eqref{conex} with $(\pi^*(\cdot),f_1^*)$ replaced by $(\check \pi(\cdot),\check f_1)$.
\end{itemize}

Further, we consider three more measures that are aimed at understanding the efficacy of these methods for the purpose of post-estimation multiple testing.

\begin{itemize}
	
	\item[(d)] Underestimation in the vector of lFDRs $(\lfdr^*_1, \dots, \lfdr^*_n)$: $\frac{1}{n} \sum_{i = 1}^n \E (\lfdr^*_i - \check \lfdr_i)_+$. In multiple testing problems, such underestimation may result in too many hypotheses being rejected which may lead to inflated measures of Type I error, such as FDR. Thus, for an efficient multiple testing procedure, we would expect this underestimation metric to be large. 
	
	\item[(e)] FDR: $\E \left[ \frac{ \mbox{Number of false rejections}}{\mbox{Total number of rejections}}\right].$
	
	\item[(f)] True Positive Rate (TPR): $\E \left[ \frac{\mbox{Number of true rejections}}{ \mbox{Total number of non-null hypotheses}} \right].$
\end{itemize}
Measures (e) and (f) can be interpreted as analogs of Type I error and power, respectively.
Note that, methods that yield higher values of TPR while keeping FDR under a certain specified threshold, should be considered more effective. 	

We consider the following choices for $\pi^*(x) \defeq [1 + \exp(-s(x))]^{-1}$: 
\vspace{-0.5cm}
\begin{itemize}
	\item[(A)] $s(x_1,x_2) = -2 + 3.5 x_1^2 - 3.5 x_2^2; \qquad \qquad\;\;$ (B) $s(x_1,x_2) = -3 + 1.5 x_1 + 1.5 x_2$;
	\item[(C)] $s(x_1,x_2) = -1 + 9(x_1 - 0.5)^2 -5|x_2|$; \qquad (D) $s(x_1,x_2) = 20(x_1 -0.75)$.
\end{itemize}
\vspace{-0.25cm}

For the non-null density $f_1^*$ we choose the following:
\vspace{-0.25cm}
\begin{itemize}
	\item[(i)] $f_1^* = 0.4 \N(-1.25, 3) + 0.2 \N(0, 5) + 0.4 \N(1.25, 3)$;
	\item[(ii)] $f_1^* = 0.3 \N(0, 1.1) + 0.4 \N(0, 2) + 0.3 \N(0, 10)$;
	\item[(iii)] $f_1^* = 2^{-1} \N(0.5, 1) + 3^{-1} \N(1, 1.1) + 6^{-1} \N(1.5, 2)$;
	\item[(iv)] $f_1^* = 0.48 \N(-2, 2) + 0.04 \N(0, 17) + 0.48 \N(2, 2)$.
\end{itemize}
\begin{figure}
	\centering
	\begin{subfigure}[b]{0.47\textwidth}
		\centering
		\includegraphics[width=\textwidth]{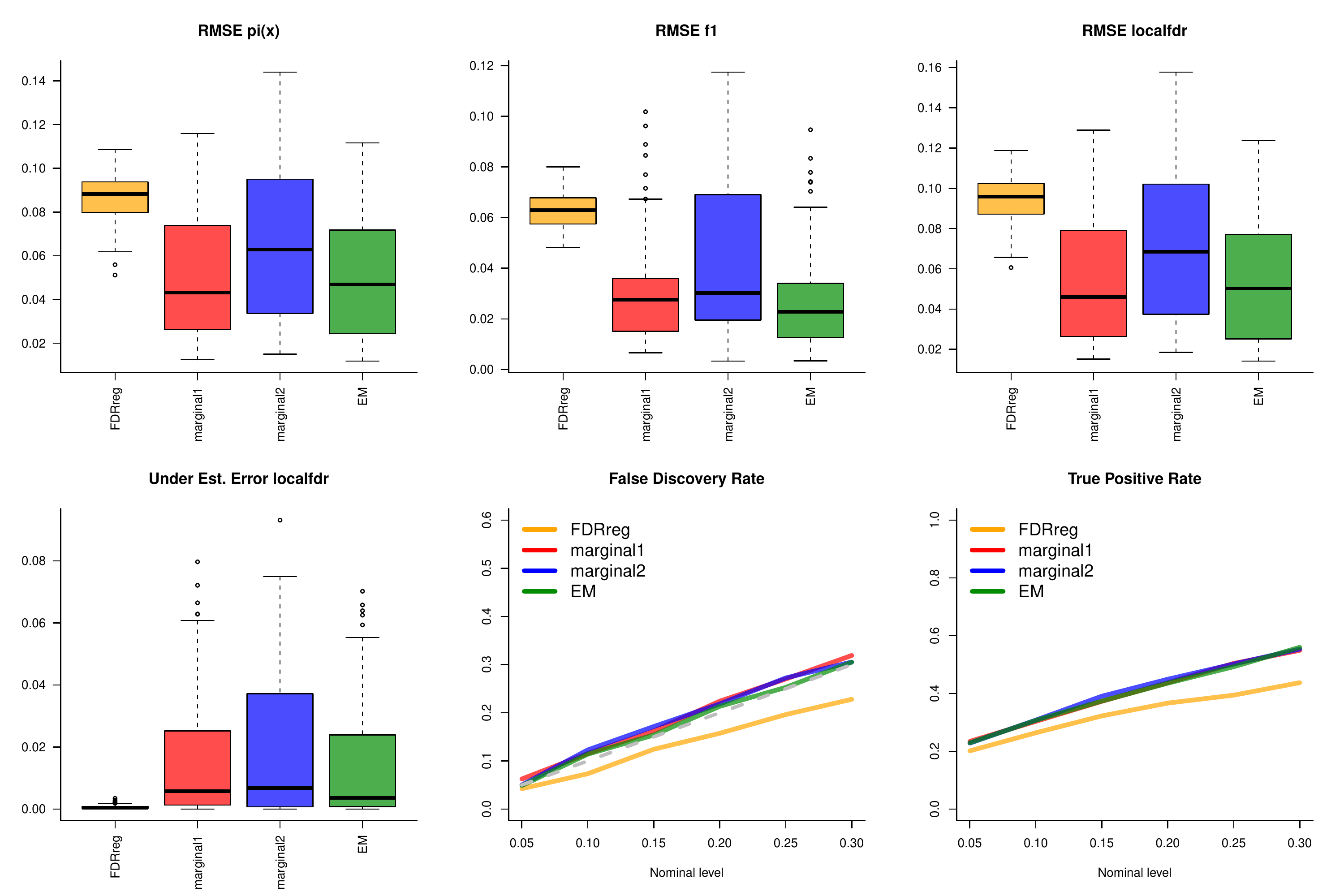}
		\caption{$f_1^*$ as in setting (i).}
	\end{subfigure}
	\hfill
	\begin{subfigure}[b]{0.47\textwidth}
		\centering
		\includegraphics[width=\textwidth]{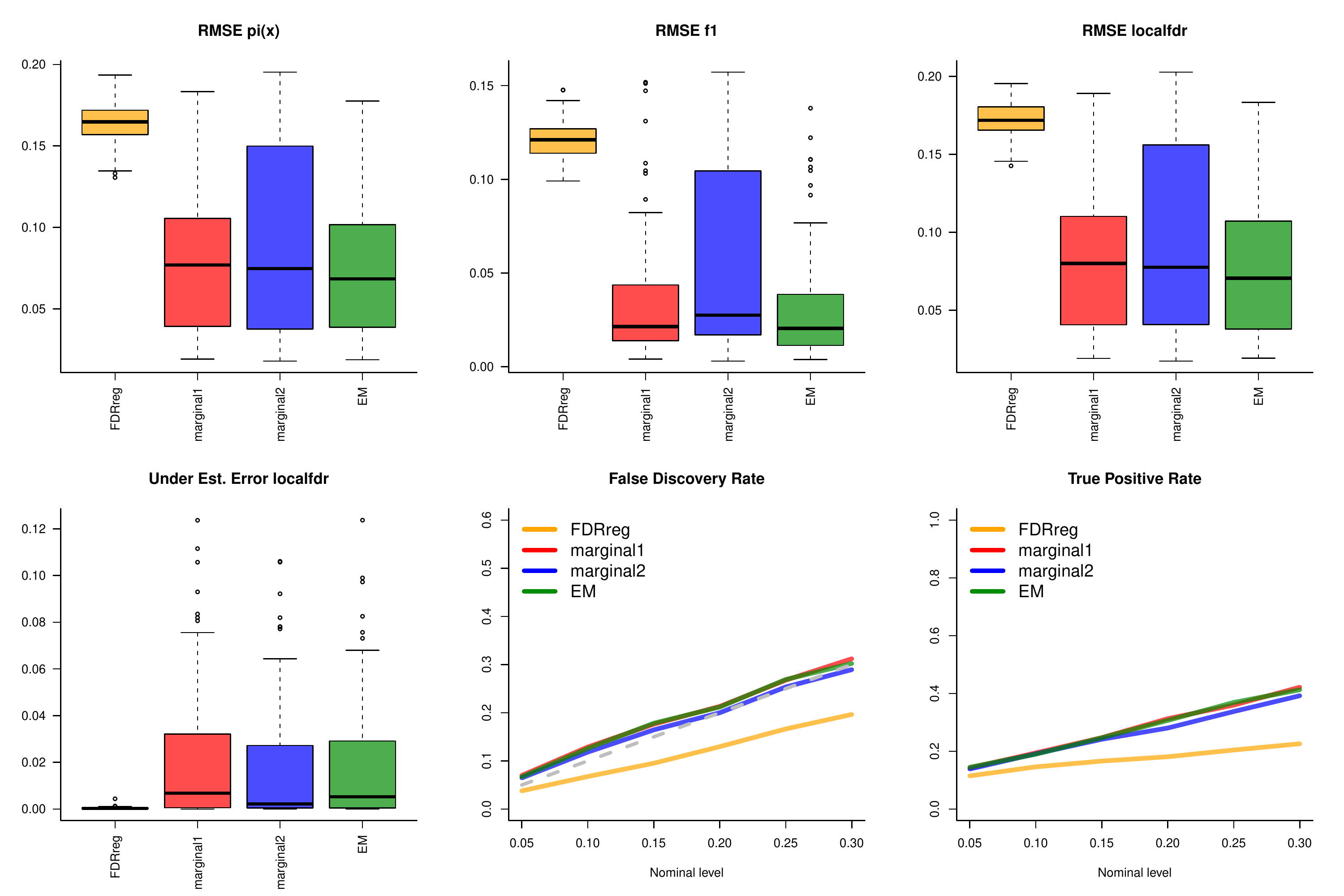}
		\caption{$f_1^*$ as in setting (ii).}
		\label{fig:pi_setting_Aii}
	\end{subfigure}
	\vspace{15mm}
	\begin{subfigure}[b]{0.47\textwidth}
		\centering
		\includegraphics[width=\textwidth]{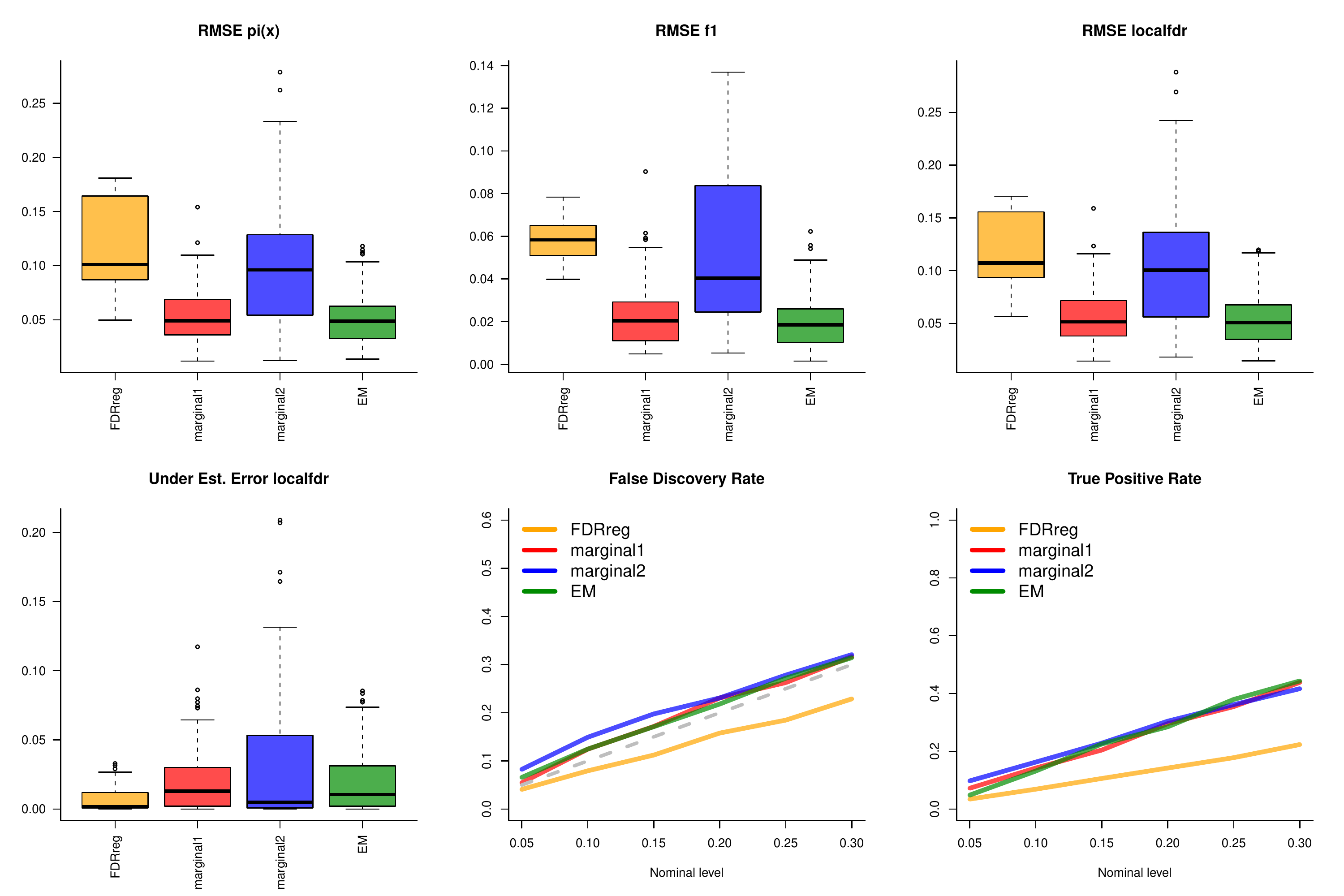}
		\caption{$f_1^*$ as in setting (iii).}
		\label{fig:pi_setting_Aiii}
	\end{subfigure}
	\hfill
	\begin{subfigure}[b]{0.47\textwidth}
		\centering
		\includegraphics[width=\textwidth]{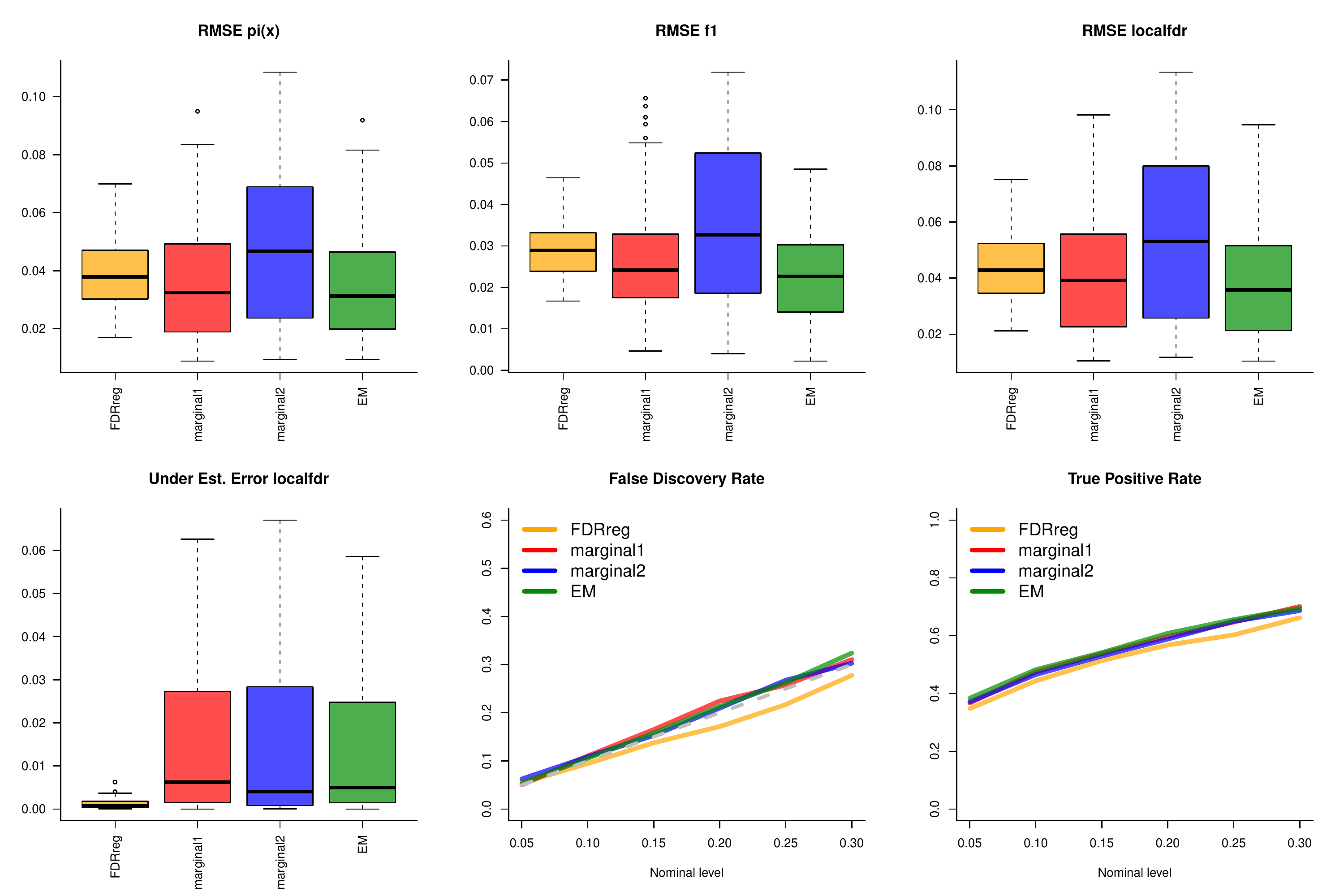}
		\caption{$f_1^*$ as in setting (iv).}
		\label{fig:pi_setting_Aiv}
	\end{subfigure}
	\vspace{-1.5cm}
	\caption{Each subplot shows the performances of FDRreg (in yellow), Marginal - I (in red), Marginal - II (in blue) and fMLE (in green) based on the six metrics (a)-(f) in row-major order. The 4 subplots are obtained for the 4 different choices of $f_1^*$, namely (i)-(iv) (in row-major order); the choice of $s(\cdot)$ was fixed at setting (A). For metrics (a)-(d), boxplots are constructed based on $200$ replicates. For metrics (e) and (f), the plots show average false discovery and true positive rates computed over $200$ replicates for a grid of nominal levels $\{0.05, 0.10, 0.15, 0.20, 0.25, 0.30\}$. In the plot depicting FDR (metric (e)), the grey dashed line indicates the nominal level.}\label{fig:pi_setting_A}
\end{figure}
Most of the settings mentioned above, in particular, (A) and (B) for $s(\cdot,\cdot)$, and (i), (ii) and (iv) for $f_1^*$, have been borrowed from~\cite{ScottEtAl15}. The settings (A) --- (D) capture a broad spectrum of relationships between the covariates and the response: for instance, the graph of $\pi^*(\cdot)$ corresponding to scenario (B) seems relatively flat as $(x_1,x_2)$ varies, whereas the graph of $\pi^*(\cdot)$ from scenario (D) shows a steep change in $\pi^*(\cdot)$ as $x_1$ exceeds $0.6$. Scenarios (A) and (C) are in between these two extremes. Through Figures~\ref{fig:pi_setting_A} and~\ref{fig:pi_setting_D}, and Figures~\ref{fig:pi_setting_B} and~\ref{fig:pi_setting_C} (see~\cref{tables} in the appendix) we illustrate the performance of  FDRreg, Marginal - I, Marginal - II and fMLE in these diverse simulation settings. We observe that our proposed methods consistently outperform FDRreg, in terms of most of the metrics ((a) --- (f)) as discussed above, more so when $\pi^*(\cdot)$ varies significantly with $(x_1,x_2)$.

For each pair of parameters $(\pi^*, f_1^*)$, we implement the methods --- Marginal - I, Marginal~-~II, fMLE, and FDRreg --- on $200$ independent replicates each with sample size $n = 10^4$. In each replicate, two-dimensional covariates $X_i = (X_{i1}, X_{i2})$, $i = 1,\dots, n$, are drawn uniformly at random from the unit square, i.e., $[0,1]^2$. Then $\{Y_i\}_{i=1}^n$ are drawn independently from the mixture density $\pi^*(X_i) f_1^* + (1-\pi^*(X_i))f_0$. In our simulations we model the covariates, expanded from two dimensions to six dimensions, via basis splines with three degrees of freedom (using a logistic link) as in~\cite{ScottEtAl15}. 
\begin{figure}
	\centering
	\begin{subfigure}[b]{0.47\textwidth}
		\centering
		\includegraphics[width=\textwidth]{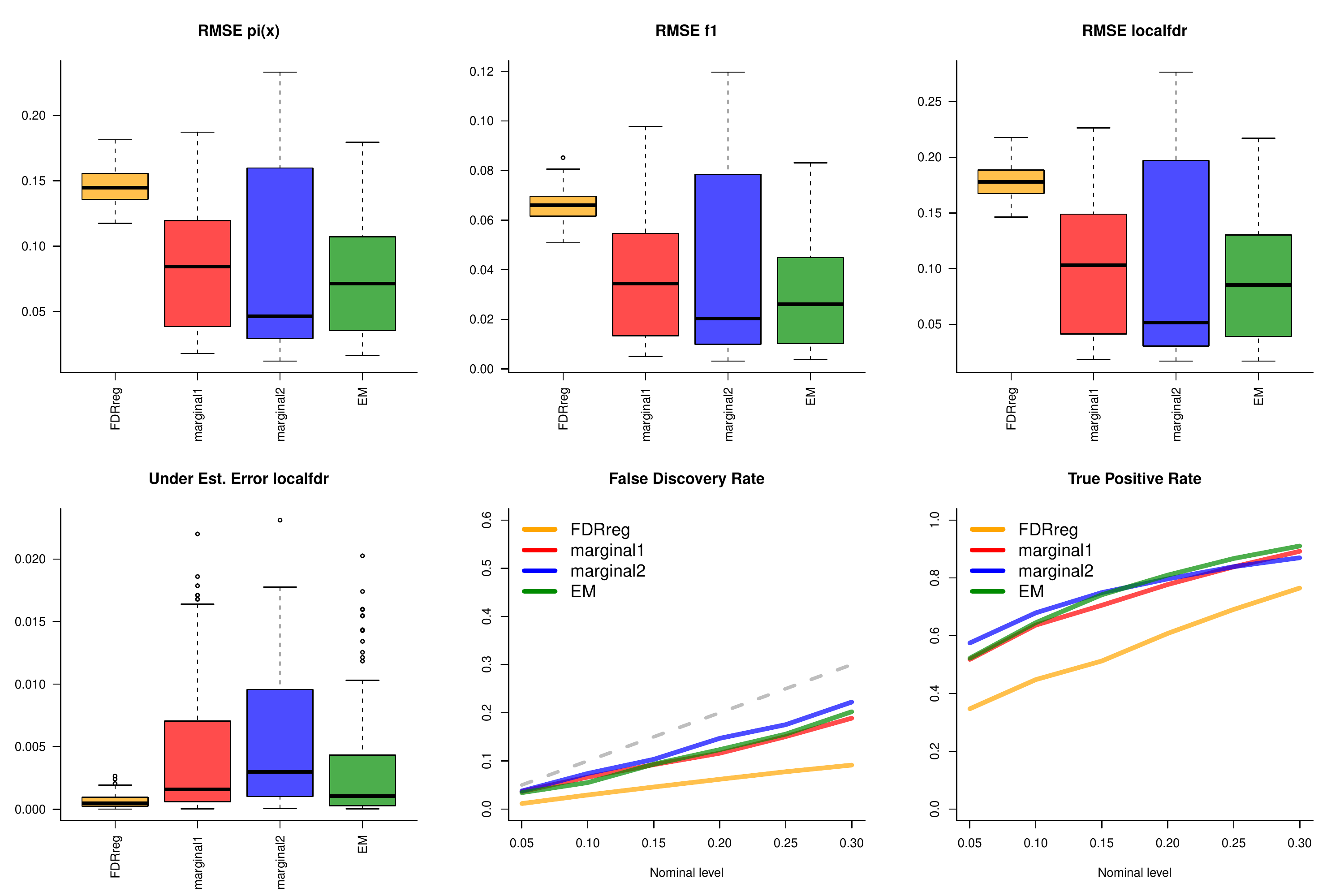}
		\caption{$f_1^*$ as in setting (i).}
		\label{fig:pi_setting_Di}
	\end{subfigure}
	\hfill
	\begin{subfigure}[b]{0.47\textwidth}
		\centering
		\includegraphics[width=\textwidth]{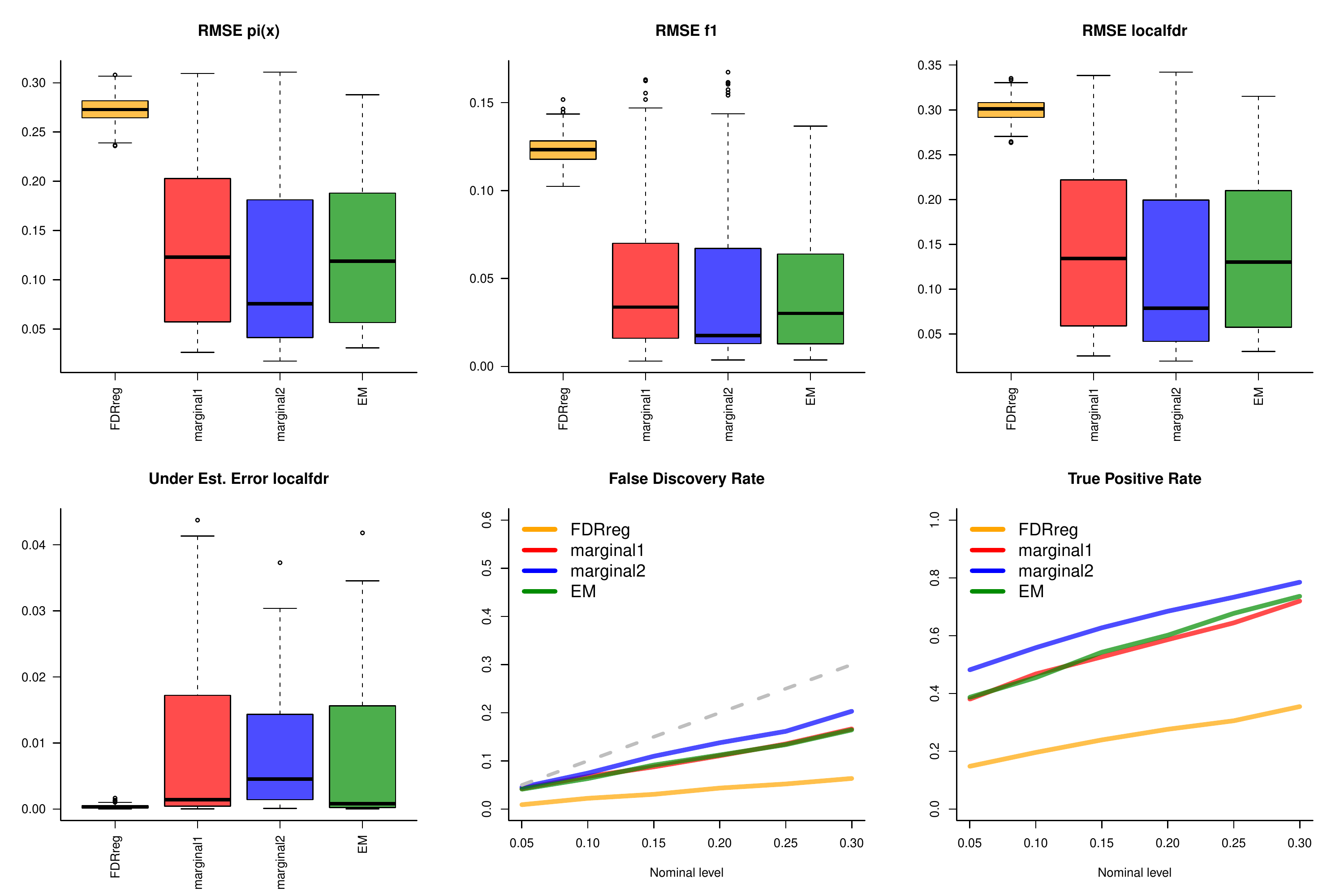}
		\caption{$f_1^*$ as in setting (ii).}
		\label{fig:pi_setting_Dii}
	\end{subfigure}
	\vspace{15mm}
	\begin{subfigure}[b]{0.47\textwidth}
		\centering
		\includegraphics[width=\textwidth]{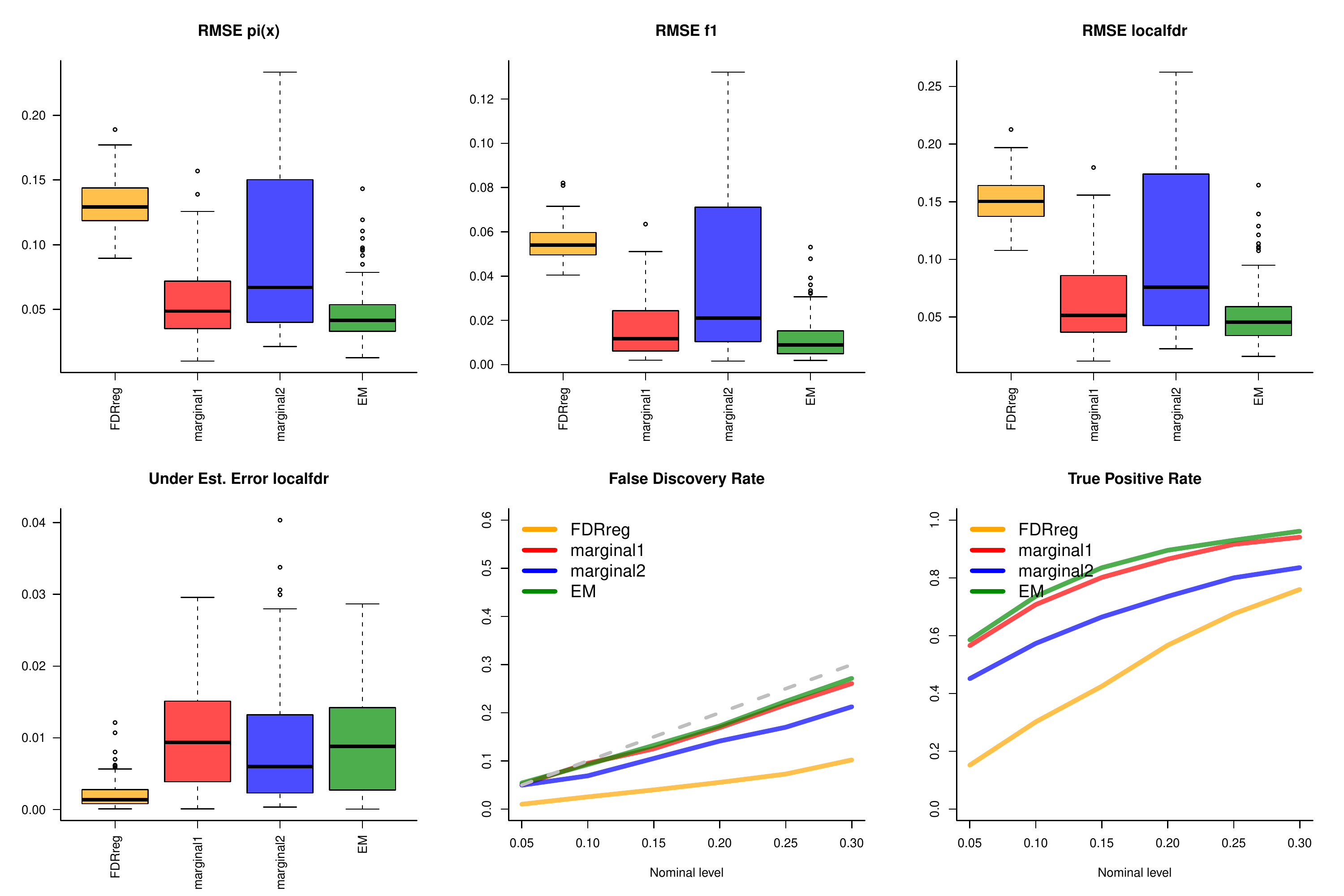}
		\caption{$f_1^*$ as in setting (iii).}
		\label{fig:pi_setting_Diii}
	\end{subfigure}
	\hfill
	\begin{subfigure}[b]{0.47\textwidth}
		\centering
		\includegraphics[width=\textwidth]{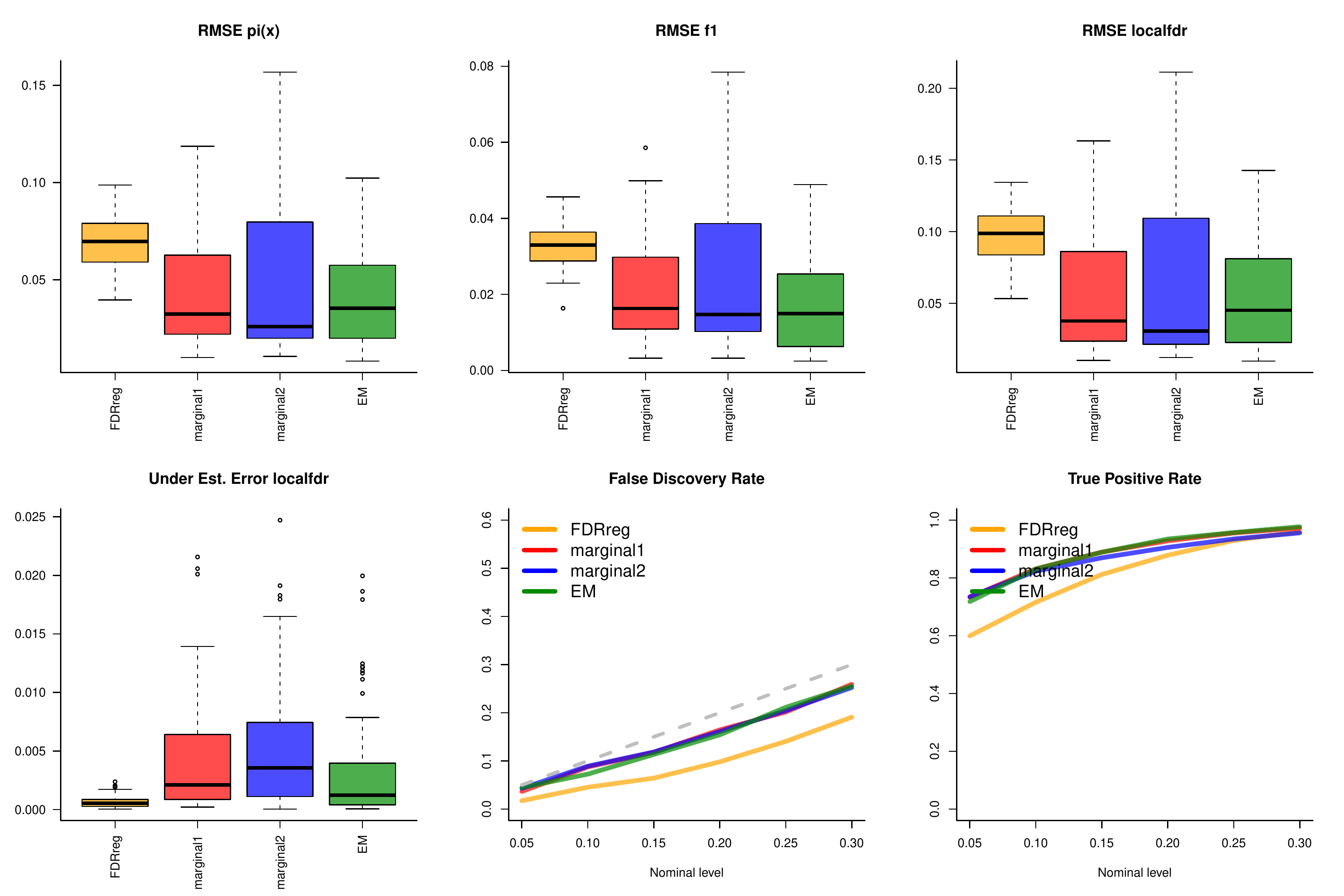}
		\caption{$f_1^*$ as in setting (iv).}
		\label{fig:pi_setting_Div}
	\end{subfigure}
	\vspace{-1.5cm}
	\caption{Figure depicts the same plots as in Figure~\ref{fig:pi_setting_A} when $s(\cdot)$ is from setting (D).}
	\label{fig:pi_setting_D}
\end{figure}


Recall that in order to compute the fMLE, one has to solve a non-convex optimization problem and a good starting point is necessary. We initialize this iterative method by choosing the estimate with the highest likelihood value obtained from the other procedures, namely, Marginal - I, Marginal - II, and FDRreg. The EM algorithm is then run for $500$ iterations or until convergence (i.e., the iterative change in the norm of the vector of the estimated lFDRs falls below $10^{-6}$). Our results are illustrated in Figures~\ref{fig:pi_setting_A} and \ref{fig:pi_setting_D} (and in Figures~\ref{fig:pi_setting_B} and \ref{fig:pi_setting_C} in the appendix,~\cref{tables}). 
In Table~\ref{table:init_proportions} (see~\cref{tables} in the appendix), we show that Marginal - I most often has the highest likelihood value and thus serves as the initializer for fMLE. With the exception of setting (D)(i), in the same table, we also note that FDRreg was rarely used to initialize fMLE. This shows that, across our simulation settings, estimates from Marginal - I and Marginal - II consistently yield higher likelihoods than those from FDRreg.

\subsubsection{Estimation of model parameters}\label{sec:emp}
We begin our discussion by considering the RMSEs in estimating the unknowns $\pi^*(\cdot)$, $f_1^*$ and $\lfdr_i^*$'s, as defined in the metrics (a)-(c); see Figures \ref{fig:pi_setting_A} and \ref{fig:pi_setting_D}. Note that, fMLE is almost always the most accurate estimator as it results in lower RMSEs (except in Figure~\ref{fig:pi_setting_Biv} in the appendix where FDRreg performs the best). Even Marginal - I and Marginal - II yield better estimates than FDRreg in most settings; except in Figures~\ref{fig:pi_setting_Biv} and~\ref{fig:pi_setting_Civ} for Marginal - I, and in Figures~\ref{fig:pi_setting_Aiv},~\ref{fig:pi_setting_Bi},~\ref{fig:pi_setting_Biii},~\ref{fig:pi_setting_Biv}, and~\ref{fig:pi_setting_Civ} for Marginal - II (see~\cref{tables} in the appendix).

In the interest of fairness however, we point out two specific caveats. Firstly, Figure~\ref{fig:pi_setting_Biv} (see~\cref{tables} in the appendix) shows an example where fMLE is outperformed by FDRreg. This figure describes the performance of the various methods for setting (B)(iv). In this setting, we see that all three of the methods Marginal - I, Marginal - II and fMLE are outperformed by FDRreg. However, a closer inspection reveals that by slightly tweaking the above simulation setting we observe a  completely different outcome, i.e., fMLE performs much better than FDRreg. This is related to a phenomenon we call ``near non-identifiability''; see~\cref{sec:non-iden} in the appendix for more details.	
Secondly, note that, although our methods outperform FDRreg in almost all the settings, they are in general more time consuming to compute than FDRreg. This is expected because FDRreg overlooks covariate information while estimating $f_1^*$ while our methods utilize covariate information, solving a more complex optimization problem in the process. In Table~\ref{table:timecomp} in~\cref{tables} (see the appendix), we tabulate the average time required for each of the methods in each of the settings mentioned above. In Table~\ref{table:timereq} in the appendix, we demonstrate the time required by each method in setting (A)(i) as $n$ varies from $10^4$ to $10^5$. Observe that Marginal - I and Marginal - II are about $15-20$ and $6-10$ times slower than FDRreg respectively. However this does not seem to be a big issue as the net time taken by Marginal - I or Marginal - II is still below $6$ minutes for a sample size of $n=10^5$. In fact, through extensive simulations, we have also checked that the computation of the fMLE requires less than an hour, for $n$ as large as $10^5$. 

\subsubsection{Multiple hypotheses testing}

Having established the superiority of fMLE for the purposes of estimating the model parameters, we now move our attention to the application of each of these methods for the purpose of multiple hypotheses testing. As described in~\cref{sec:testing} in the appendix, multiple hypotheses testing is conducted in these settings by estimating the lFDR of each observation and then constructing a set of rejections based on these lFDRs. An overwhelming observation based on the metrics (d)-(f) is the conservatism of FDRreg. In this context, conservatism refers to whether a method leads to substantially lower false rejections than the nominal FDR level it has been set to, and consequently suffers a loss in power. Indeed, in most of the simulations, the underestimation corresponding to FDRreg is almost zero, implying that it regularly overestimates the true lFDRs. As such, false null hypotheses are often accepted by FDRreg, leading to low power (TPR). Thus, based on these simulations it is evident that the FDRreg method frequently produces heavily biased estimates of lFDR, with the bias directed such that FDR control is satisfied but TPR is low. 

In contrast, fMLE and the marginal methods do not exhibit such a behavior. Indeed, in all figures except in Figures~\ref{fig:pi_setting_Bi},~\ref{fig:pi_setting_Bii}, and~\ref{fig:pi_setting_Biii} in the appendix, Marginal - I, Marginal - II and fMLE maintain (or only marginally exceed) the nominal level in FDR and are further able to correctly reject more false hypotheses (higher TPR) as compared to FDRreg. We reiterate that one of our goals in the investigation of likelihood based methodology in model~\eqref{denmodel}, beyond the estimation of model parameters, is to construct more powerful multiple testing procedures utilizing the information present in the covariates. As such, we conclude that in most settings, fMLE provides a valid, more powerful multiple testing procedure than FDRreg. Nevertheless, Figure \ref{fig:pi_setting_B} in the appendix shows instances where the likelihood based methods may exceed nominal FDR levels. This is most evident in Figure \ref{fig:pi_setting_Biii}. As stated before, we believe that this behavior is related to the \emph{near non-identifiability} phenomenon which we shall explore in further details in~\cref{sec:non-iden} (see the appendix).
It may be worth pointing out that although we are unable to guarantee finite sample FDR control for our methods, we do however expect asymptotic FDR control (based on the accuracy in estimating lFDRs as observed in~\cref{sec:emp} and~\cref{sr1}).

\subsection{Related discussions and recommendations}\label{reldec}
In addition to the discussions in this section so far, there are two important observations which we believe augment the utility and reliability of our methods. Firstly, recall the statement of~\cref{sr1}. The near parametric rates that we derive there, for estimating the conditional distribution of $Y$ given $X$, hold for all AMLEs. A natural question arises: ``Do our proposed methods yield AMLEs in practice?". In~\cref{appMLEsim} in the appendix, we show results from extensive simulations that illustrate the consistency with which all of our methods (particularly fMLE) result in AMLEs. Secondly, recall the statement of~\cref{nonconv} which highlights that the fMLE method solves a non-convex optimization problem. Therefore, a natural question to ask during implementation is whether the proposed iterative (EM) algorithm is sensitive to the proposed starting points (Marginal - I, Marginal - II or FDRreg). In~\cref{initfMLE} (see the appendix), our simulations demonstrate that the fMLE approach yields estimates which are mostly stable across the suggested initializations.

Based on our detailed simulation studies (and theoretical results), we would recommend the fMLE method to estimate the unknowns in~\eqref{denmodel} and consequently address the multiple hypotheses testing problem, especially for moderate sample sizes (at least up to $n=10^5$). We believe that Marginal - I is possibly the most reliable candidate for producing estimates that may be used to initialize the EM algorithm for computing the fMLEs. It must be pointed out though that we expect the estimates from Marginal - I  and Marginal - II initializations of the EM algorithm to be pretty similar; see~\cref{initfMLE} for details. For very large datasets ($n$ over a million), we suggest using Marginal - I instead of fMLE.

Note that, if the two-groups model~\eqref{eq:TwoComp} is adequate for the data, the estimates produced by fMLE  (and also FDRreg) can be unreliable, due to identifiability issues (as discussed in~\cref{secIdenti}). Therefore we recommend using the distance covariance based method (see~\cref{sec:hypotest}) first, in order to understand whether model~\eqref{eq:TwoComp} is adequate, before proceeding with our proposed methodology. However even under identifiability, estimates from model~\eqref{denmodel} (based on Marginal - I, Marginal - II, fMLE, FDRreg) may turn out to be highly variable (unless $n$ is very large) if the model is nearly non-identifiable (see~\cref{sec:non-iden} in the appendix for some discussion on this issue).

\section{Real data example} \label{realdata}

\subsection{Neuroscience application} \label{realdata:neuro}
\begin{figure} 
	\centering
	\includegraphics[width=20cm,height=6cm,keepaspectratio]{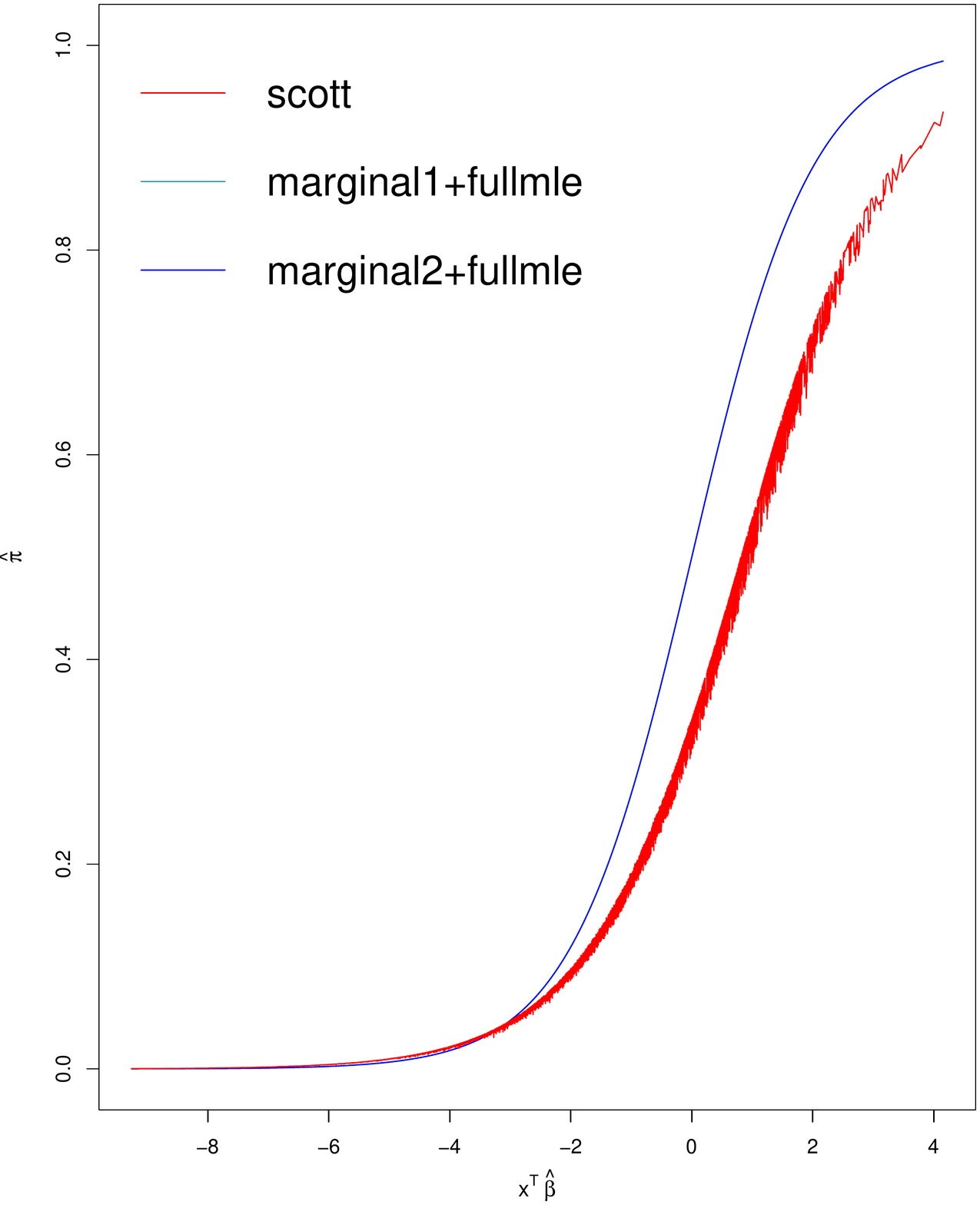}
	\includegraphics[width=20cm,height=6cm,keepaspectratio]{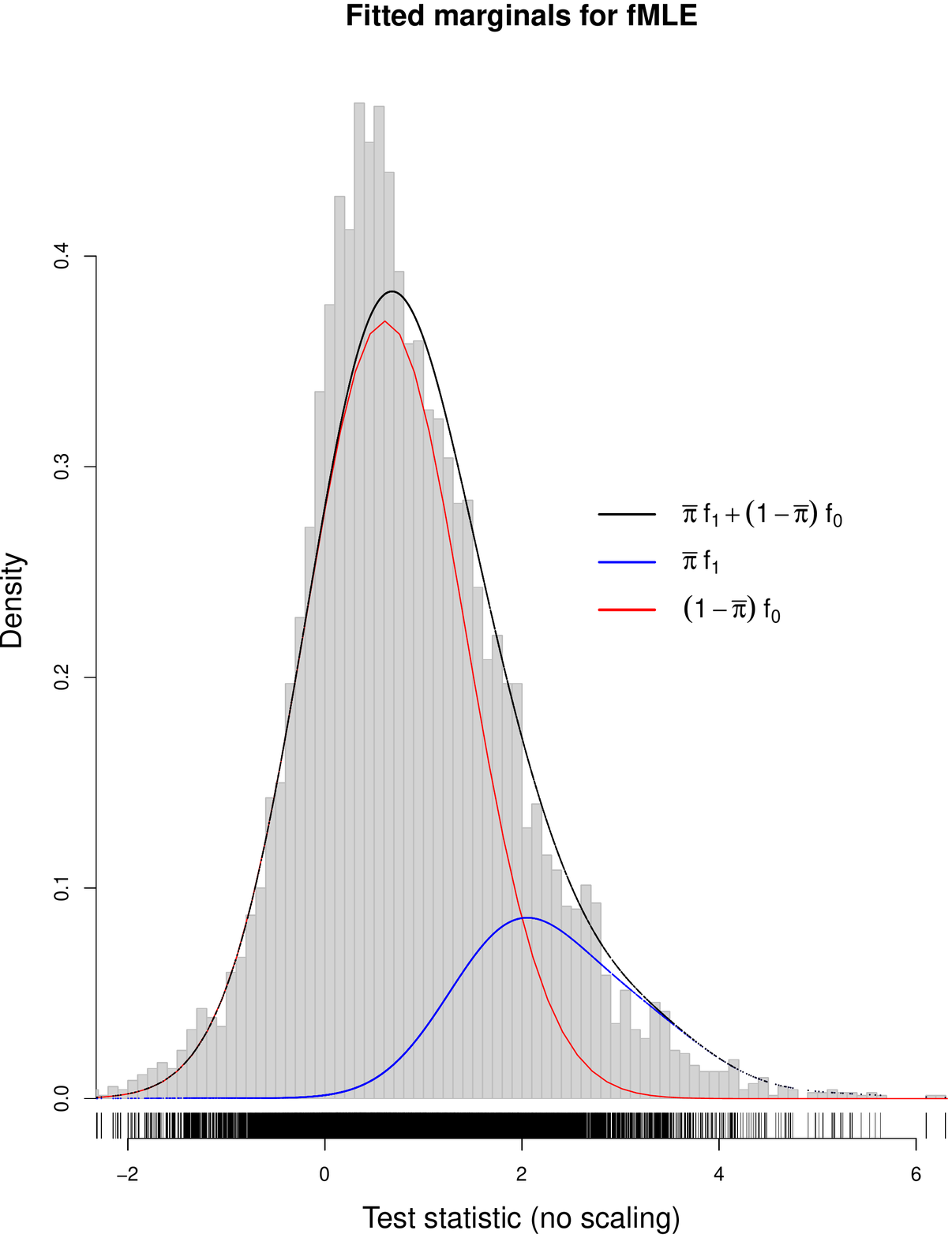}
	\includegraphics[width=20cm,height=6cm,keepaspectratio]{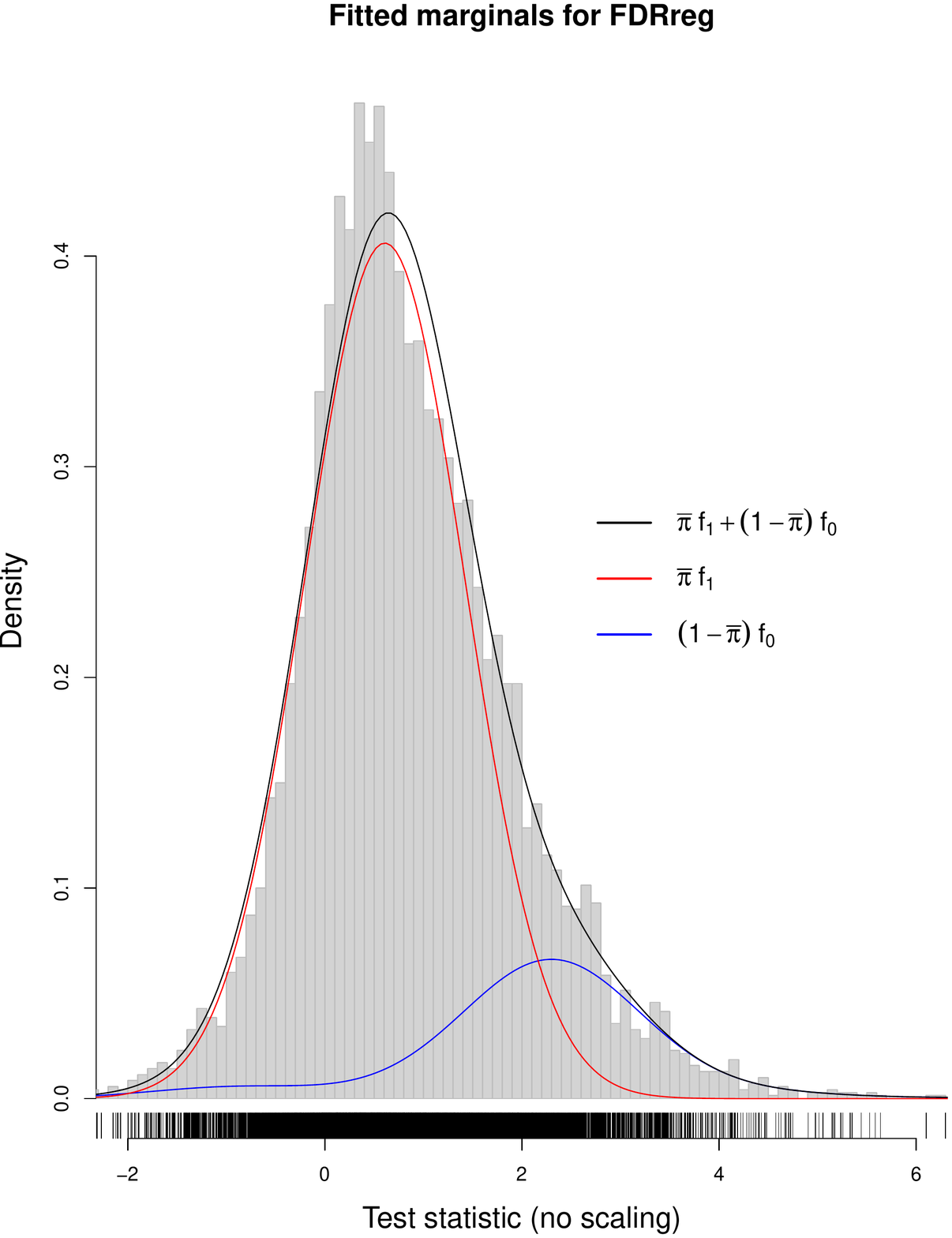}
	\includegraphics[width=20cm,height=6cm,keepaspectratio]{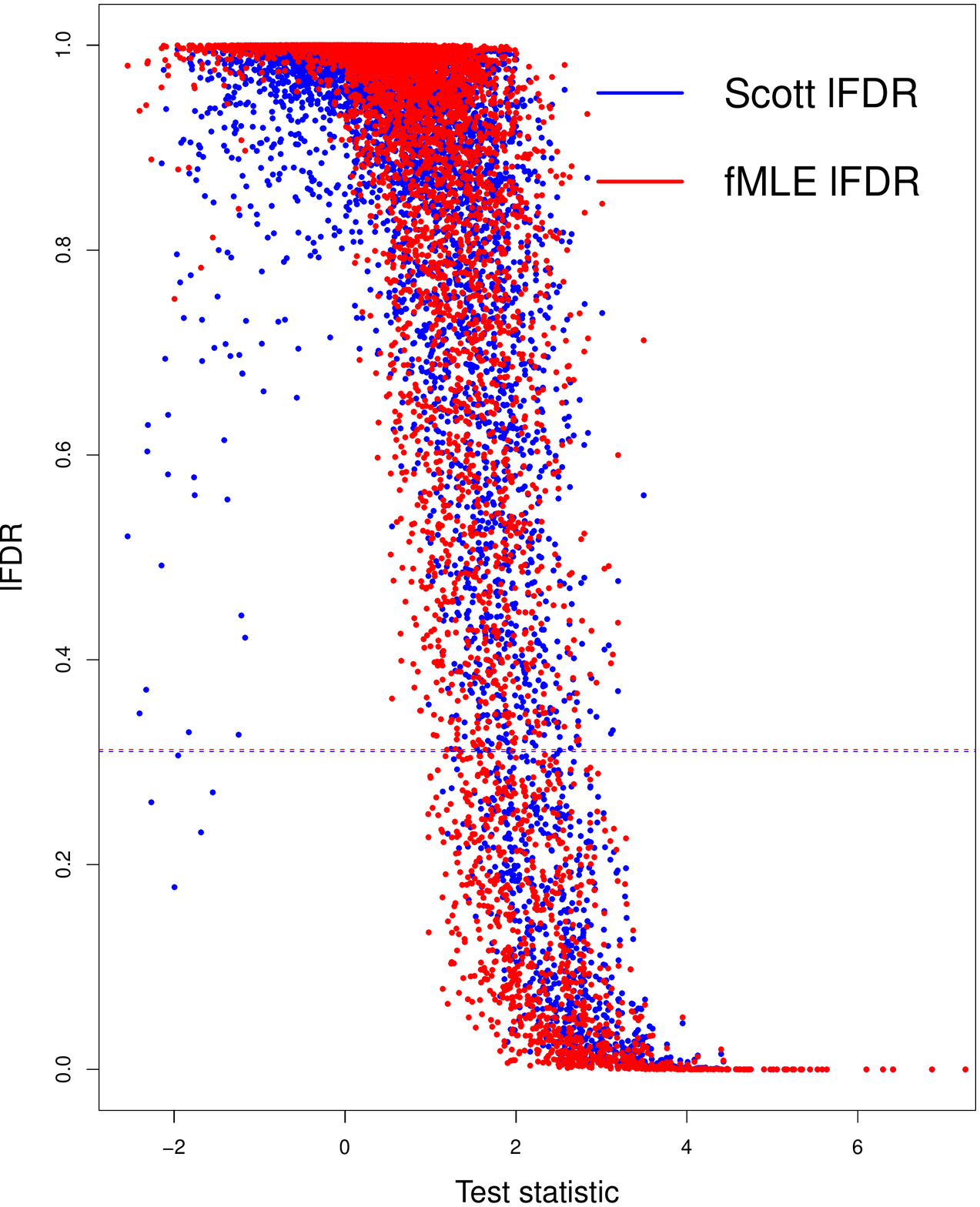}
	\includegraphics[width=20cm,height=6cm,keepaspectratio]{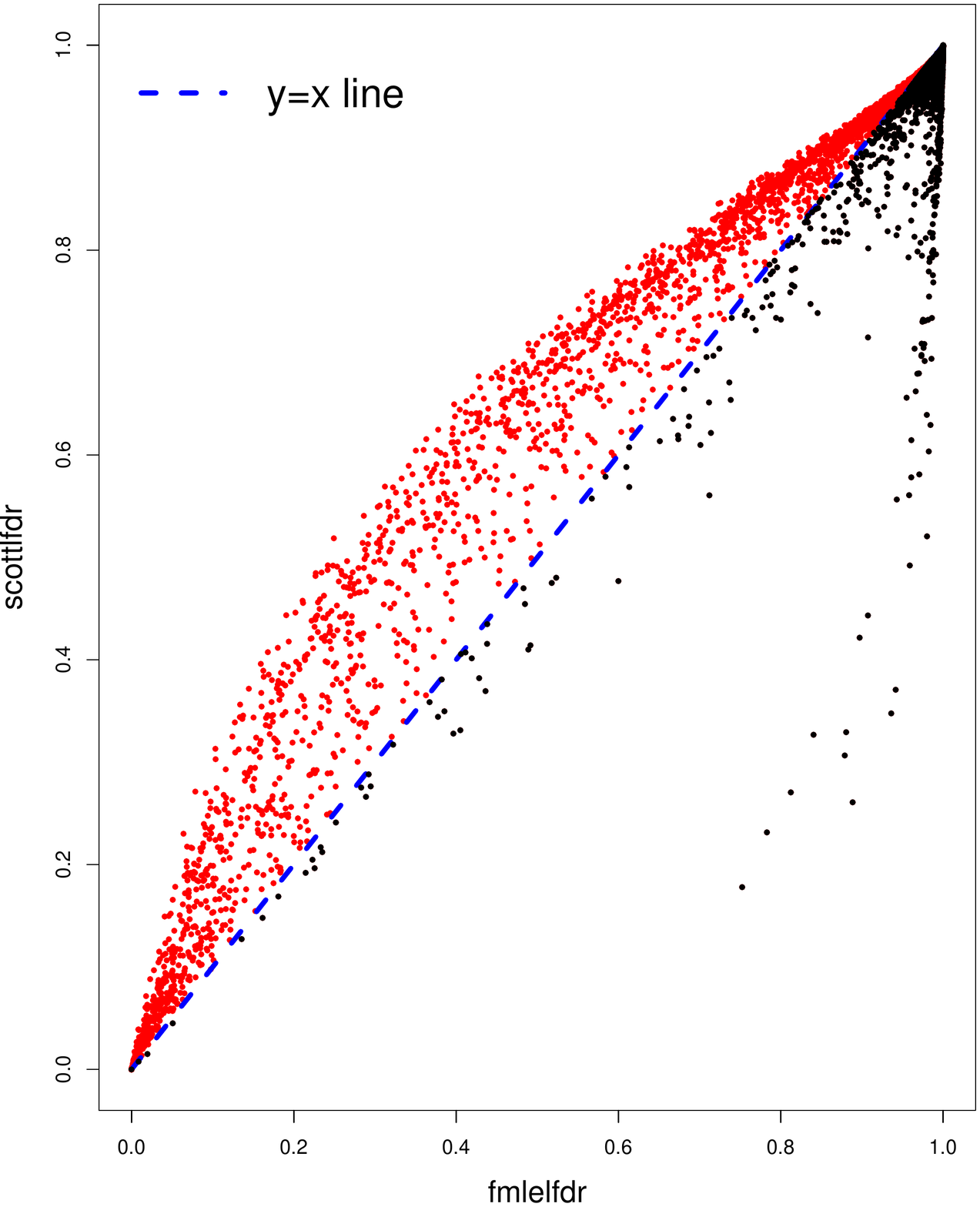}
	\includegraphics[width=20cm,height=6cm,keepaspectratio]{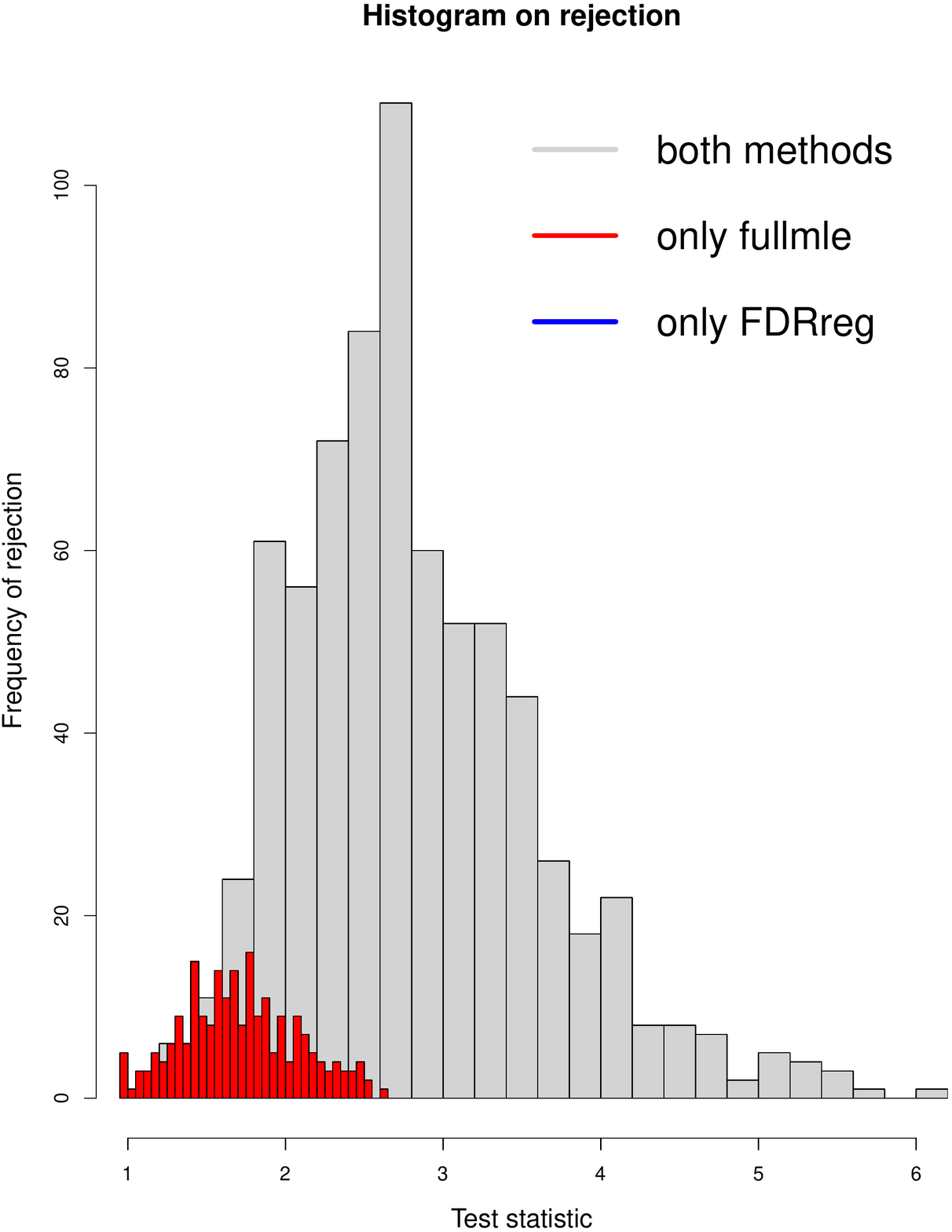}
	\caption{Top left panel: The plot of $\hat \pi^*(\cdot)$ against $x^\top\hat\beta$ (obtained from fMLE) for the two methods: FDRreg and fMLE (with initializations Marginal - I and Marginal - II) where $\hat\beta$ is computed using fMLE. Top center and top right panels: Plots of fitted marginal densities for fMLE and FDRreg respectively. Bottom left panel: The plot of lFDRs from FDRreg and fMLE with the test statistics plotted along the $x$-axis. The horizontal line indicates the threshold for rejection for the two methods (which are essentially the same $\approx 0.31)$. Bottom center panel: Plot of lFDRs from FDRreg versus the same from fMLE (points above and below the $y=x$ line have been colored using red and black respectively). Bottom right panel: Plot shows the rejection sets plotted across the test statistic for fMLE and FDRreg.}
	\label{fig:neural_analysis}
\end{figure}
Recall the multiple testing problem discussed in~\cref{ex:Neuro} where we have data arising from the firing rate of $128$ V1 neurons in an anesthetized monkey in response to a visual stimulus (see \url{https://github.com/jgscott/FDRreg/data}). The data consists of $7004$ test statistics, each one corresponding to a test of the null hypothesis of no interaction between a neuron pair. The dataset also includes two interesting covariates which capture the spatial and functional relationships among neurons: (a) distance between units, and (b) tuning curve correlation between units; for a more detailed understanding of this experiment see~\cite{Kelly07}. The primary goal of this study was to detect spiking synchrony among neuron pairs. 

For our analysis, we will use the same data processing as has been thoroughly outlined in~\cite[Sections 4.2 and 5]{ScottEtAl15}. In particular, we use a basis spline expansion on the covariates and model the null distribution as a Gaussian with mean and variance estimated using Efron's method of maximum likelihood, see e.g.,~\cite{efron2004large}. The estimates turn out to be $\mu$ (mean) = $0.61$ and $\sigma^2$ (variance) = $0.66$. We model the joint distribution of $(Z,X)$ (here $Z\defeq(Y-\mu)/\sigma$ denotes the centered and scaled test statistic and $X$ denotes the covariate) as in~\eqref{mdl} with $F_0=\Phi(\cdot)$, $\pi^*(\cdot)\in\logi$ and $F_1^*\in\Fga$. This is slightly different from the approach in~\cite{ScottEtAl15} where the authors directly model $Y$ --- they take $F_0$ as $\mathcal{N}(\mu,\sigma^2)$ and $F_1^*(y)\defeq \int\Phi\left(\frac{y-\mu-\theta}{\sigma}\right)\,dG(\theta)$, where $\mu$ and $\sigma$ are the same as above and $G$ is an unknown DF. To estimate the parameters in our model, we use the methods discussed in Sections~\ref{sec:EM-Gauss}, \ref{sec:MargMeth-I}, and~\ref{sec:MargMeth-II}. We then apply the multiple testing proposal from~\cref{sec:testing} with these estimates. We use a nominal level of $\alpha=0.1$ in our analysis (same as in~\cite{ScottEtAl15}). Figure~\ref{fig:neural_analysis} illustrates our findings.

The top left panel in Figure~\ref{fig:neural_analysis} shows that the estimate of $\pi^*(\cdot)$ from fMLE is in general higher than that from FDRreg. 
From the top center and right panels it looks as though the marginally fitted density from FDRreg fits the data slightly better. However, on observing the test statistic values between $-1$ and $-2$, we find that FDRreg estimates a non-trivial contribution of the signal density in that region (see the blue solid line in that region). This leads to smaller lFDRs corresponding to $Y$ values between $-1$ and $-2$ (see the bottom left panel) which seems rather counterintuitive. The bottom center panel offers more insight into this observation. Among the $7004$ test statistics, the lFDR estimates corresponding to fMLE actually turn out to be higher in over $4000$ cases compared to those from FDRreg. However, almost all these cases correspond to points in the top right corner of the bottom center panel (below the $y=x$ line). So, the fMLE procedure essentially yields higher lFDRs for test statistics which are highly unlikely to be signals. Correspondingly, the lFDRs based on fMLE are smaller (compared to FDRreg) in the more critical regions (i.e., where both lFDR estimates are small). Also, in the same plot, observe a sparse cluster near the lower right corner. These points correspond to test statistics between $-1$ and $-2$ for which FDRreg yields much lower lFDRs as compared to fMLE. 

The plot in the bottom right panel illustrates the rejection sets from the two methods. Observe that fMLE admits more rejections than FDRreg. In particular, fMLE rejects $220$ more hypotheses (all in the range of test statistics values between $1$ and $3$). FDRreg rejects $5$ more hypotheses all of which correspond to $Y$ values between $-1$ and $-2$, which as we mentioned before, seems somewhat counterintuitive. Overall, the fMLE procedure rejects $970$ hypotheses out of $7004$, at a nominal level of $0.1$, whereas FDRreg rejects $755$.
\begin{spacing}{0.98}
	{\small{
			\bibliographystyle{chicago}
			\bibliography{SS}}}
\end{spacing}

\appendix

\section{Algorithmic details}\label{sec:appen}

\subsection{Projected gradient descent algorithm to solve~\eqref{fdem}}\label{sec:A-PGD}
The pseudocode of the projected gradient descent algorithm that can be used to solve~\eqref{fdem} is described below. 
\begin{algorithm}[H]
	\caption{Projected gradient descent to solve \eqref{fdem}}
	\label{em_mixture}
	\begin{algorithmic}
		\STATE Input $\{Y_i\}_{i = 1}^n$
		\STATE $p_j^{(0)} \gets 1/m$ for $j = 1, \dots, m$
		\STATE $k \gets 1$
		\REPEAT 
		\STATE Compute update direction along gradient at $\mathbf{p}^{(k-1)}$ \begin{equation*}
		u_j \gets \frac{1}{n} \sum_{i = 1}^n w_i \frac{p_{j}^{(k-1)} \phi(y_i - a_{j})}{ \sum_{j'} p_{j'}^{(k-1)} \phi(y_i - a_{j'})}
		\end{equation*}
		\STATE Set initial step size $\alpha \gets 1$
		\REPEAT
		\STATE $\mathbf{p}^{(k)} \gets \text{Proj}(\mathbf{p}^{(k-1)} + \alpha\mathbf{u})$
		\STATE $\alpha \gets \alpha/2$
		\UNTIL $\frac{1}{n} \sum_{i=1}^n w_i \log \left[\sum_{j=1}^m p_j^{(k)} \phi(Y_i - a_j) \right] > \frac{1}{n} \sum_{i=1}^n w_i \log \left[\sum_{j=1}^m p_j^{(k-1)} \phi(Y_i - a_j) \right]$
		\STATE $k \gets k + 1$
		\UNTIL convergence of $\mathbf{p}$
	\end{algorithmic}
\end{algorithm}
For $\mathbf{x} \in \R^m$,  define the projection operator onto the probability simplex $\mathcal{P}_m$ as
\begin{equation*}
\text{Proj}(\mathbf{x}) = \argmin_{\mathbf{y} \in \mathcal{P}_m} \|\mathbf{x} - \mathbf{y}\|.
\end{equation*}
where $\|\cdot\|$ denotes the usual Euclidean norm in $\R^m$.
\citet{wang2013projection} provides a simple subroutine to solve this problem with $\bigO(m \log m)$ complexity.

\subsection{When $\mathfrak{F} = \Fde$ in~\eqref{fem}}\label{deninc}
In this case, optimization problem \eqref{fem} is similar to the computation of the Grenander estimator (the NPMLE of a nonincreasing density on $[0,\infty)$) which can be solved efficiently via the PAVA (see~\citet[Chapter
2]{groeneboom2014nonparametric}). We argue below that~\eqref{fem} can be solved by a modification of the usual PAVA.  

For the purposes of this optimization, we can normalize the
weights $w_i$ so that $\sum_{i=1}^n w_i = n$ (this does not change problem \eqref{fem} as the optimization is over
$f_1$). The first observation is that the maximizer of \eqref{fem}
will be a density that is piecewise constant with possible jumps only at the data
points $Y_1, \dots, Y_n$ (this follows from a similar fact for the
Grenander estimator; see \citet[Lemma
2.2]{groeneboom2014nonparametric}). As a result, one can restrict
$f_1$ in \eqref{fem} to such piecewise constant densities. Thus,~\eqref{fem} can be solved via the finite-dimensional optimization problem:  
\begin{equation*}
\hat {f}_{\text{EM}}(\mathbf{w},\mathfrak{F}_\downarrow) \defeq \argmax_{f_1 \geq f_2 \geq \dots \geq f_n \geq 0 : \sum_{i=1}^n f_i
	\left(Y_{(i)} - Y_{(i-1)} \right) = 1}  \frac{1}{n} \sum_{i=1}^n
w_i \log f_i
\end{equation*}
where $0 = Y_{(0)} < Y_{(1)} < \dots < Y_{(n)}$ denote the
order statistics of $Y_1, \dots, Y_n$. It can then be shown (see e.g.,~\citet[Exercise 2.5]{groeneboom2014nonparametric}) that the 
above optimization problem has the same solution (remember that the
weights $w_{i}$ are scaled so that $\sum_{i=1}^n w_i = n$)
as the following weighted least squares problem:  
\begin{equation*}
\argmin_{f_1 \geq f_2 \geq \dots \geq f_n} \sum_{i=1}^n
\left(\frac{w_i}{n(Y_{(i)} - Y_{(i-1)})}   - f_i
\right)^2 \left(Y_{(i)} - Y_{(i-1)} \right). 
\end{equation*}
This weighted least squares problem under the monotonicity constraint can be
solved efficiently via PAVA (see \citet[Chapter 1]{RWD88}). 


\subsection{Nonparametric function classes}\label{pinonpara} 
Note that the objective function in~\eqref{piem} is concave in $\pi(\cdot)$. Therefore as long as $\Pi$ is a convex class of functions (e.g., $\Pi = \Pide$),~\eqref{piem} is a convex optimization problem. In particular,~\eqref{piem} is easily converted to the following
finite-dimensional convex optimization problem:   
\begin{equation}\label{piem1}
\argmax_{(\pi_1, \dots, \pi_n) \in \Pi_n} \frac{1}{n}
\sum_{i=1}^n \left\{w_i \log \pi_i + (1 - w_i) \log \left(1 -
\pi_i \right) \right\} 
\end{equation}
where
$\Pi_n \defeq \left\{(\pi(X_1), \dots, \pi(X_n)) : \pi \in \Pi \right\}$
and $\pi_i=\pi(X_i),i=1,2,\ldots ,n$.
For many natural nonparametric convex classes $\Pi$, the set $\Pi_n$
can be written in a tractable fashion and the resulting
finite-dimensional convex optimization problem \eqref{piem1} can be
solved via standard techniques; see e.g., \citet[Chapters 6, 8, 9, 12 and 17]{NW06} and \citet[Chapters 5, 9, 10 and 11]{boyd04}. In certain specific cases, we can also employ specialized techniques to solve the above problem. As an example, let us consider the case when $\Pi = \Pide$. We assume here that the observations are re-ordered so that $X_i$'s are in increasing order. Then,  
in view of \citet[Theorem 1.5.1]{RWD88}, problem~\eqref{piem1} is equivalent to
\begin{equation*}
\argmax_{0 \le \pi_1 \le \pi_2 \le \dots \le \pi_n \le 1} \frac{1}{n}
\sum_{i=1}^n \left( w_i - \pi_i \right)^2
\end{equation*}
which is the standard (bounded) isotonic regression problem; see e.g.,~\cite{barlow72}. Many standard implementations are available to solve the above problem, e.g., the pool-adjacent-violators algorithm (PAVA), which can solve the above optimization problem in $O(n)$ time; see~\citet[pp.~9-10]{RWD88},~\cite{Stout14}.

\section{Are the covariates at all important?} \label{sec:hypotestcont}
We briefly describe the ``distance covariance" test for independence between $X$ and $Y$. Given the random sample $\{(X_i,Y_i)\}_{i=1}^n$ where $X_i\in\R^p$ and $Y_i\in\R$, we compute the Euclidean
distance matrices $(a_{kl})_{k,l=1}^n = (\|X_k - X_l\|)$ and $(b_{kl}) = (|Y_k - Y_l|)$. We further define $A_{kl} \defeq a_{kl} - \bar a_{k\cdot} - \bar a_{\cdot l} + \bar a_{\cdot\cdot}$, for $k,l = 1,\ldots, n$, where $\bar a_{k\cdot} \defeq n^{-1}\sum_{l=1}^n a_{k l}$, $\bar a_{\cdot l} \defeq n^{-1} \sum_{k=1}^n  a_{k l}$, and $\bar a_{\cdot\cdot} \defeq n^{-2}\sum_{k,l=1}^n a_{k l}$. Similarly, we define $B_{kl} : = b_{kl} - \bar b_{k\cdot} - \bar b_{
	\cdot l} + \bar b_{\cdot \cdot}$, for $k,l = 1,\ldots,n$. Then the sample {\it distance covariance} is defined as $$\mbox{dCov}_n(X,Y) \defeq \frac{1}{n^2} \sum_{k,l=1}^n A_{k l} B_{k l}.$$ We reject the null hypothesis of statistical independence between $X$ and $Y$ if the value of observed $\mbox{dCov}_n$ is significant. The null distribution of $\mbox{dCov}_n$ is also easy to find using a permutation test, i.e., we permute the $Y_i$'s, compute the test statistic again with the permuted $Y_i$'s and repeat these two steps multiple times to obtain the required null distribution. A rejection of the null hypothesis (at a prespecified level of significance) indicates that model~\eqref{eq:TwoComp} is inadequate and this suggests that we should model the effect of the covariates.

\section{Multiple testing with local FDR}\label{sec:testing}
In this section our primary goal is to develop a multiple testing procedure. In particular, we discuss the optimal multiple testing procedure in the setup of~\eqref{denmodel} and develop estimation strategies  based on the methods discussed so far (in Sections~\ref{sec:EM-Gauss},~\ref{sec:MargMeth-I} and~\ref{sec:MargMeth-II}). Suppose that we wish to test $n$ independent null hypotheses $\{H_{0i},i=1,2,\ldots ,n\}$ versus the corresponding alternative hypotheses $\{H_{1i},i=1,2,\ldots ,n\}$ simultaneously, based on $n$ $p$-values (or $z$-scores or some other test statistic) denoted by $Y_1,Y_2,\ldots ,Y_n$. Assume that we have covariate information $X_i\in \mathbb{R}^p$ available for each test statistic $Y_i$, for $i=1,\ldots, n$. Denote by $T_i \defeq (X_i,Y_i)$, $i=1,2,\ldots ,n$, and let $T\defeq (T_1,T_2,\ldots ,T_n)$. Consider unobserved latent (Bernoulli) variables $Z_1,Z_2,\ldots ,Z_n$ such that $Z_i=0$ if $H_{0i}$ is true and $Z_i=1$ if $H_{1i}$ is true. The main goal of multiple hypotheses testing is to recover the $Z_i$'s from the observations $\{(X_i,Y_i):i=1,\ldots, n\}$. A natural model to use in this scenario is
\begin{equation*}
X \sim m, ~~~~~~ Z | \left\{ X = x \right\}  \sim \text{Ber}(\pi^*(x)), ~~~~~~ Y | \left\{ Z = z, X = x \right\} \sim \begin{cases}
f_0 &\text{ if } z = 0 \\
f^*_1 &\text{ if } z = 1
\end{cases},
\end{equation*}
which is precisely a reformulation of~\eqref{denmodel}. This model and its variants have been recently used in many multiple testing problems, see e.g.,~\cite{li2016} and~\cite{lei2016adapt}. We refer to $\{(X_i,Y_i)\}_{i : Z_i = 0}$ as \emph{null} observations and the complement $\{(X_i,Y_i)\}_{i : Z_i = 1}$ as \emph{non-null} or \emph{significant} observations. 

Let us first introduce some notation. Let $\alpha \in (0,1)$ denote the prespecified significance level. Suppose $\mathcal{D}$ is the class of all decision rules $\bm{\delta} \defeq(\delta_1(\cdot),\ldots, \delta_n(\cdot))$ where $\delta_i: T \mapsto \{0,1\}$ is a function that decides to accept/reject (i.e., $\delta_i = 0$ or 1) the $i^{th}$ null hypothesis. Define the false discovery rate FDR$^*$ corresponding to a decision rule $\bm{\delta}$ as
\begin{equation}\label{newFDR}
\mbox{FDR}^*(\bm{\delta})\defeq\frac{\mathbb{E}\left[\sum\limits_{i=1}^n (1-Z_i)\delta_i\right]}{\mathbb{E}\left[\sum\limits_{i=1}^n \delta_i\right]}.
\end{equation}
Observe that this definition is slightly different from the definition of FDR used in~\cite{ben1997}. In~\citet[Supplementary Section D]{basu2018weighted}, the authors prove an asymptotic equivalence between the two definitions. Finally, define the expected true positive ETP for a decision rule $\bm{\delta}$ as
\begin{equation}\label{usualETP}
\mbox{ETP}(\bm{\delta})\defeq\mathbb{E}\left(\sum\limits_{i=1}^n Z_i\delta_i\right).
\end{equation}
Note that in both~\eqref{newFDR} and~\eqref{usualETP} the expectation is taken over the joint distribution of $(Y_i,Z_i)$ conditioned on $X_i$, for $i=1,\ldots,n$. In this multiple testing framework, FDR$^*$ and ETP are analogues of level and power of a statistical hypothesis test. Motivated by this intuition~\citet[equation 2.7]{basu2018weighted} propose the following optimization problem:
\begin{equation*}
\mbox{maximize ETP}(\bm{\delta})\quad \mbox{ subject to the constraint FDR}^*(\bm{\delta})\leq\alpha\mbox{ over }\bm{\delta}\in\mathcal{D}.
\end{equation*}
\citet{basu2018weighted} characterized the optimal testing procedure for this problem at every level $\alpha \in (0,1)$; their result is an extension of a similar result in the two-groups model~\eqref{eq:TwoComp} by~\cite{tang2005bayes}. This optimal decision rule is easily described in terms of oracle quantities called local false discovery rates (lFDRs) corresponding to each observation. For the $i^{th}$ observation $(X_i,Y_i)$, where $i = 1,\dots, n$, its lFDR is the posterior probability of the $i^{th}$ hypothesis being null, given the data, i.e., 
\begin{equation}\label{lFDRtrue}
\lfdr^*_i \defeq \mathbb{P}[Z_i=0|X_i,Y_i] = \frac{ (1-\pi^*(X_i)) f_0(Y_i) }{(1-\pi^*(X_i)) f_0(Y_i) + \pi^*(X_i) f^*_1(Y_i) }\;\;.
\end{equation}
Let $\lfdr^*_{(1)}, \dots, \lfdr^*_{(n)}$ denote the order statistics of $\lfdr^*_{1}, \dots, \lfdr^*_{n}$ and define \newline $k^*(\alpha)~\defeq~\max \left\{ k : \frac{1}{k} \sum_{i = 1}^k \lfdr^*_{(i)} \leq \alpha \right\}$ where the maximum of an empty set is interpreted as $0$. Given $\alpha\in (0,1)$, define  $\Rs^*(\alpha) \defeq \left\{ i : \lfdr^*_{i} \leq \lfdr^*_{(k^*(\alpha))} \right\}$. Then the optimal decision rule $\bm{\delta}^*$, which maximizes ETP (i.e., power) subject to FDR$^*$ control, is given by the vector which takes the value $1$ in its $i^{th}$ entry if $i\in\Rs^*(\alpha)$ and $0$ otherwise (see~\citet[Theorem 1]{basu2018weighted}). In other words, the $i^{th}$ hypothesis will be rejected if $\lfdr_{i}^*\leq \lfdr_{(k^*(\alpha))}^*$.

Although $\Rs^*(\alpha)$ describes the optimal test in this setup, it is not obvious how to practically implement it as the above quantities are determined in terms of the unknown parameters $(\pi^*(\cdot), f^*_1(\cdot))$. The methods we propose in Sections~\ref{sec:EM-Gauss} and \ref{sec:MargMeth} readily yield estimates of $\lfdr^*_i,i=1,2,\ldots ,n$, which consequently lead  to a data-driven plug-in procedure for multiple testing. We describe our method below.

First we obtain estimates $(\hat{\pi},\hat{f}_1)$ of $(\pi^*,f_1^*)$, by using the joint maximum likelihood approach (or either one of the marginal methods as discussed previously). These estimates can then directly be  used to obtain a plug-in estimate of $\lfdr^*_i$ as
\begin{equation*}
\hat{\lfdr}_i \defeq \frac{ (1-\hat{\pi}(X_i)) f_0(Y_i) }{(1-\hat{\pi}(X_i)) f_0(Y_i) + \hat{\pi}(X_i) \hat{f}_1(Y_i) }\;\;.
\end{equation*}
Denote the order statistics of $\hat{\lfdr}_1,\ldots, \hat{\lfdr}_n$ by $\hat{\lfdr}_{(1)},\ldots,\hat{\lfdr}_{(n)}$. Define $\widehat \Rs(\alpha) \defeq \left\{ i : \hat \lfdr_i \leq \hat \lfdr_{(\hat k(\alpha))} \right\}$ where $\hat k(\alpha)$ is defined as $\hat k(\alpha) \defeq \max \left\{ k : \frac{1}{k} \sum_{i = 1}^k \hat \lfdr_{(i)} \leq \alpha \right\}.$ Thus, we reject the $i^{th}$ hypothesis if $\hat{\lfdr}_{i}\leq \hat{\lfdr}_{(\hat{k}(\alpha))}$.

It is natural to investigate whether $\hat{\lfdr}_i$ is a good approximation of $\lfdr_i^*$, $i=1,2,\ldots,n$, which are the oracle quantities that determine the optimal rule. In~\cref{sec:sim}, we use extensive simulations to illustrate that our proposed method estimates the oracle lFDRs accurately.	
\begin{remark}\label{mono}
	Suppose $Y_1,\ldots,Y_n$ denotes $p$-values from $n$ independent hypotheses. Let \newline $Y_{(1)},\ldots ,Y_{(n)}$ denote the corresponding order statistics. Multiple testing procedures that do not take into account auxiliary information (e.g.,~\cite{ben1997}) usually satisfy the following {\it monotonicity} property: if the hypothesis corresponding to $Y_{(j)}$ is rejected, then the hypotheses corresponding to $Y_{(1)},\ldots,Y_{(j-1)}$ will also be rejected. Although this is a reasonable principle to go by in the absence of covariates, there is a priori no reason why one should adhere to this property when useful covariate information is  available; in fact our proposed methodology (and others, see e.g.,~\cite{lei2016adapt},~\cite{li2016} and~\cite{ScottEtAl15}) does not necessarily satisfy this property. 
\end{remark}

\section{Astronomy application} \label{realdata:astro}
We consider Example~\ref{ex:Astro} with data from the Carina dSph galaxy (see \url{http://vizier.cfa.harvard.edu/viz-bin/VizieR?-source=J/AJ/137/3100}). Recall that this is a contamination problem  where the sample of observations (on stars) from the Carina dSph is contaminated with foreground Milky Way stars in the field of view. The main problem is to identify and separate the Carina stars, based on the line-of-sight velocity ($Y$), along with covariate information, which in this case is the distance $X$ (in parsec) of the stars from the center of the dSph galaxy (which is also the center of the field of view). We model the conditional distribution $Y | X = x$ as in~\eqref{denmodel}. Note that this problem may be rephrased as a multiple hypotheses testing problem, where the null and alternative hypotheses, for $i=1,2,\ldots,n$, are $H_{0i}:$ $i^{th}$ star is a member of the Milky Way, and $H_{1i}:$ $i^{th}$ star is a member of the Carina dSph.

As the Carina dSph is quite far from the Milky Way, most of the stars in the center of the field of view are  expected to be from Carina, whereas the foreground Milky Way stars should be uniformly distributed over  the entire field of view. We leverage this information to model the prior probability $\pi^*(\cdot)$ of observing a Carina star as a nonincreasing function of $X$. The relevant `null' distribution here is the distribution of the line-of-sight velocity of stars in the Milky Way galaxy. This information is available to us from the \textit{Besancon model}, as discussed in~\cite{RobinEtAl03}. As a result, the density $f_0$ is hereafter treated as known. 

The plots in Figure \ref{fig:Carina} demonstrate that $X$ is informative for $Y$ and gives us an intuition towards modeling $f_1^*$. From the density estimate in Figure \ref{fig:Carina}, the distribution of $Y$ for stars in Carina can be observed to be unimodal and bell-shaped. As in~\cite{walker09stellar} and~\cite{walker2009clean}, we model $f_1^*$ as $N(\mu, \sigma^2)$ for some unknown $\mu$ and $\sigma^2$. Thus, our model can be written as
\begin{equation}\label{model:astro}
Y | X = x \sim \pi^*(x) N(\mu, \sigma^2) + (1-\pi^*(x)) f_0
\end{equation}
where $\pi^*(\cdot)$ is a nonincreasing function. The density $f_0$ (although we do not mention the form explicitly) is such that, any convex combination of $f_0$ and a Gaussian density, is no longer Gaussian. This implies that model~\eqref{model:astro} is identifiable. Note that, for simplicity, in the above modeling, we have ignored measurement errors in the line-of-sight velocities and other interesting covariates such as the magnesium index; for a detailed description of this dataset see~\cite{walker2009clean}.

\begin{figure}[h] 
	\centering
	\includegraphics[width=22cm,height=6cm,keepaspectratio]{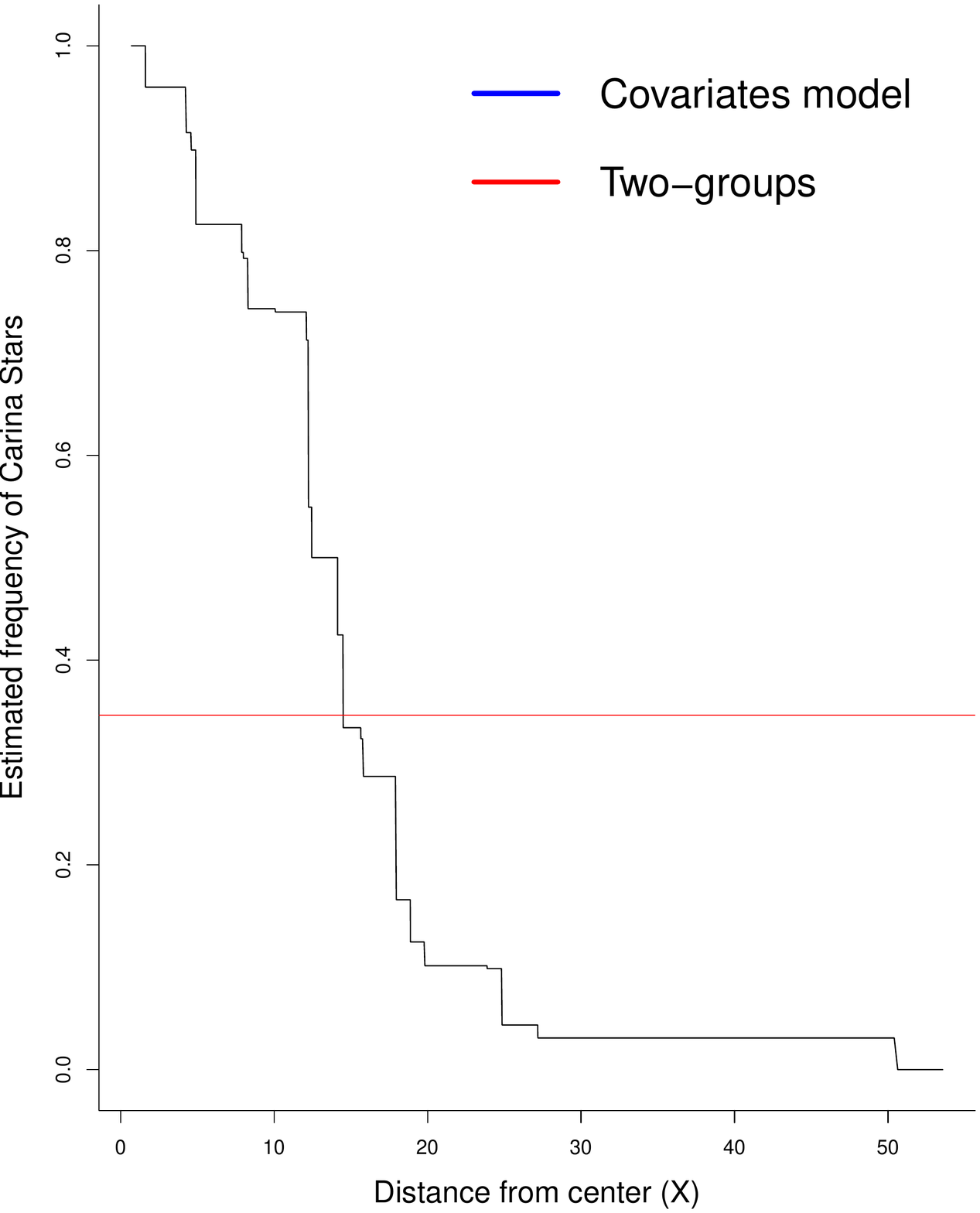}
	\includegraphics[width=22cm,height=6cm,keepaspectratio]{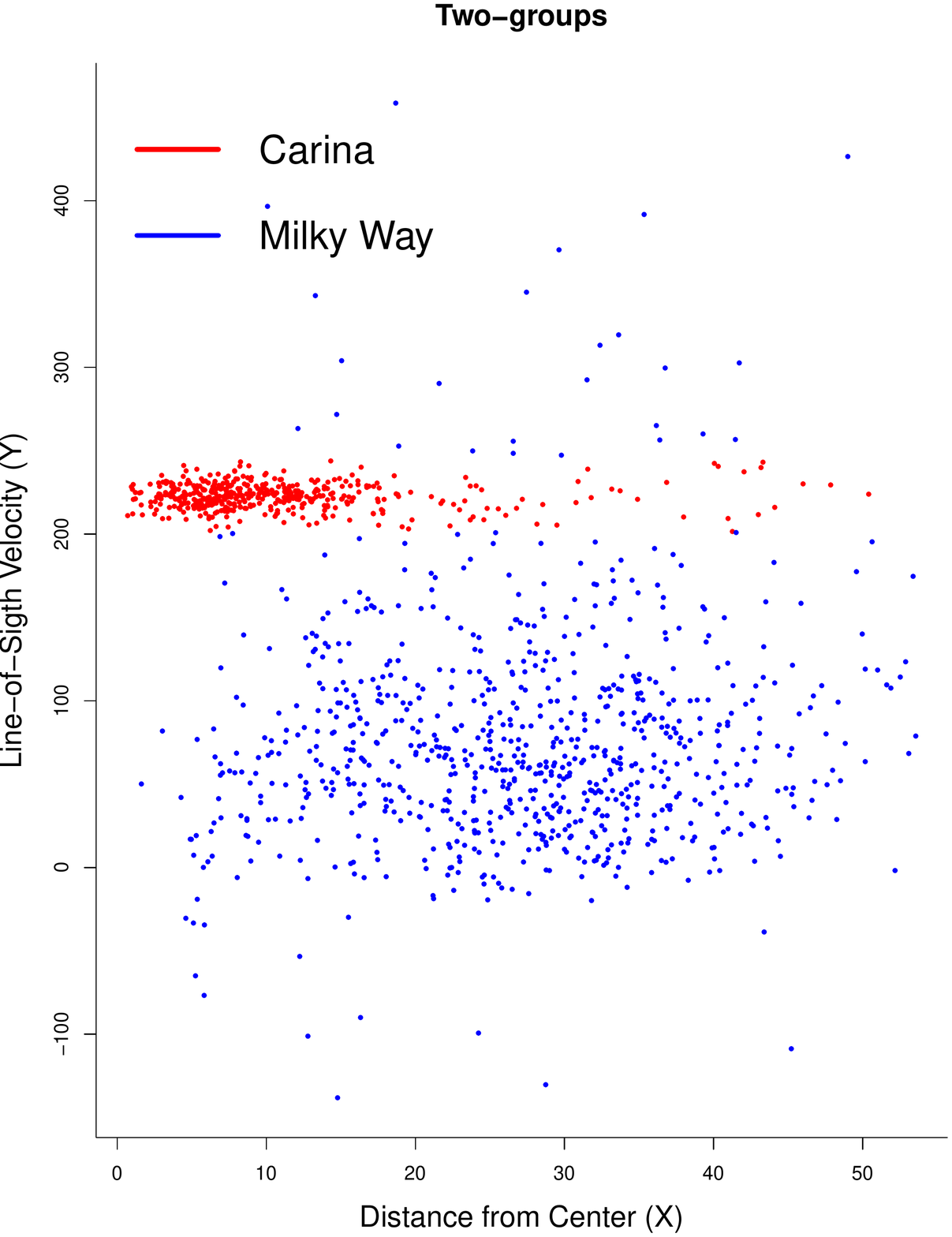}
	\includegraphics[width=22cm,height=6cm,keepaspectratio]{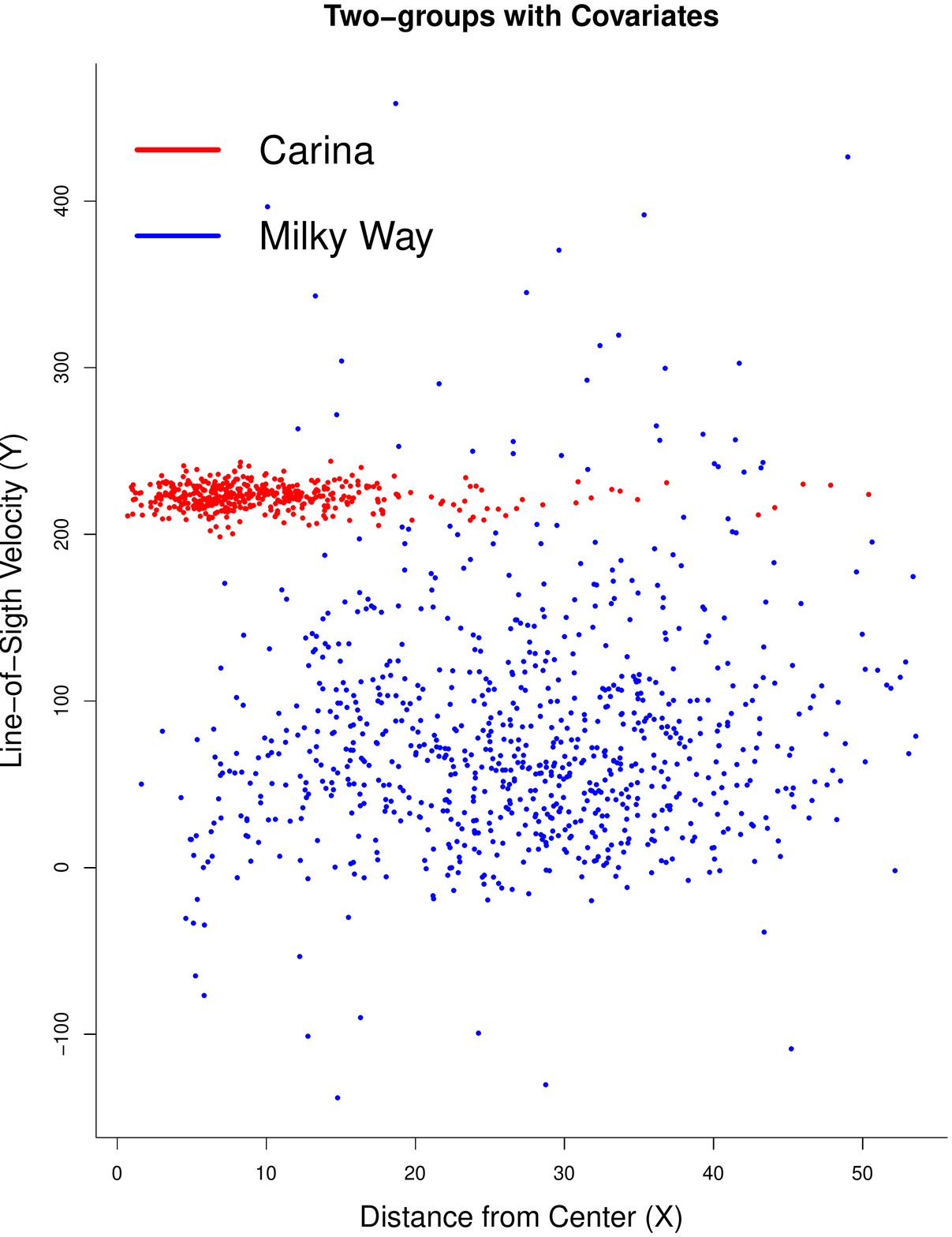}
	\caption{The left panel shows how the estimate of $\pi^*(\cdot)$ varies with $X$. The red horizontal line denotes the estimated $\bar\pi$ ($\approx 3.4$). The center and right panels illustrate the rejections from the two-groups model and our model respectively.}
	\label{fig:carina_analysis}
\end{figure}

For the purpose of estimating $\mu$, $\sigma^2$ and $\pi^*(\cdot)$, we use the EM algorithm along with PAVA (see~\cref{pinonpara}  for the details). Once we have the required estimates, we then compute plug-in estimates for the lFDRs and classify stars with estimated lFDR $\leq 0.5$ as being members of the Carina dSph. The left panel of Figure~\ref{fig:carina_analysis} clearly shows that the estimated proportion of Carina stars varies with $X$, which is not captured by the two-groups model. The other two plots in Figure~\ref{fig:carina_analysis} show a comparison between our analysis and that from using a standard two-groups model. We see that, according to our model (see~\eqref{model:astro}), some stars with slightly larger line-of-sight velocities (see the right panel in Figure~\ref{fig:carina_analysis}) that have large $X$ values are not detected as Carina members. In fact, our proposed method detects $14$ fewer stars as members of the Carina dSph, compared to the two-groups model, all of which have large $X$ values. 

\section{General discussions}\label{sec:gen}
In this section, we discuss certain aspects of this paper which are interesting in their own right but not directly related to our contributions. We further present some recommendations with respect to our proposed methods that may be of interest to the practitioner.

\subsection{Effect size}\label{subsec:esize}
To further motivate the study of $\Fga$, consider the following real example on prostate data investigated by many authors, see e.g.,~\cite{Efron10},~\cite{patra2016estimation}. The prostate data contains genetic expression levels for $n= 6033$ genes for $m = 102$ men, $m_1 = 50$ normal control subjects and $m_2 = 52$ prostate cancer patients. Without going into the biology involved, the principal goal of the study was to discover a small number of  ``interesting'' genes, i.e., genes whose expression levels differ between the cancer and control patients. Such genes, once identified, might be further investigated for a causal link to prostate cancer development.
The prostate data is a 6033 $\times$ 102 matrix $\mathbb{X}$ having entries $x_{ij}=\mbox{expression level for gene } i \mbox{ on patient }j$, $i=1,2,\ldots,n$, and $j=1,2, \ldots,m,$ with $j=1,2,\ldots,50$, for the normal controls, and $j=51,52,\ldots,102$, for the cancer patients. Let $\bar{x}_i(1)$ and $\bar{x}_i(2)$ be the averages of $x_{ij}$ for the normal controls and for the cancer patients, respectively, for gene $i$. The two-sample $t$-statistic for testing significance of gene $i$ is $$t_i=\{\bar{x}_i(1)-\bar{x}_i(2)\}/s_i,$$ where $s_i $ is an estimate of the standard error of  $\bar{x}_i(1)-\bar{x}_i(2)$, i.e.,
$s_i^2 = (1/50 + 1/52) [\sum_{j=1}^{50} \{x_{ij}-\bar{x}_i(1)\}^2 + \sum_{j=51}^{102} \{x_{ij}-\bar{x}_i(2)\}^2 ]/100.$ Note that $$t_i=\{\bar{x}_i(1)-\bar{x}_i(2)\}/s_i \approx Z_i + \frac{\mu_i(2) - \mu_i(1)}{\sigma_i/10}$$ where $Z_i \sim N(0,1)$, and  $\mu_i(1)$ and $\mu_i(2)$ denote the population means of expression levels for the normal and cancer patients, respectively, for gene $i$. Letting $\Delta_i \defeq {\mu_i(2) - \mu_i(1)}/{\sigma_i}$ denote the {\it effect size} for the $i$'th gene, we can assume that $t_i$'s are approximately normal with mean $10 \Delta_i$ and variance 1. Under the assumption that $\Delta_i$'s are i.i.d.~(with $10 \Delta_i \sim G$) and that $t_i$'s are (approximately) independent, the $t_i$'s are (approximately) i.i.d.~having density $$f(z) \defeq \bar \pi \int \phi(z-u) dG(u) + (1 - \bar \pi) \phi(z),\qquad z \in \R$$ which is a two-groups model (as~\eqref{eq:TwoComp}) where $F_1^*$ has a density that belongs to $\Fga$ (and $\bar \pi \in [0,1]$ denotes the proportion of non-null genes). 

\subsection{More on identifiability in the presence of discrete covariates}\label{sec:idenmore}

\begin{corollary}\label{revide}
	Consider the model $\Ps(\Pi, \F)$, with $\Pi\defeq \Pig$ where $g(x)=(1+\exp(x))^{-1}$ or $g(x)=\Phi(x)$, and $\mathcal{F}\defeq \Fga$ or $\Fde$. Suppose $\samp=\samp_1\times\samp_2$ where $\samp_1\in\mathbb{R}^{p_1}$, $\samp_2\in\mathbb{R}^{p_2}$, $p_1+p_2=p$, and the probability measure $m$ induces a marginal distribution $m_1$ on $\samp_1$ where $m_1$ is supported on a non-empty countable subset of $\samp_1$. Let $m_{2|1}(A_2,a_1)$ denote the conditional probability of $A_2\subseteq\samp_2$ given $a_1\in\samp_1$ under $m$. Assume that there exists $\tilde{x}_1\in\samp_1$ and a non-empty open set  $\samp'_2\subseteq\samp_2$ such that $m_1(\{\tilde{x}_1\})>0$  and $m_{2|1}(\cdot,\tilde{x}_1)$ assigns strictly positive probability to every open ball in $\samp'_2$. Also suppose that $F_0\neq F_1^*$ and $\pi^* \in  \Pig$ is given by $\pi_1^*(x)\defeq g(\beta_0^* + (\beta_1^*)^\top x_1+(\beta_2^*)^\top x_2)^{-1}$ for $(x_1,x_2) \in \samp_1\times\samp_2$ and some $(\beta_0^*,\beta_1^*,\beta_2^*) \in \mathbb{R}\times \R^{p_1}\times \R^{p_2}$ and non-zero vector $\beta_2^* \in \R^{p_2}$. Then $P_{(\pi^*, F_1^*)}$ is identifiable if $\beta_2^* \ne 0$.
\end{corollary}

The assumption on $m$ in the above corollary is perhaps a bit difficult to parse. So we consider this very simple example as an explanation. Suppose a multiple testing experiment has been conducted which has yielded $p$-values from $n$ different independent hypotheses with available covariates $(X_{i1},X_{i2})$ corresponding to the $p$-value $p_i$, for $1\leq i\leq n$. Assume $X_{i1}$ takes values in the set $\{0,1\}$ and $X_{i2}$ is a real-valued random variable. Then~\cref{revide} is applicable if, given $X_{i1}=0$ (or $1$), the conditional distribution of $X_{i2}$ is absolutely continuous. We believe that in light of the discussion immediately preceding~\cref{revide}, this assumption is pretty natural and one that analysts will be willing to make. The following simple example shows that without the condition $\beta_2^*\neq 0$, model~\eqref{mdl} may fail to be identifiable.

\begin{remark}\label{logitnoniden}
	Let $(Y_i,X_i)\in\R^2$, for $i=1, 2,\ldots, n$, be random variables drawn according to~\eqref{mdl} where $m$ is supported on $\{0,1\}$, $\Pi\defeq\logi$, $F_0(\cdot)=\Phi(\cdot)$ and $F_1^*\in\F\defeq \Fga$. Also suppose $\pi^*(x)\defeq \pi^*_{\beta_0,\beta_1}(x)=(1+\exp{(\beta_0+\beta_1x)})^{-1}\in\logi$, with $\beta_0=\log{2}$ and $\beta_1=-\log{2}$. Next, define $\tilde{\beta}_0=-\log{2}$, $\tilde{\beta}_1=\log{(2/3)}$ and note that $\pi^*_{\tilde{\beta}_0,\tilde{\beta}_1}(\cdot)$ satisfies $\pi^*_{\beta_0,\beta_1}(x)=2\pi^*_{\tilde{\beta}_0,\tilde{\beta}_1}(x)$ for $x\in\{0,1\}$.
	This implies that condition~\eqref{bid.eq1} holds in~\cref{lem:bid} with $c=0.5$. Observe that, as $F_0\in\Fga$ and $\Fga$ is a convex class of densities, condition~\eqref{bid.eq2} in~\cref{lem:bid} is trivially satisfied. Therefore, by an application of~\cref{lem:bid}, we get that model~\eqref{mdl} is non-identifiable in this setting.
\end{remark}  

\subsection{Estimation of \texorpdfstring{$f_1^*$} ~:~role of covariates (further details)}\label{sec:rolecov}

\begin{figure}
	\centering
	\begin{subfigure}[b]{0.45\textwidth}
		\centering
		\includegraphics[width=\textwidth]{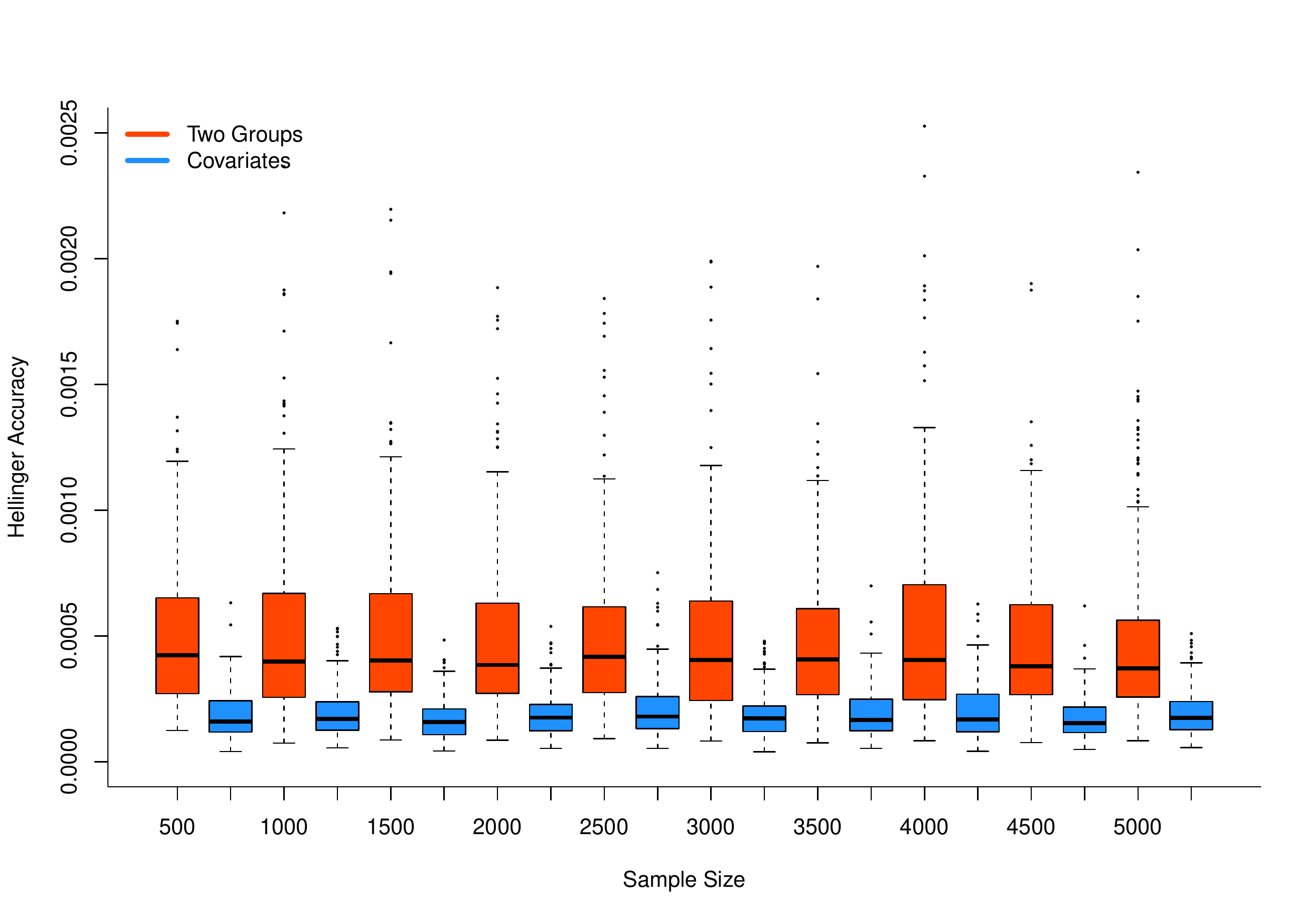}
		\caption{ }
		\label{fig:f1_estimation_11}
	\end{subfigure}
	\hfill
	\begin{subfigure}[b]{0.45\textwidth}
		\centering
		\includegraphics[width=\textwidth]{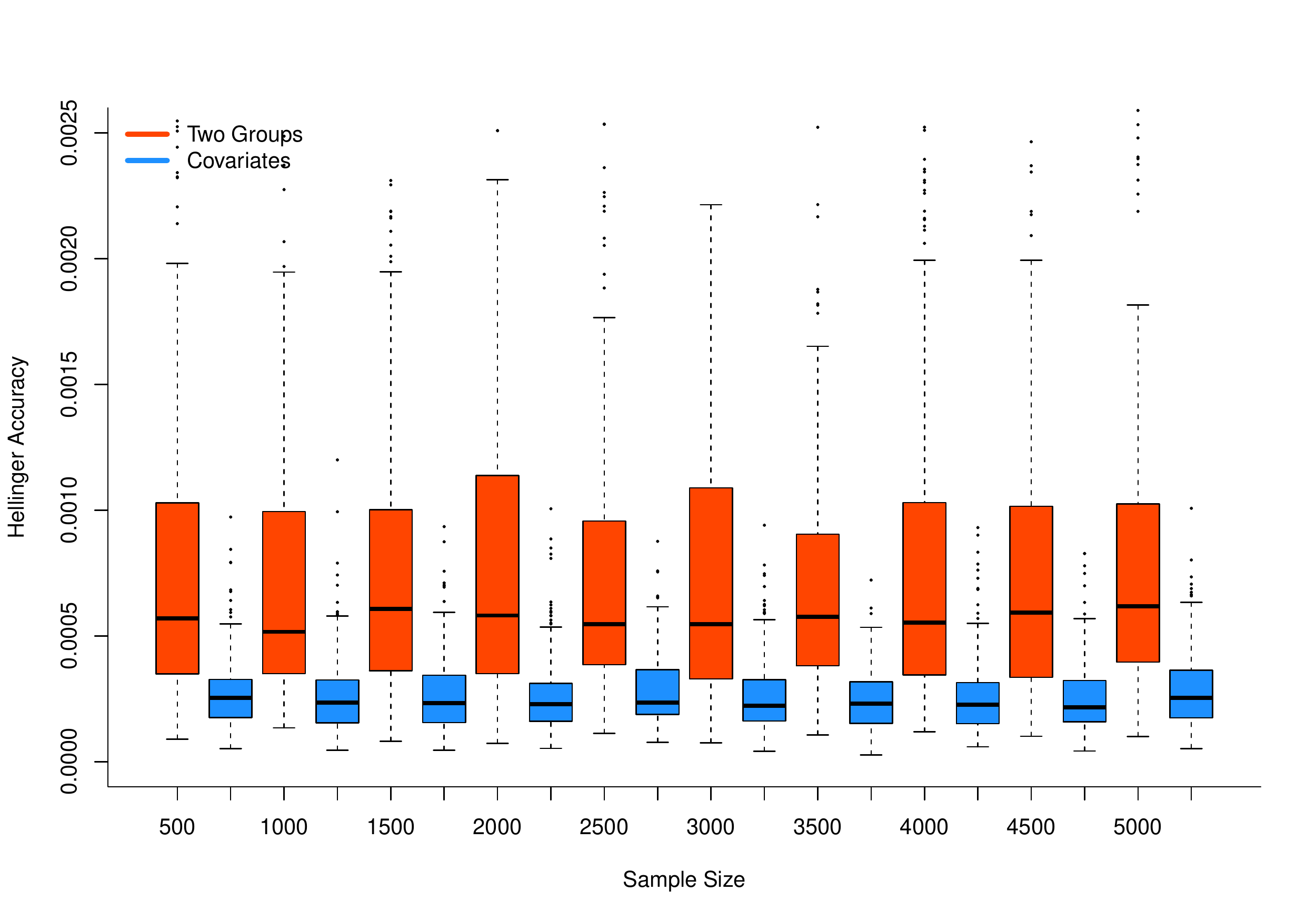}
		\caption{ }
		\label{fig:f1_estimation_12}
	\end{subfigure}
	\vspace{5mm}
	\begin{subfigure}[b]{0.45\textwidth}
		\centering
		\includegraphics[width=\textwidth]{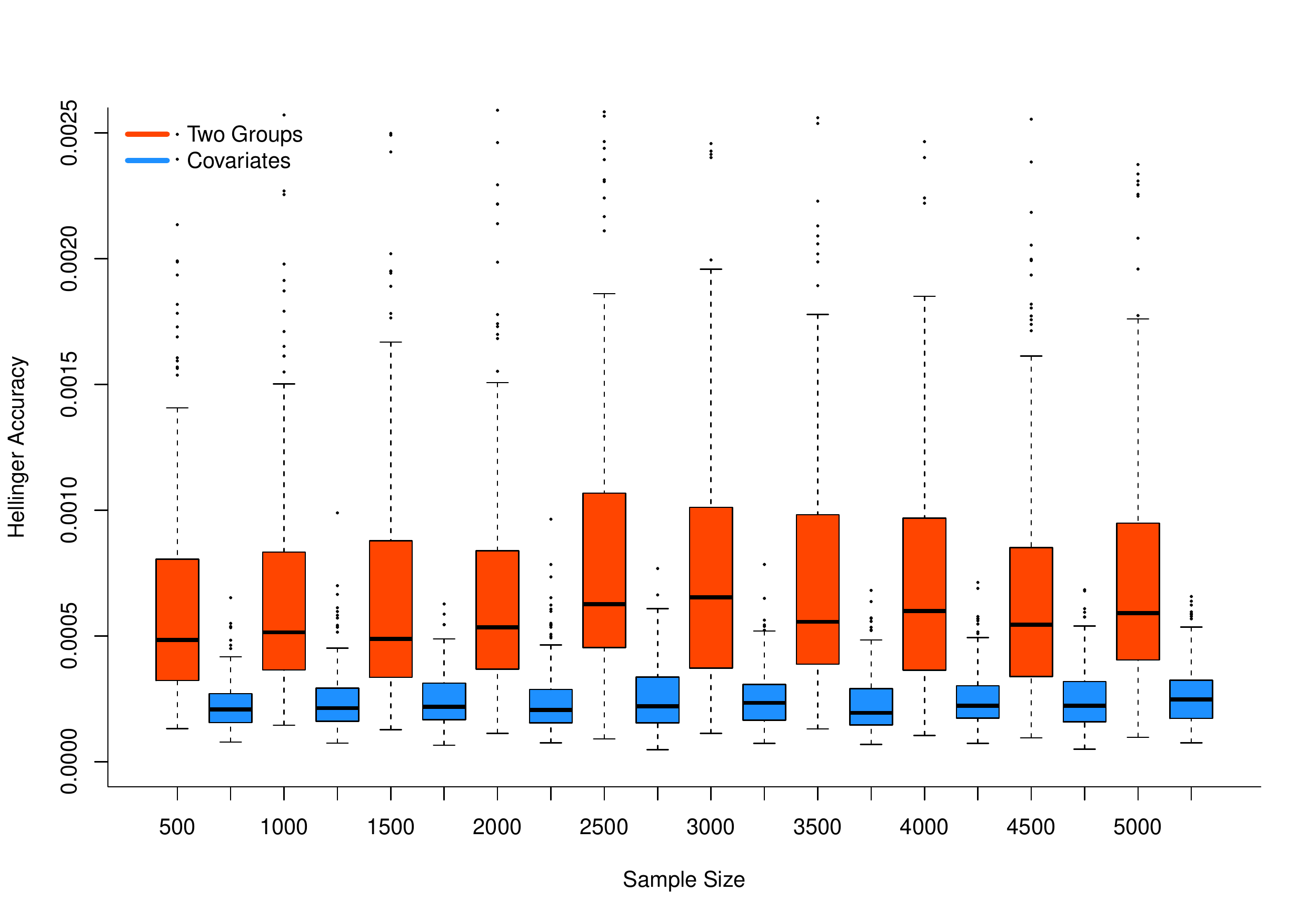}
		\caption{ }
		\label{fig:f1_estimation_13}
	\end{subfigure}
	\hfill
	\begin{subfigure}[b]{0.45\textwidth}
		\centering
		\includegraphics[width=\textwidth]{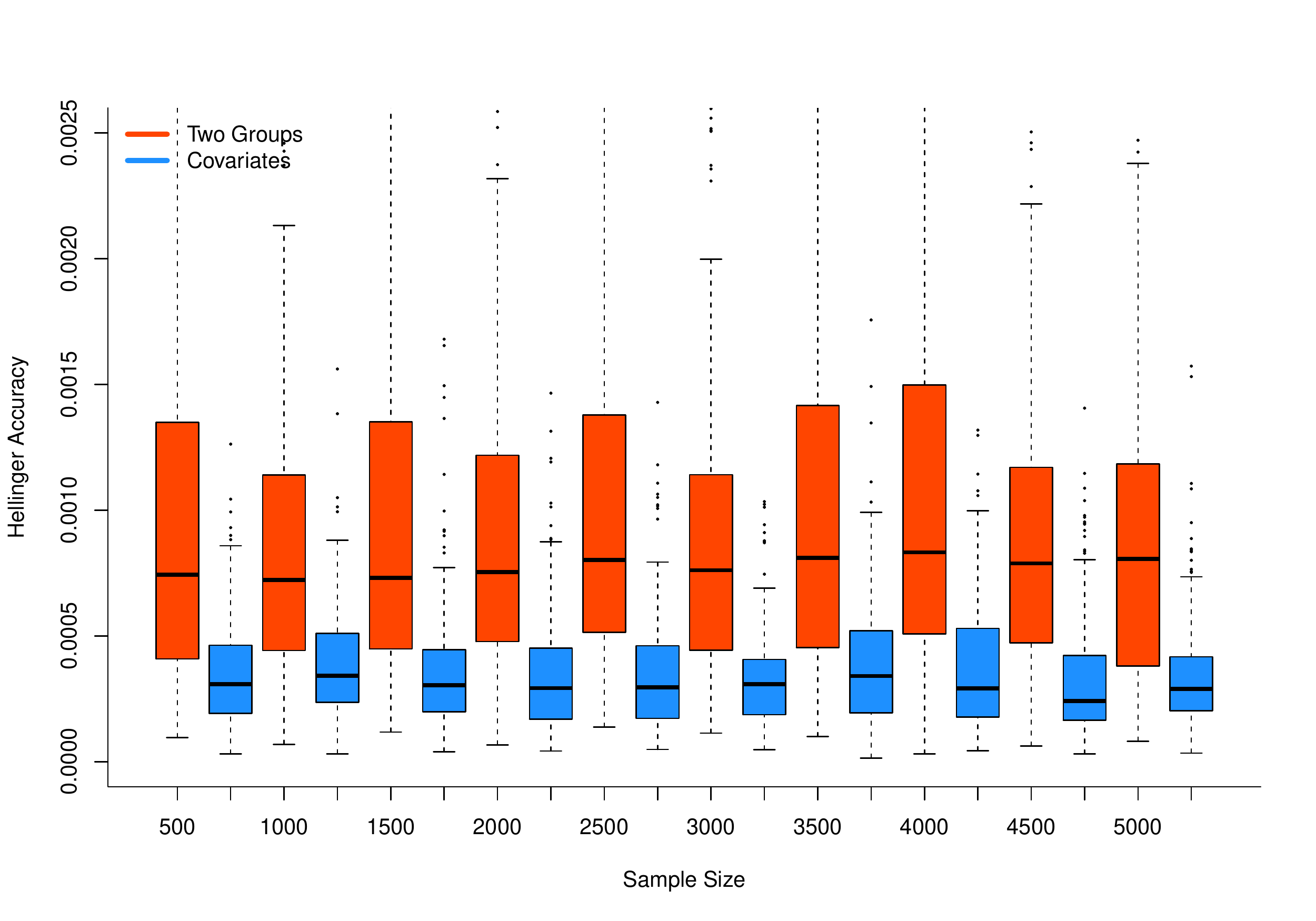}
		\caption{ }
		\label{fig:f1_estimation_14}
	\end{subfigure}
	\vspace{-0.5cm}
	\caption{Boxplots of the Hellinger distances in the estimation of $f_1^*$ via the two-groups model~\eqref{eq:TwoComp} (in red) and the two-groups model with covariates~\eqref{denmodel} (in blue). The four plots correspond to settings (A).(1)-(A).(4) in~\citet[Section 3]{ScottEtAl15}.}
	\label{fig:f1_estimation}
\end{figure}

Note that the marginal density of $\{Y_i\}_{i=1}^n$ under model~\eqref{denmodel} is identical to the well-known two-groups model~\eqref{eq:TwoComp}; recall the definition of $\bar\pi$ from~\cref{sec:MargMeth-I}. Through the following simulations we exhibit that utilizing covariate information indeed leads to more accurate estimation of $f_1^*$. For this demonstration, we generate datasets from four simulation settings, where $\pi^*(\cdot)$ is generated using setting (A) and $f_1^*$ is generated using settings (1)-(4) from \citet[Section 3]{ScottEtAl15}. First, we compute the MLE $\hat f_1^{\text{(TG)}}$ of $f_1^*$ based on the two-groups model~\eqref{eq:TwoComp} assuming $\bar \pi$ is known. We compare this with the MLE of $f_1^*$ based on model~\eqref{denmodel} assuming the function $\pi(\cdot)$ is known (call this estimate $\hat{f}_1^{\text{(cov)}}$). In particular, we estimate the Hellinger distance between each of these two estimators and $f_1^*$,  via Monte Carlo simulations. The resulting plots are shown in Figure \ref{fig:f1_estimation}. In our simulations both the above estimators were computed using the same strategy outlined in~\cref{dengauss} (with obvious modifications).

As can be clearly seen from the boxplots in Figure \ref{fig:f1_estimation}, using the information present in the covariates leads to significantly more accurate estimates of $f_1^*$. While similar simulations were done for many more settings, the conclusions remained the same and we therefore refrain from presenting all of our results.

\subsection{Do we get approximate MLEs?}\label{appMLEsim}
Once again, recall the statement of~\cref{sr1}. Note that we have derived near parametric rates of convergence for any approximate MLE (AMLE as defined in~\cref{sec:GaussMix}) for the conditional density of $Y|X$, see~\eqref{denmodel}. Therefore, it is natural to inspect whether our proposed methods yield AMLE's for this conditional density. Table~\ref{table:Likcomp} addresses this question.
\begin{table}[h]
	\centering
	\begin{tabular}{| l | r | r | r | r | r |}
		\hline
		Settings & \texttt{M - I} & \texttt{M - II} & \texttt{fMLE + M - I} & \texttt{fMLE + M - II} & \texttt{fMLE + FDRreg} \\
		\hline
		(A)(i) & 1.00 & 0.95 & 1.00 & 1.00 & 1.00 \\ 
		(A)(ii) & 1.00 & 1.00 & 1.00 & 1.00 & 1.00 \\ 
		(A)(iii) & 1.00 & 0.65 & 1.00 & 1.00 & 1.00 \\ 
		(A)(iv) & 1.00 & 0.97 & 1.00 & 1.00 & 1.00 \\ 
		\hline
		(B)(i) & 1.00 & 1.00 & 1.00 & 1.00 & 1.00 \\ 
		(B)(ii) & 1.00 & 1.00 & 1.00 & 1.00 & 1.00 \\ 
		(B)(iii) & 1.00 & 0.85 & 1.00 & 1.00 & 1.00 \\ 
		(B)(iv) & 1.00 & 0.95 & 1.00 & 1.00 & 1.00 \\ 
		\hline
		(C)(i) & 1.00 & 0.98 & 1.00 & 1.00 & 1.00 \\ 
		(C)(ii) & 1.00 & 1.00 & 1.00 & 1.00 & 1.00 \\ 
		(C)(iii) & 0.96 & 0.77 & 1.00 & 1.00 & 1.00 \\ 
		(C)(iv) & 1.00 & 1.00 & 1.00 & 1.00 & 1.00 \\ 
		\hline
		(D)(i) & 0.91 & 0.74 & 1.00 & 1.00 & 1.00 \\ 
		(D)(ii) & 0.98 & 0.96 & 1.00 & 1.00 & 1.00 \\ 
		(D)(iii) & 0.76 & 0.34 & 1.00 & 1.00 & 1.00 \\ 
		(D)(iv) & 0.98 & 0.88 & 1.00 & 1.00 & 1.00 \\ 
		\hline
	\end{tabular} 
	\caption{Likelihood comparisons: Each column denotes the proportion of times the corresponding method yields an AMLE. The simulation parameters are identical to those described in the previous section, e.g., in~Figure~\ref{fig:pi_setting_A}. Here M - I and M - II represent Marginal - I and Marginal - II.} 
	\label{table:Likcomp}
\end{table} 
It is clear from Table~\ref{table:Likcomp} that the fMLE (with either one of the three possible initializations) is very reliable in terms of yielding AMLEs. The method marginal 1 seems to be the next best candidate. The performance of marginal 2 is somewhat fluctuating in this regard, which may be because of the fact that it arrives at estimates of $\pi^*(\cdot)$ by solving a non-convex optimization problem. 

\subsection{More on initialization of fMLE}\label{initfMLE}
\begin{figure}
	\centering
	\begin{subfigure}[b]{0.47\textwidth}
		\centering
		\includegraphics[width=\textwidth]{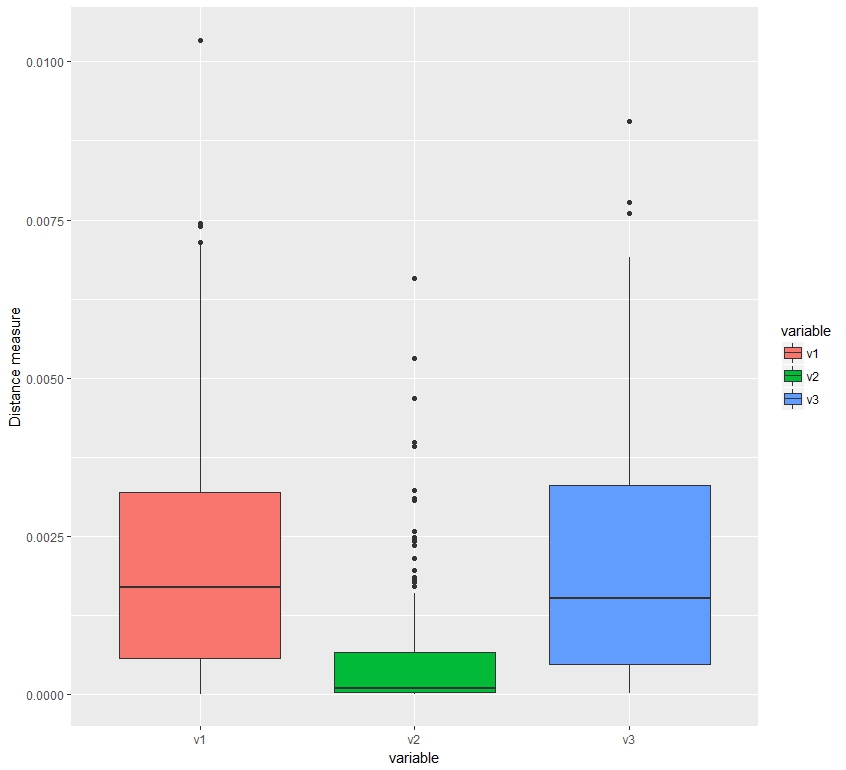}
		\caption{$f_1$ as in setting (i).}
		\label{fig:signaldenproximi}
	\end{subfigure}
	\hfill
	\begin{subfigure}[b]{0.47\textwidth}
		\centering
		\includegraphics[width=\textwidth]{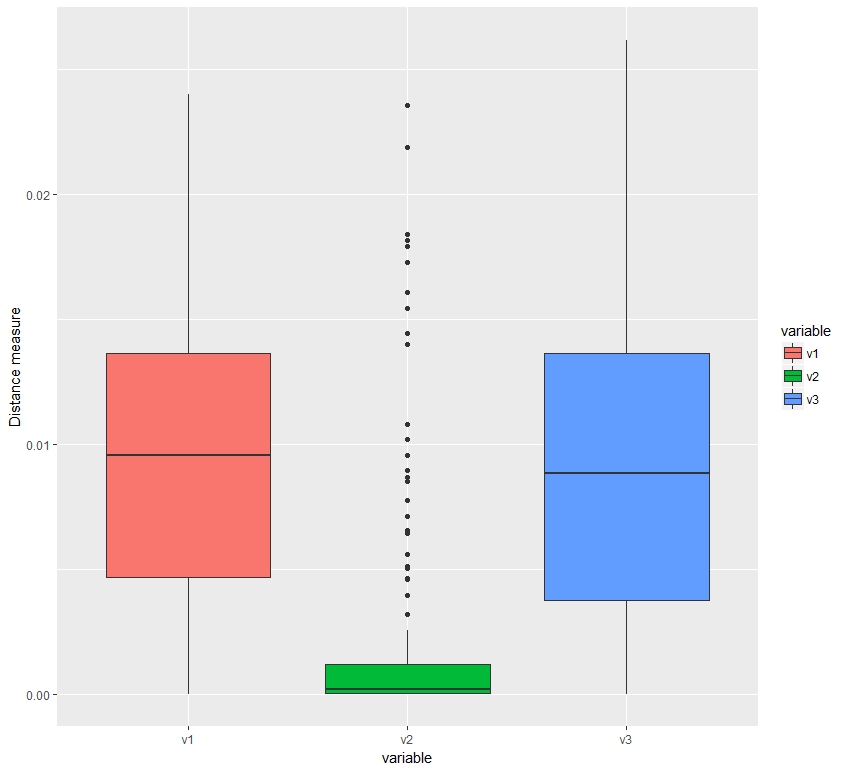}
		\caption{$f_1$ as in setting (ii). }
		\label{fig:signaldenproximii}
	\end{subfigure}
	\vspace{15mm}
	\begin{subfigure}[b]{0.47\textwidth}
		\centering
		\includegraphics[width=\textwidth]{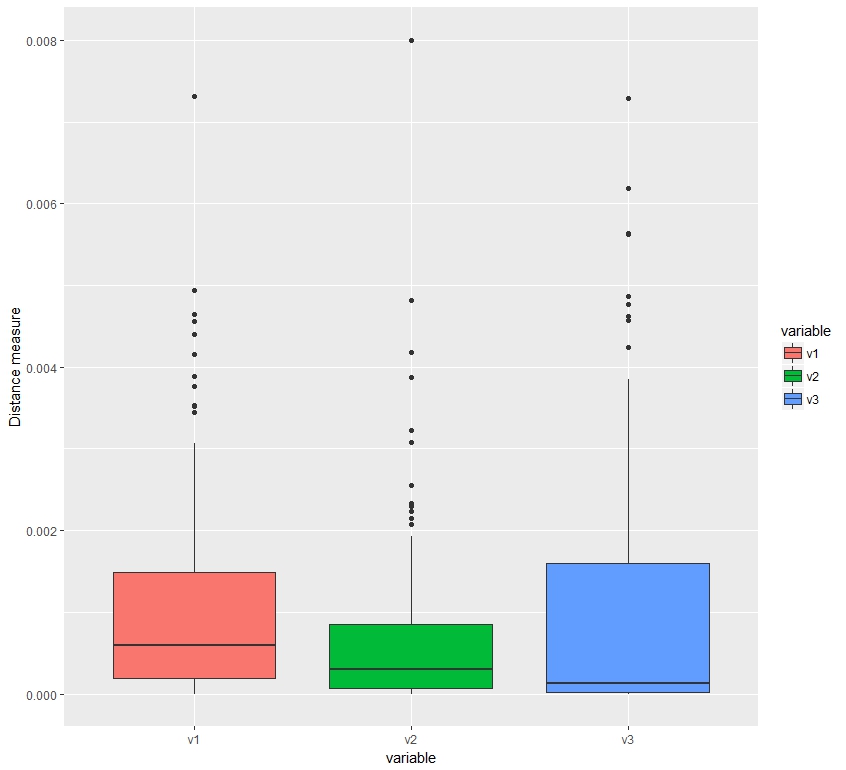}
		\caption{$f_1$ as in setting (iii).}
		\label{fig:signaldenproximiii}
	\end{subfigure}
	\hfill
	\begin{subfigure}[b]{0.47\textwidth}
		\centering
		\includegraphics[width=\textwidth]{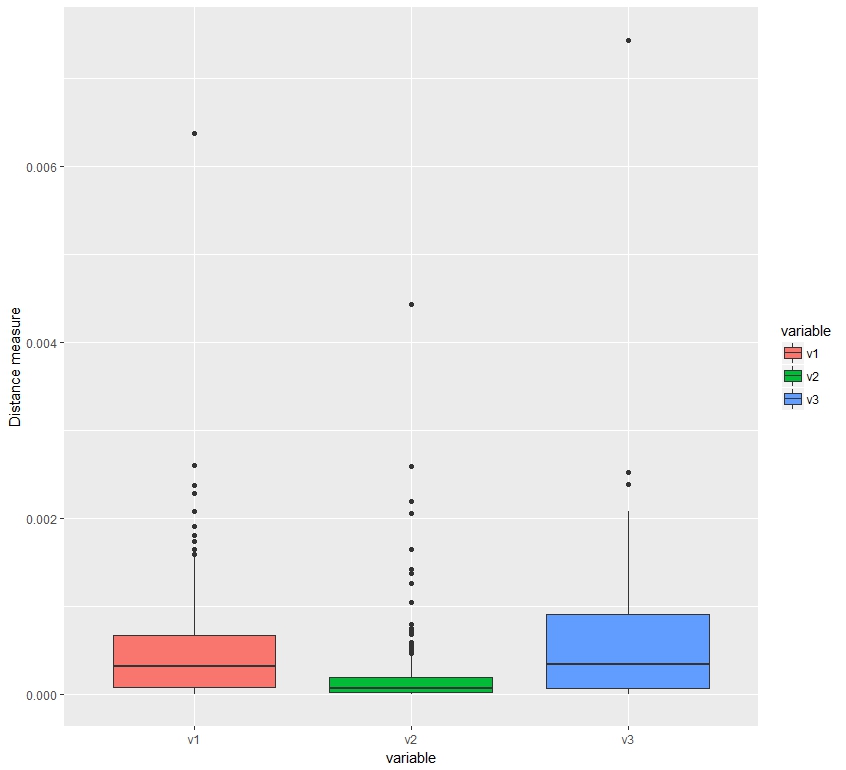}
		\caption{$f_1$ as in setting (iv). }
		\label{fig:signaldenproximiv}
	\end{subfigure}
	\vspace{-1.5cm}
	\caption{Each subplot shows the proximity between estimates of $f_1^*$ when fMLE is initialized with (v1) Marginal - I and FDRreg, (v2) Marginal - I and Marginal - II, and (v3) Marginal - II and FDRreg, based on the metric introduced in~\eqref{nmetric}. Here $s(\cdot)$ is from setting (A). }\label{fig:signaldenproxim}
\end{figure}
Recall Table~\ref{table:init_proportions} which illustrates that Marginal - I is the most common candidate for initializing the EM algorithm for the fMLE approach, based on our simulation settings. However, in practice, when we have a particular dataset, we will need to understand which initialization best suits the dataset at hand. In such a situation, a natural question to ask is --- ``how sensitive is the fMLE approach to its initialization?''. In other words, is it reasonable to expect that all $3$ initializations will yield {\it nearly} the same estimates for $(\pi^*(\cdot),f_1^*)$? Based on our simulations, we do not have a very definitive answer to this question. While the estimates do seem rather close for all $3$ initializations, we believe that the estimates from Marginal - I and Marginal - II initializations are much closer to each other than those with FDRreg initialization. To illustrate this observation, we use  
\begin{equation}\label{nmetric}
d(\hat{f}_1,\tilde{f}_1)=\frac{1}{n}\sum\limits_{i=1}^n \left(\hat{f}_1(Y_i)-\tilde{f}_1(Y_i)\right)^2
\end{equation}
to measure the proximity between $\hat{f}_1$ and $\tilde{f}_1$ which, in this context, would refer to estimates from two of the three possible initializations. We present some of the results in Figure~\ref{fig:signaldenproxim}. The main observation is that the distance (based on~\eqref{nmetric}) between estimates of $f_1^*$ from Marginal - I and Marginal - II are in general closer to $0$, have less variability as opposed to the other possible combinations. The observations are mostly similar across our chosen simulation settings, so we refrain from presenting all of them. 	

\subsection{Near non-identifiability (at moderate sample sizes)}\label{sec:non-iden}
\begin{figure}[H]
	\centering
	\begin{subfigure}[b]{0.47\textwidth}
		\centering
		\includegraphics[width=\textwidth]{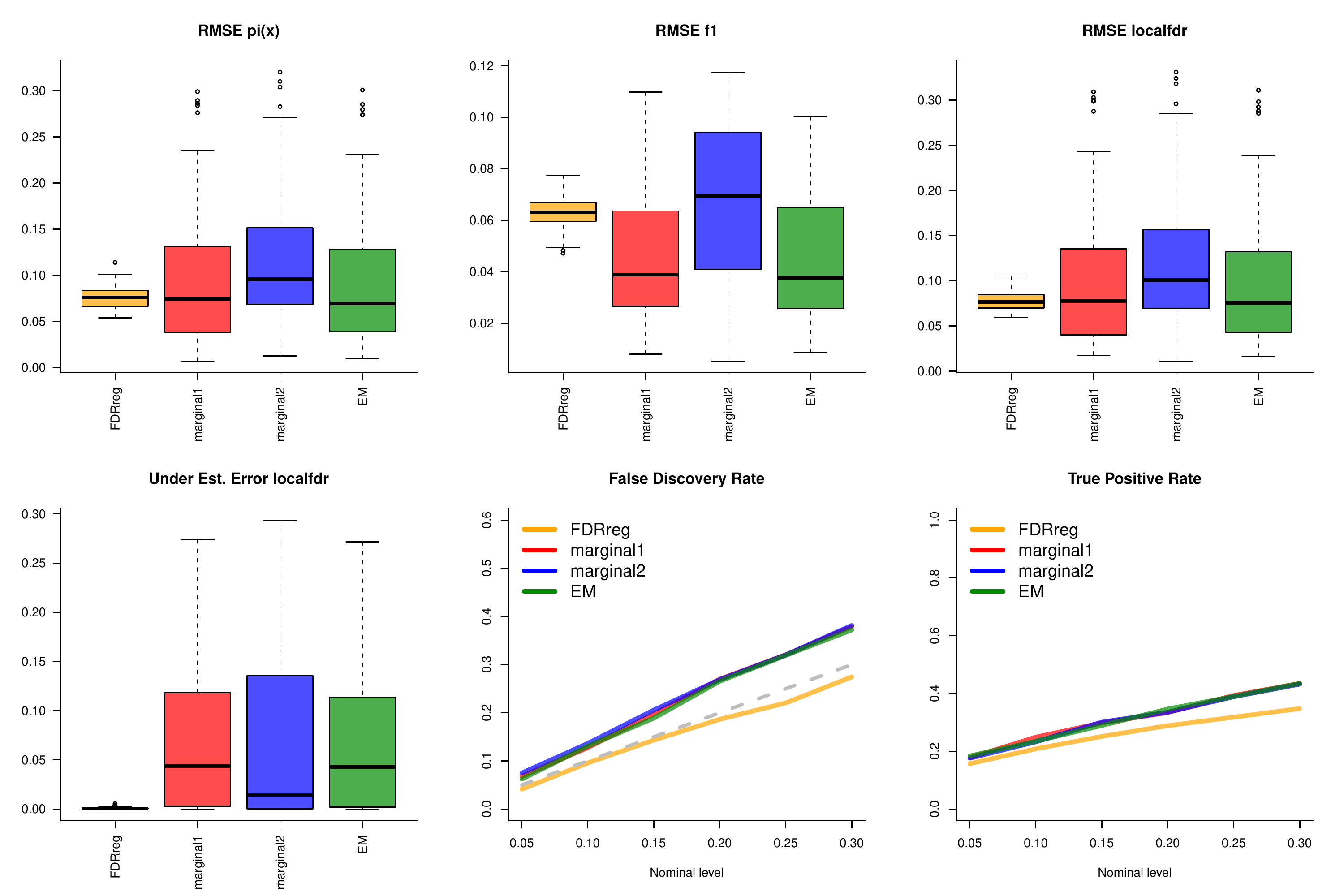}
		\caption{$f_1^*$ as in setting (i).}
		\label{fig:pi_setting_Bi}
	\end{subfigure}
	\hfill
	\begin{subfigure}[b]{0.47\textwidth}
		\centering
		\includegraphics[width=\textwidth]{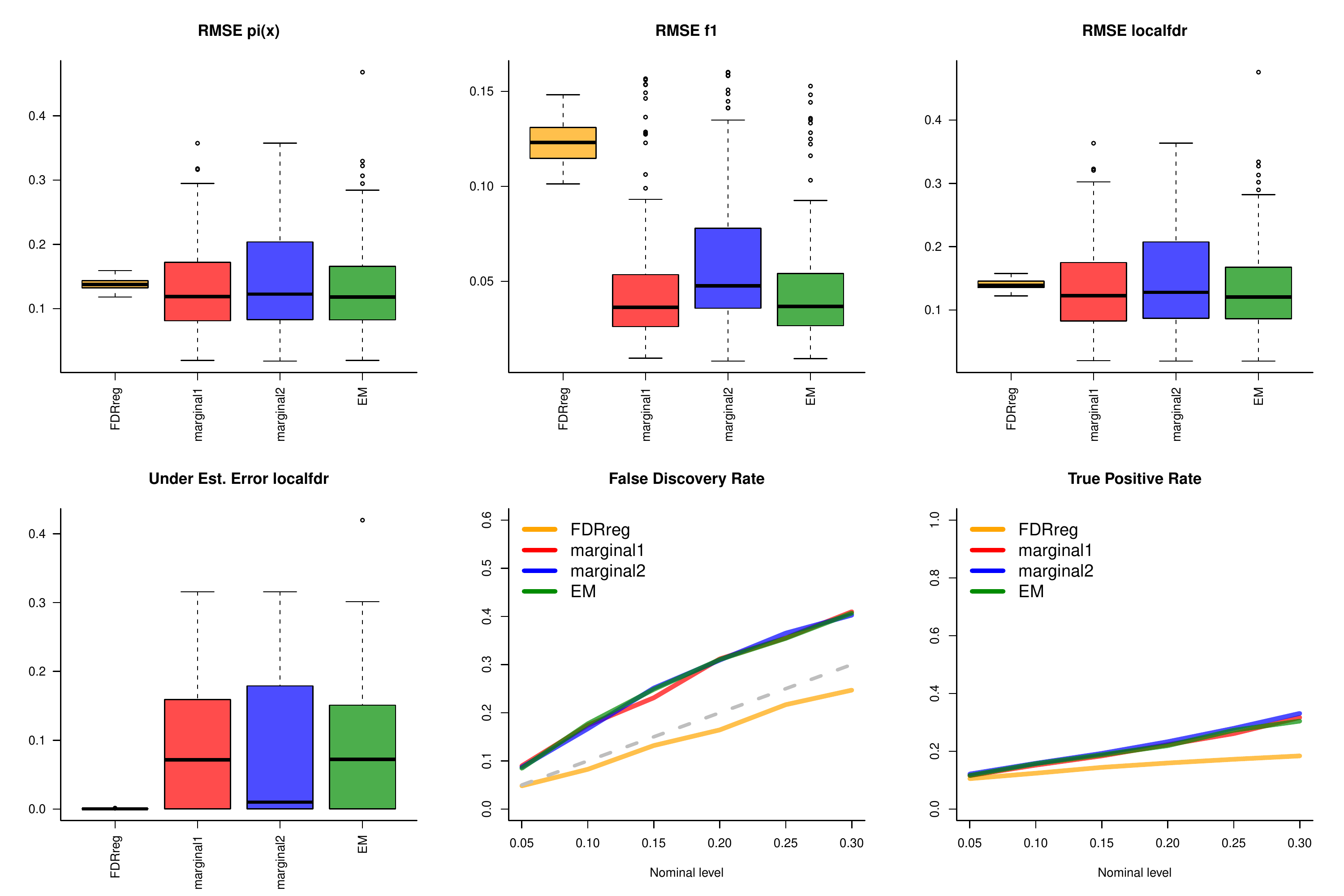}
		\caption{$f_1^*$ as in setting (ii).}
		\label{fig:pi_setting_Bii}
	\end{subfigure}
	\vspace{15mm}
	\begin{subfigure}[b]{0.47\textwidth}
		\centering
		\includegraphics[width=\textwidth]{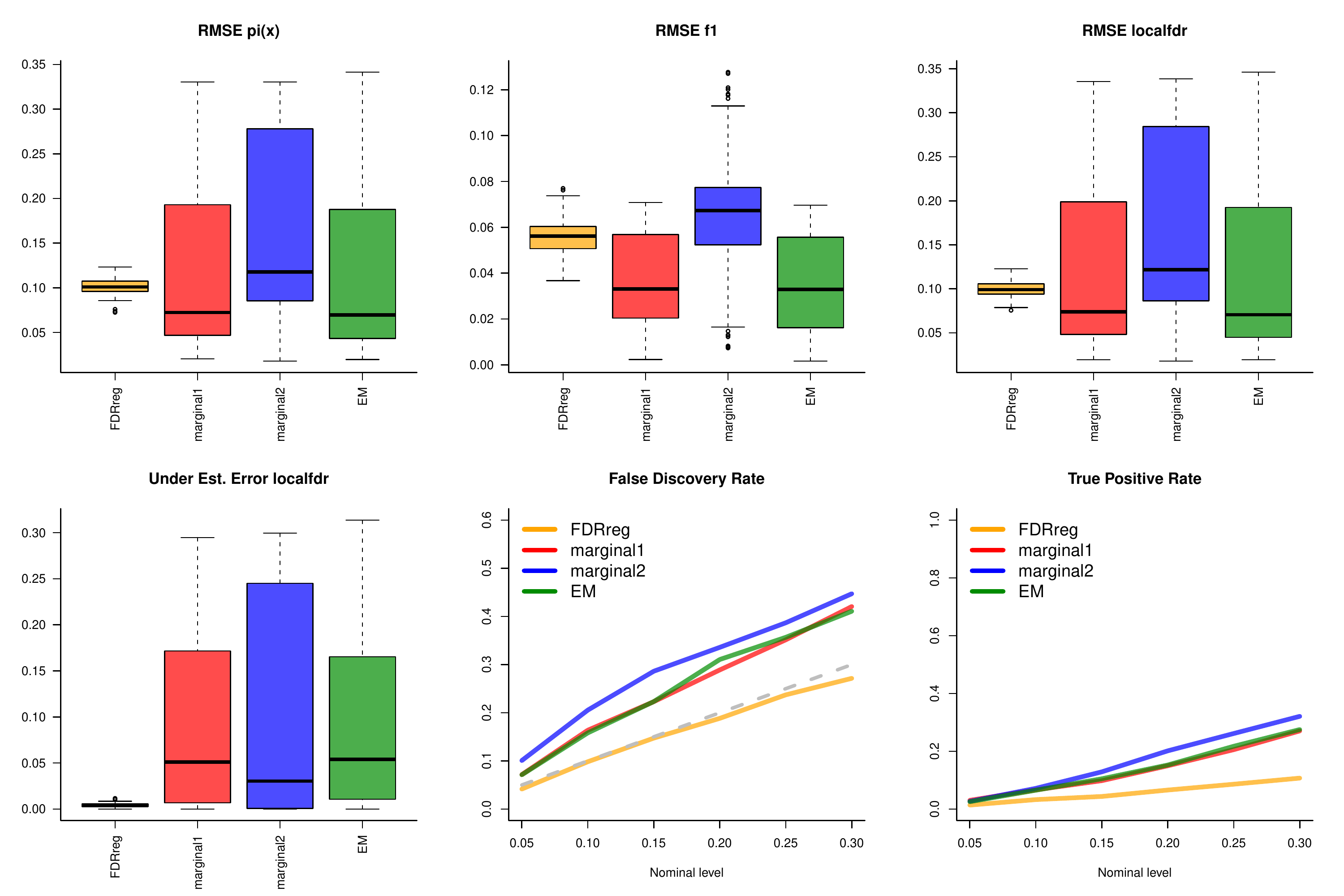}
		\caption{$f_1^*$ as in setting (iii). }
		\label{fig:pi_setting_Biii}
	\end{subfigure}
	\hfill
	\begin{subfigure}[b]{0.47\textwidth}
		\centering
		\includegraphics[width=\textwidth]{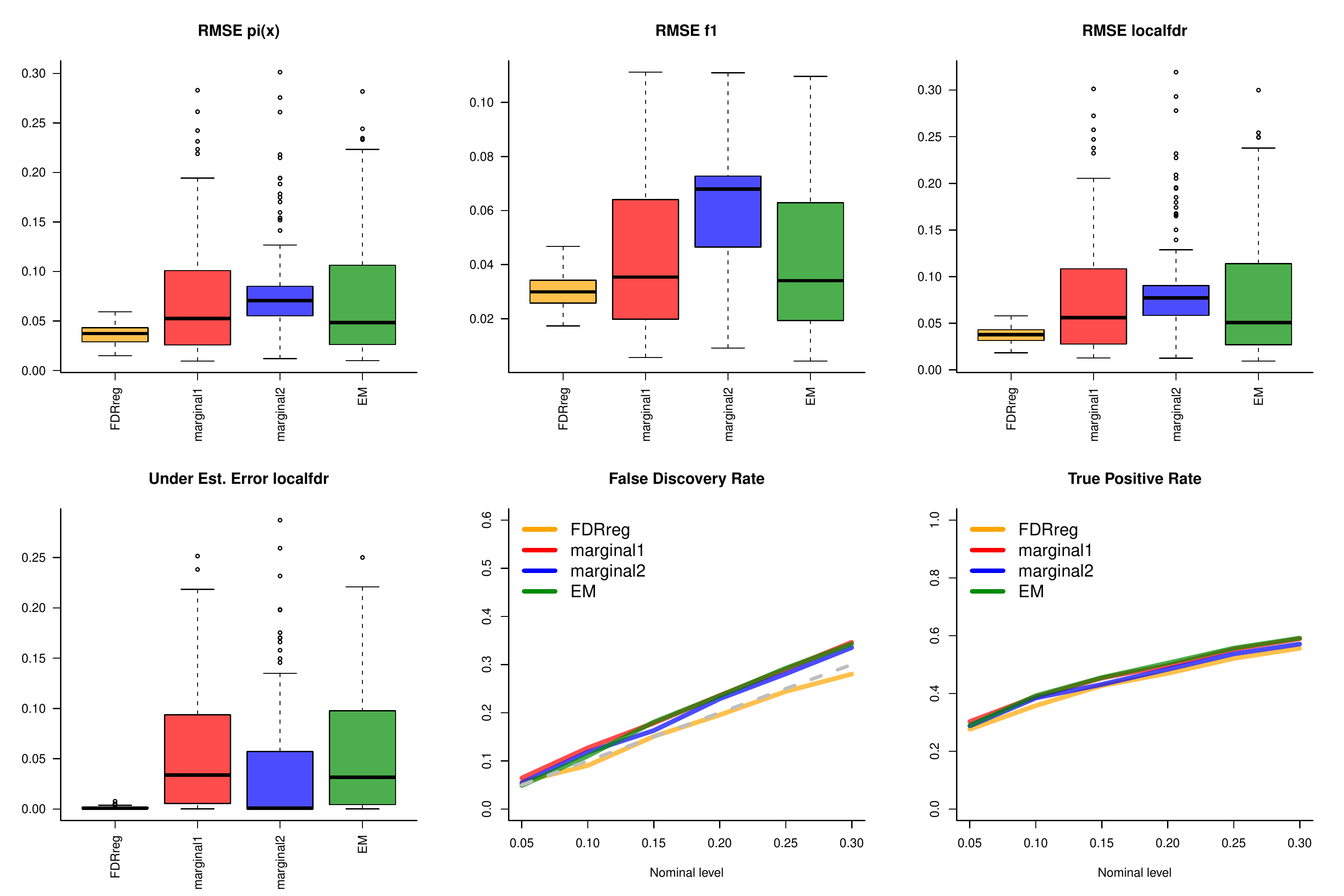}
		\caption{$f_1^*$ as in setting (iv).}
		\label{fig:pi_setting_Biv}
	\end{subfigure}
	\vspace{-1.5cm}
	\caption{Figure depicts the same plots as in Figure~\ref{fig:pi_setting_A} when $s(\cdot)$ is from setting (B).}
	\label{fig:pi_setting_B}
\end{figure}

Suppose $\{(Y_i,X_i)\}_{i=1}^n$ are i.i.d. samples from model~\eqref{denmodel} where $\pi^*\in\logi$ and $f_1^*\in\dga$, such that model~\eqref{mdl} is identifiable (as per~\cref{def:Iden}). 	As seen in Figure~\ref{fig:pi_setting_Biv} the fMLE does not perform well when $s(\cdot)$ is from setting (B) and $f_1^*(\cdot)$ is from setting (iv). In this section, we try to shed some light on scenarios where our methods may fail for small (to moderate) sample sizes, even under identifiability. 
We refer to this phenomenon as \textit{near non-identifiability}. The discussion in this section is mostly heuristic and we hope that a deeper and more rigorous understanding of this phenomenon will be developed in future work. 

For some parameter settings, our likelihood based estimators might not perform well for smaller sample sizes. This instability seems to arise from the possibility that there might be ``alternative pairs" of parameters $(\check \pi(\cdot), \check f_1) \in \Pi_{\text{logit}}\times \dga$ such that  $(1- \check\pi (x))\phi(\cdot) + \check\pi(x) \check f_1^{(\alpha)}(\cdot)$ is ``close'' to $(1- \pi^{*}(x))\phi(\cdot) + \pi^{*}(x) \check f_1(\cdot)$ (for every $x \in \mathcal{X}$). By ``closeness'' here we mean that their likelihood values are similar and is determined in a stochastic sense by $\pi^*(\cdot), f_1^*(\cdot), \mathcal{X}$ and $n$.

To illustrate this intuition, let us work with a simple parametric submodel of $\Pi_{\text{logit}}\times \dga$. Note that by~\cref{lem:bid}, there does not exist $\alpha\neq 1$ such that $\alpha\pi^*(\cdot)\in\logi$. Further, observe that for $\alpha \in A \defeq (0, (\sup_{x \in \mathcal{X}} \pi^*(x))^{-1}]$, $\alpha \pi^*(x) \in [0,1]$ for all $x\in \mathcal{X}$. We will refer to $A$ as the \textit{feasible set}. For some $\alpha$ in this feasible set $A$, we may be able to create an ``alternative pair" of parameters $(\check \pi^{(\alpha)}(\cdot), \check f_1^{(\alpha)}) \in \Pi_{\text{logit}}\times \dga$ such that  $(1- \check\pi^{(\alpha)}(x))\phi(\cdot) + \check\pi^{(\alpha)}(x) \check f_1(\cdot)$ is ``close'' to $(1- \pi^{*}(x))\phi(\cdot) + \pi^{*}(x) \check f_1^{(\alpha)}(\cdot)$ (for every $x \in \mathcal{X}$). Further, as $\alpha$ varies in $A$, especially when $\alpha$ is close to $1$, our likelihood based methods may not be able to distinguish between $(\check \pi^{(\alpha)}(\cdot), \check f_1^{(\alpha)})$ and the true parameter $(\pi^{*}(\cdot), f_1^{*})$. 
Next, we formalize the above through a few examples. We also describe the construction of $(\check \pi^{(\alpha)}(\cdot), \check f_1^{(\alpha)})$ for $\alpha \in A$.
\begin{figure}[H]
	\centering
	\includegraphics[width=\textwidth]{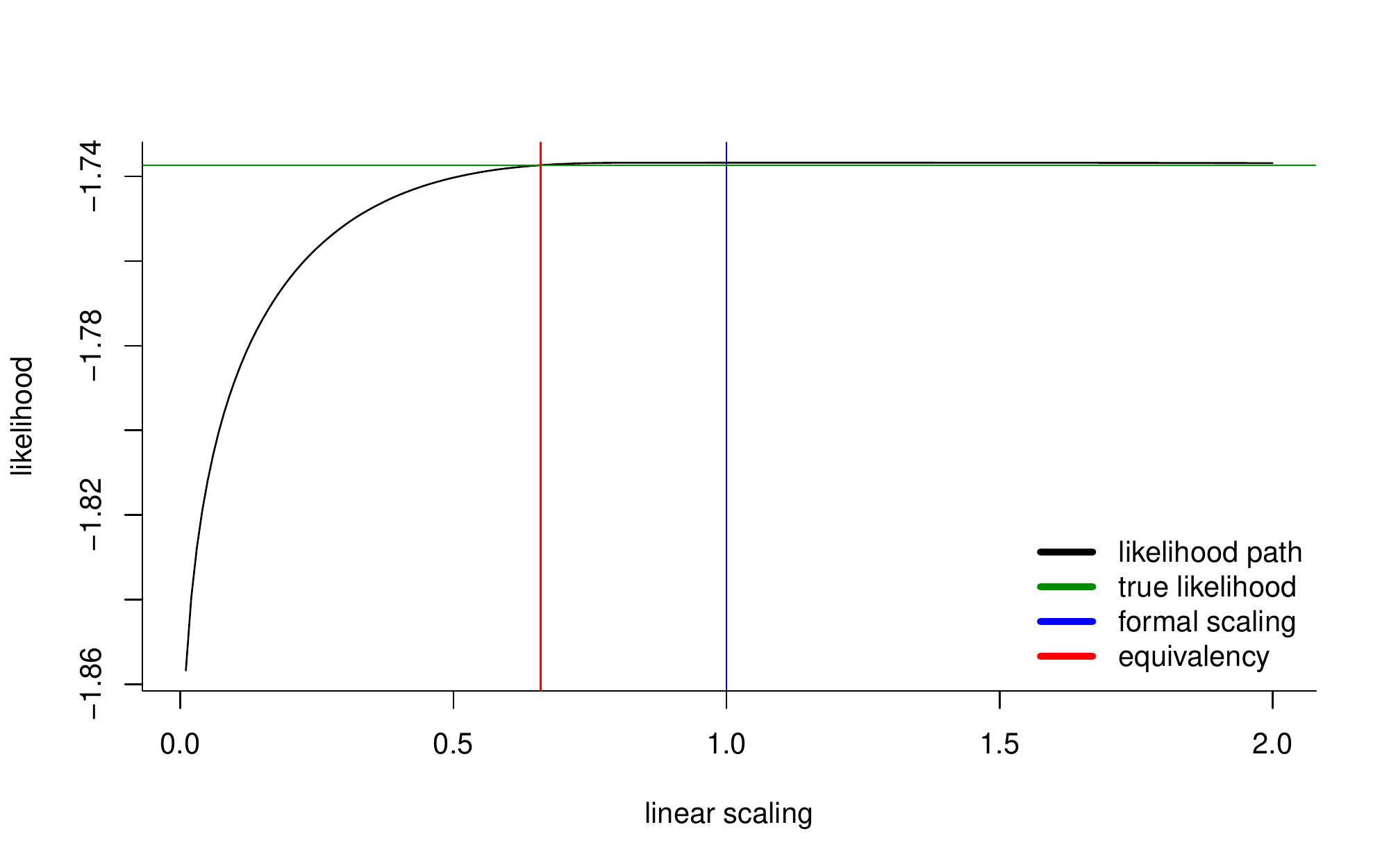}
	\caption{This plot shows $\alpha$ on the $x$-axis and the resulting likelihood value $\ell(\alpha)$ on the $y$-axis. The true normalized log-likelihood of the data is plotted as the green horizontal line. The true value of $\alpha$ (which corresponds to $\pi^*(\cdot)$) is of course $1$ (blue line). From the plot it is evident that this model could equivalently be expressed in terms of any $\alpha \in [2/3, 2]$; as depicted by the red line where $\ell(\alpha)$ initially rises above the likelihood for the true parameter.}
	\label{fig:lik_path}
\end{figure}
Consider the following simulation setting:
\begin{equation}\label{nonidsim}
s(x_1,x_2) = -3 + 1.5x_1 + 1.5x_2, ~~~~~~ g_1 = 0.48 N(-2, 1) + 0.04 N(0, 16) + 0.48 N(2, 1),
\end{equation}
where $\pi^*(x) = (1 + \exp\{ -s(x) \})^{-1}$ and $f_1^*(\cdot) = \int \phi(\cdot - \theta) g_1(\theta)d\theta$. The bivariate covariates $(X_1,X_2)$ are drawn uniformly from $\mathcal{X} = [0,1]^2$. For this setting one can check that $A = [0,2]$. For a fine grid of points $\alpha \in A$, at a resolution of $10^{-2}$, we now attempt to find parameter settings $(\check \pi^{(\alpha)}, \check f_1^{(\alpha)})$ which have ``similar'' likelihood values as the truth (for a given sample size). This can be approximately obtained in two easy steps:
\begin{itemize}
	\item[(i)] $\hat \beta^{(\alpha)}$ can be taken as the vector of least squares regression coefficients obtained by regressing $\text{logit}(\alpha \pi^*(X))$ on $X$. Analogously,  $\hat{\pi}_n^{(\alpha)}(X_i) = 1/(1 + \exp(-X_i^{\top}\hat \beta^{(\alpha)}))$.
	\item[(ii)] $\hat f_{1}^{(\alpha)}$ can be taken to be the constrained KWMLE:
	\begin{equation*}
	\argmax_{f \in \dga} \sum_i \log\{ (1- \hat\pi^{(\alpha)}(X_i))\phi(Y_i) + \hat\pi^{(\alpha)}(X_i) f(Y_i) \}.
	\end{equation*}
\end{itemize}
Note that all we have attempted to do above is to ``project'' $\alpha \pi^*(\cdot)$ onto $\Pi_{\text{logit}}$ to obtain $\check \pi^{(\alpha)}$, for each $\alpha \in (0,2]$. Then, fixing $\check \pi^{(\alpha)}$, $\check f_1^{(\alpha)}$ is defined as the ``closest'' (in terms of likelihood) member of $\dga$. This gives us a parametric submodel $\{(\check \pi^{(\alpha)}, \check f_1^{(\alpha)}): \alpha \in A\}$ in $\Pi_{\text{logit}}\times \dga$ and we can define the likelihood along this path as:
\begin{equation*}
\ell(\alpha) \defeq \frac{1}{n} \sum_{i=1}^n \log\{  (1- \hat\pi^{(\alpha)}(X_i))\phi(Y_i) + \hat\pi^{(\alpha)}(X_i) \hat f_{1}^{(\alpha)}(Y_i) \}, \qquad \mbox{for } \alpha \in A;
\end{equation*}
see Figure~\ref{fig:lik_path}. It is clear from the figure  that in terms of likelihood the ``alternate" parameter settings when $\alpha \in [2/3, 2]$, is practically indistinguishable from the truth, at this sample size, i.e., $n=10^4$.
\begin{figure}[H] 
	\centering
	\includegraphics[width = 0.9\textwidth]{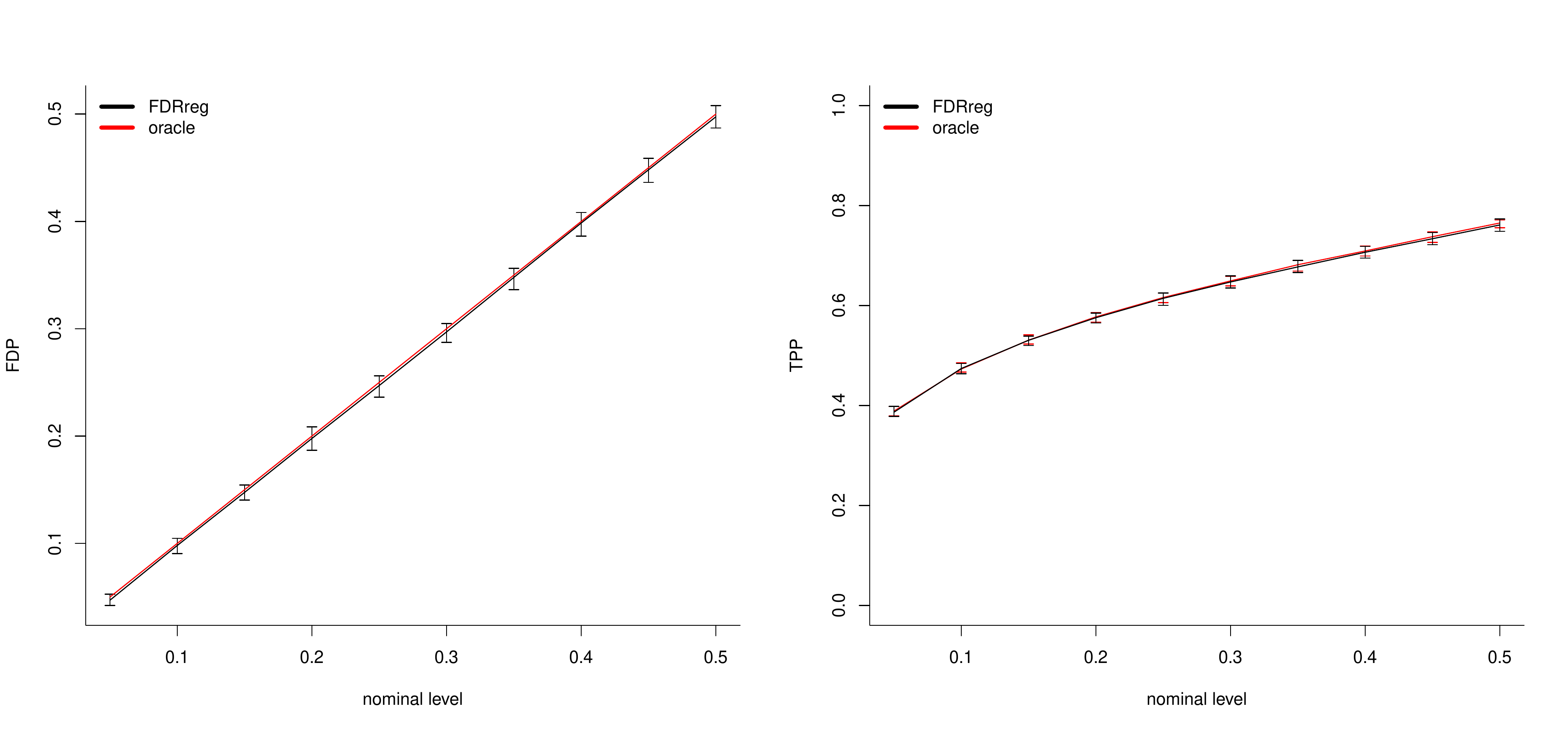}
	
	\includegraphics[width = 0.45\textwidth]{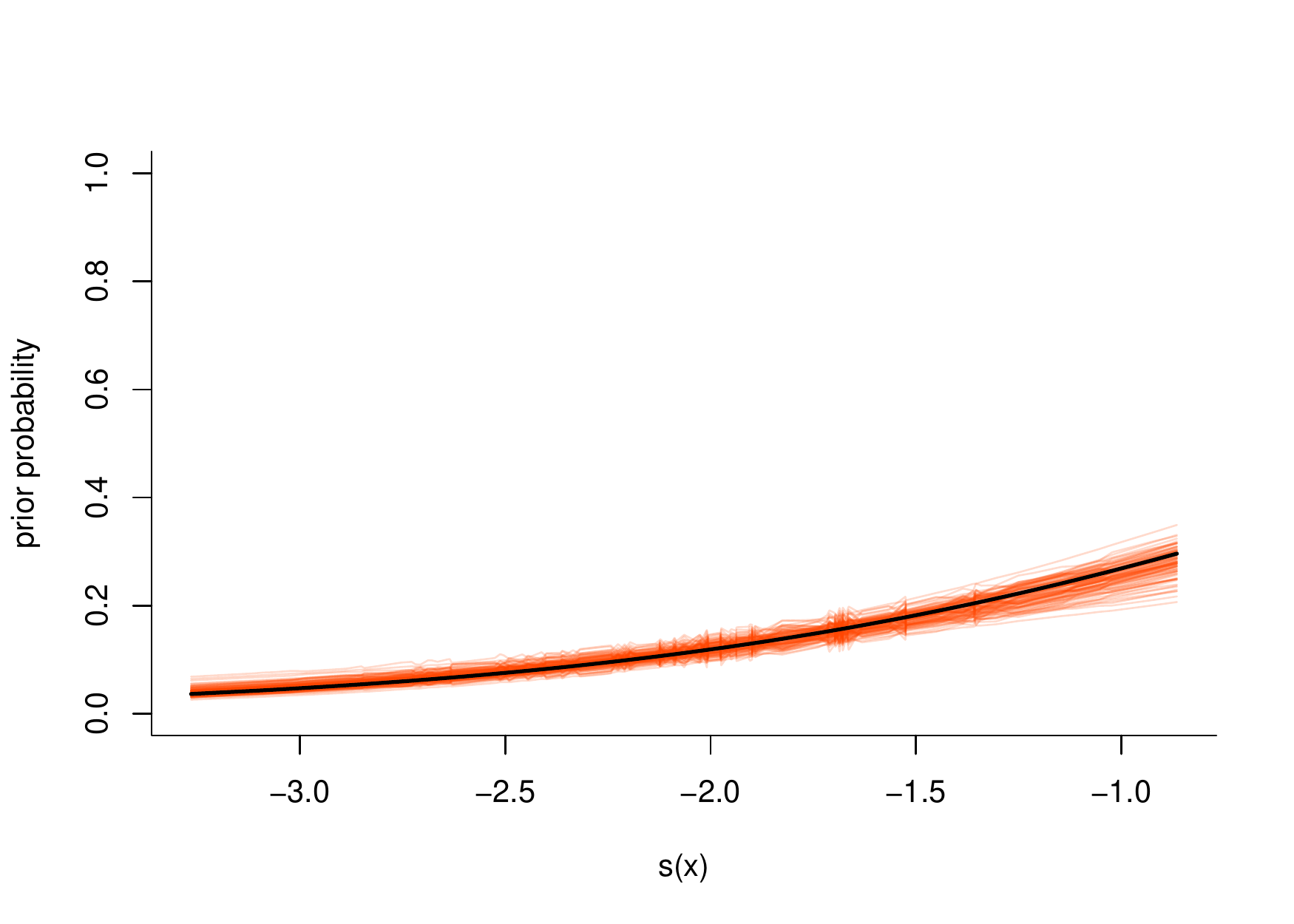}
	\includegraphics[width = 0.45\textwidth]{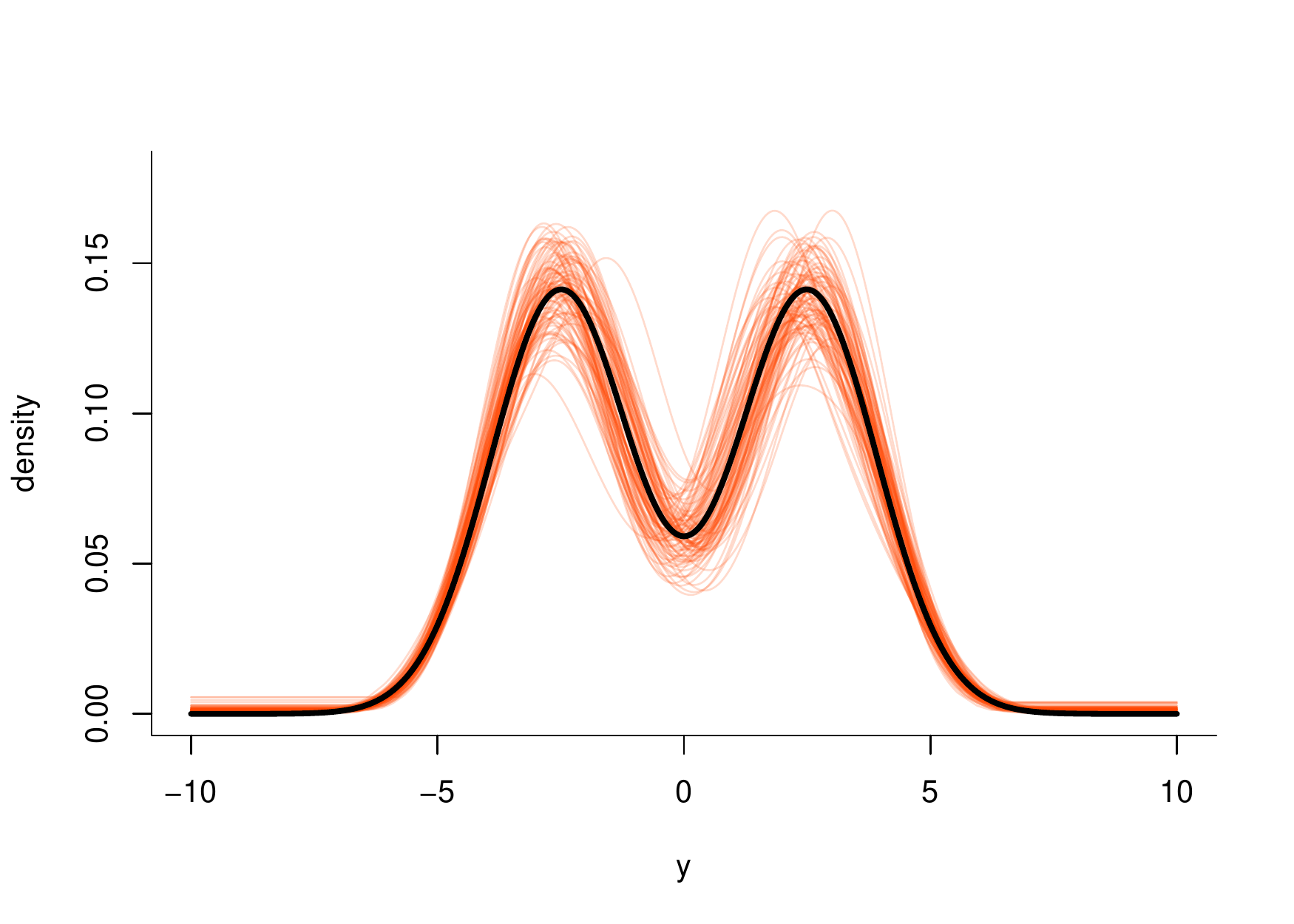}
	\caption{The above plots show the estimates obtained by FDRreg in $100$ replicates for data drawn from the following perturbation of setting~\eqref{nonidsim}: $s^{(2/3)}$ and $g_1^{(2/3)}$ with $n=10^4$. For the bottom two plots, translucent orange lines show estimates of $\pi^*(\cdot)$ and $f_1^*$ obtained by FDRreg in each of the $100$ replicates.}
	\label{fig:fdrreg_alpha_2on3}
\end{figure}

Let $\ell^*$ denote the true likelihood (i.e.,~\eqref{eq:full_MLE} evaluated at $\pi^*(\cdot),f_1^*$) and recall that $A = (0, (\sup_{x \in \X} \pi(x))^{-1})$. We define the two end points (conservative and anti-conservative respectively) as:
\begin{align*}
\alo\defeq \inf \left\{ \alpha \in A : \ell(\alpha) \geq \ell^* \right\} \qquad \mbox{and} \qquad
\ahi\defeq \sup \left\{ \alpha \in A : \ell(\alpha) \geq \ell^* \right\}.
\end{align*}
For this example therefore, $\alo=2/3$ and $\ahi=2$, which describes a near non-identifiable likelihood region indexed by $(\underline{\alpha}, \overline{\alpha})$ (along this simple one dimensional submodel). Of course, when $s, g_1$ and $\mathcal{X}$ are kept fixed, we expect $\underline{\alpha}\rightarrow 1$ and $\ahi\rightarrow 1$ as $n \rightarrow \infty$.
\begin{figure}[H]
	\centering
	\includegraphics[width = 0.9\textwidth]{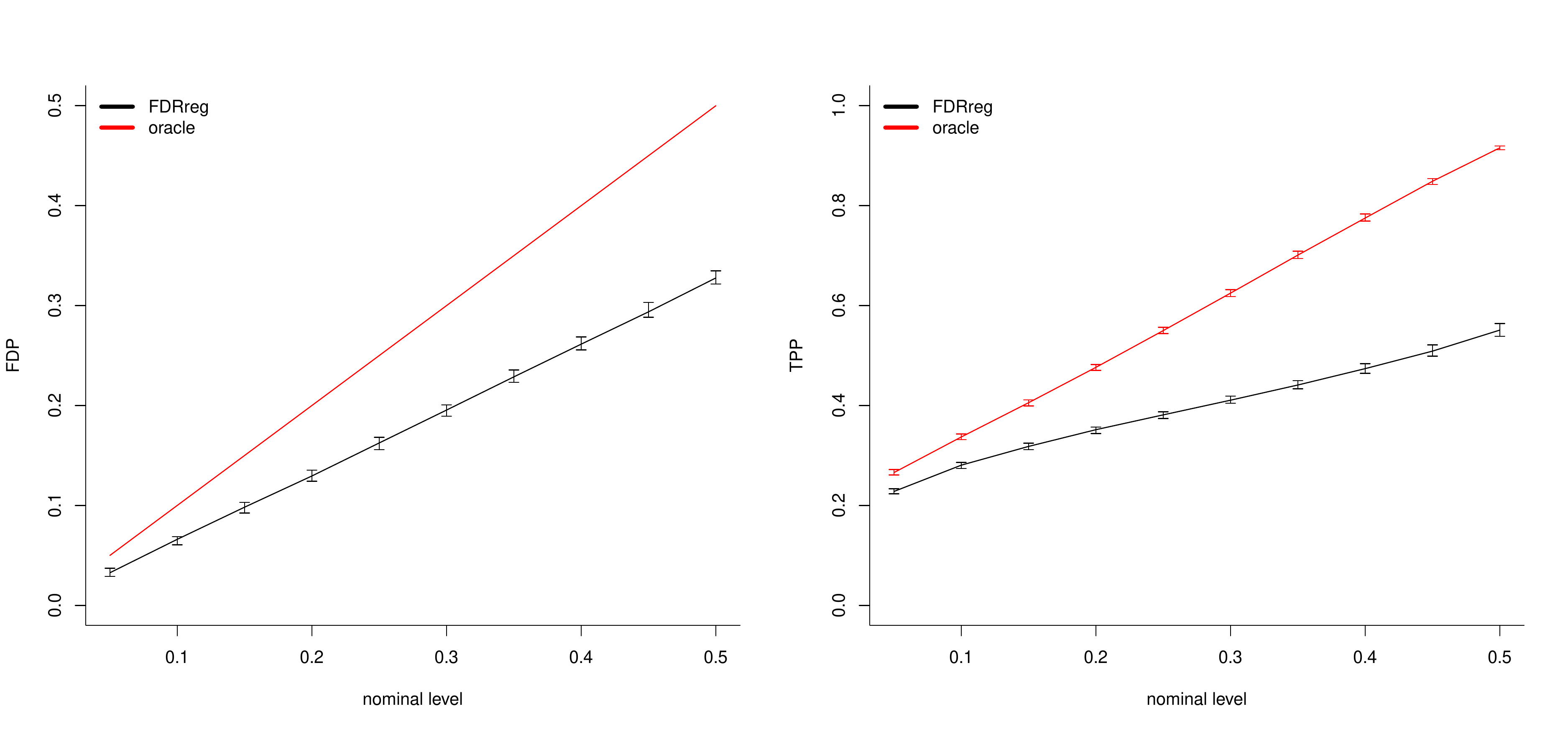}
	
	\includegraphics[width = 0.45\textwidth]{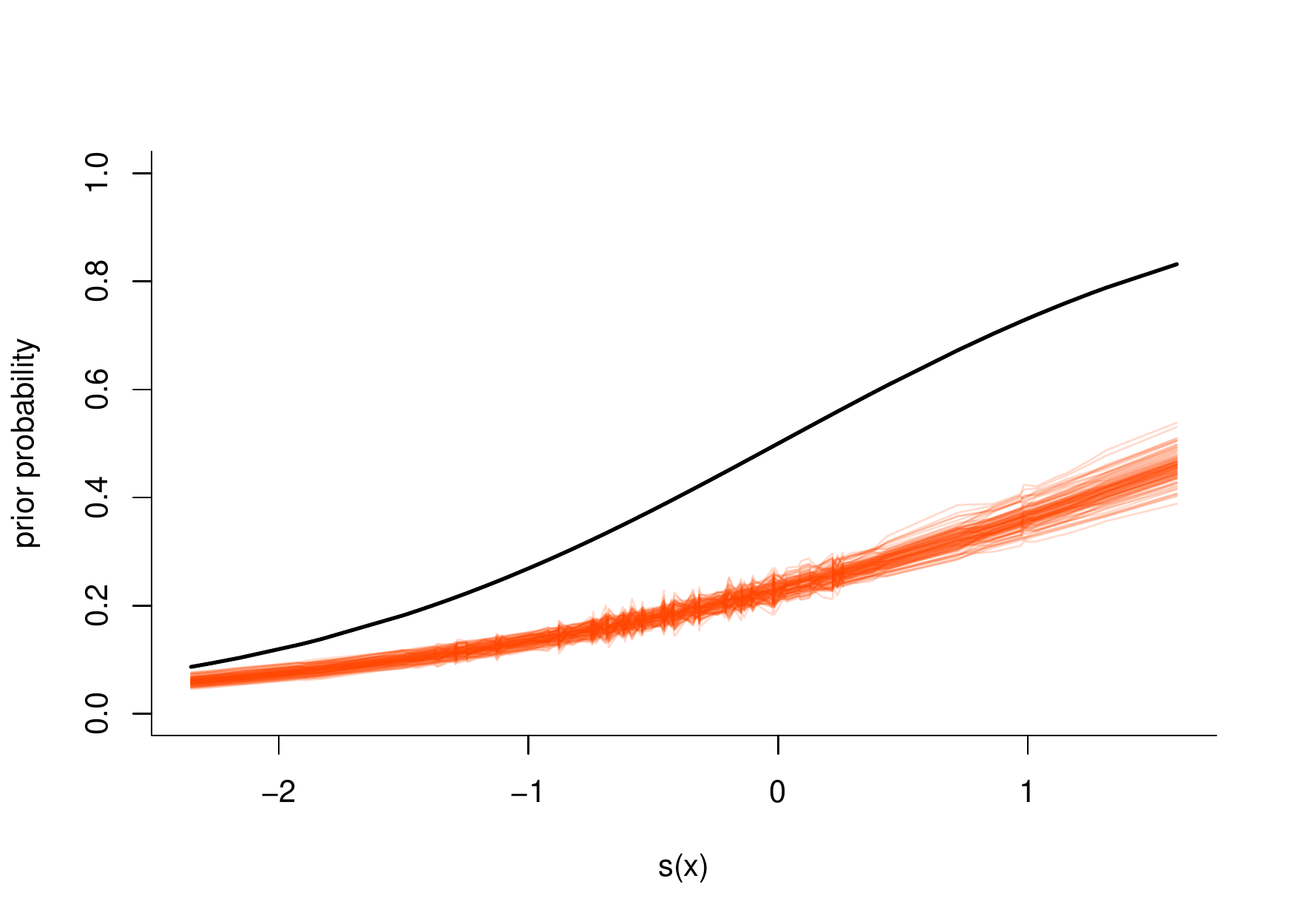}
	\includegraphics[width = 0.45\textwidth]{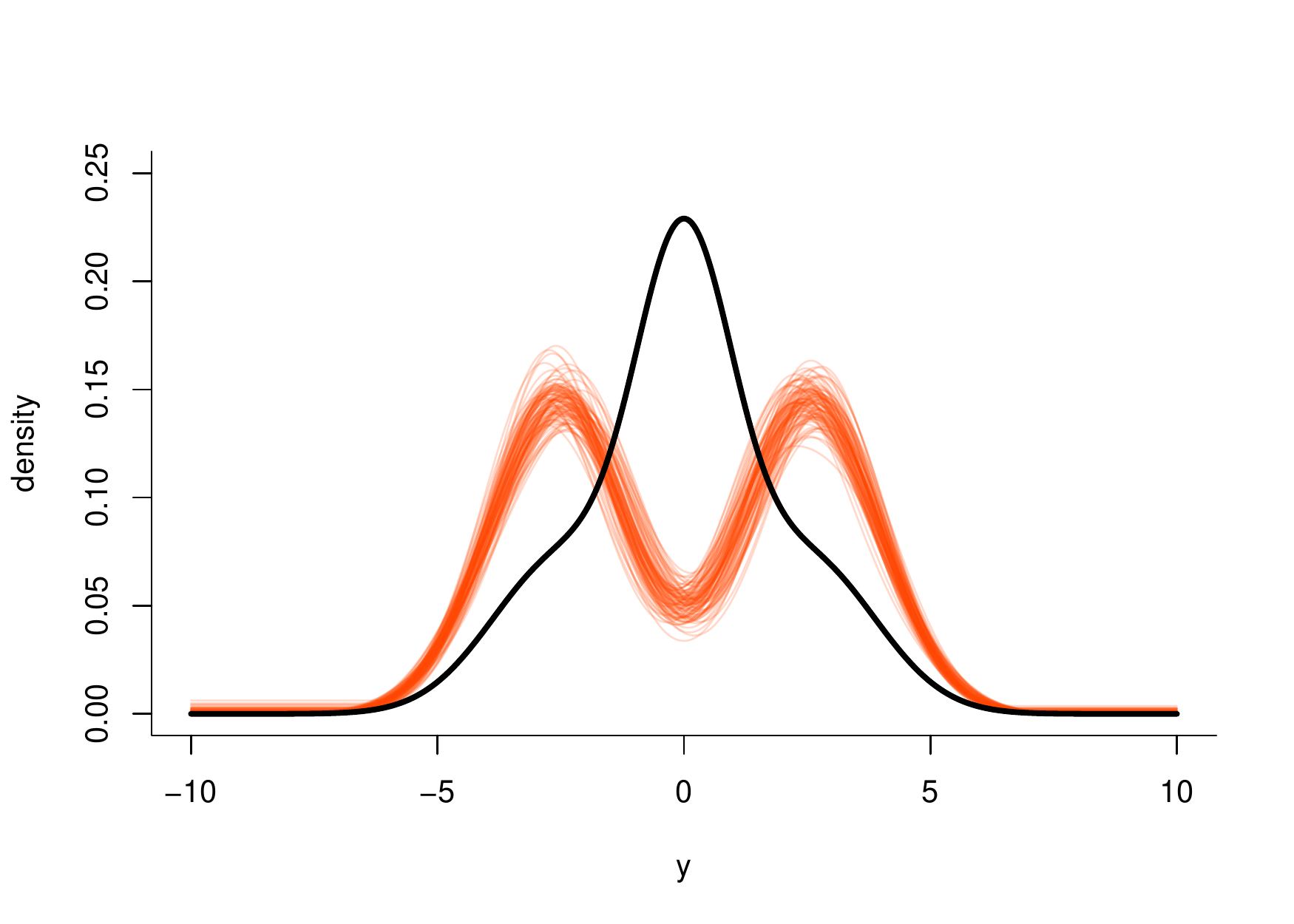}
	\caption{The same plots as in Figure~\ref{fig:fdrreg_alpha_2on3} when data is generated from $s^{(2)}$ and $g_1^{(2)}$ with $n=10^4$.}
	\label{fig:fdrreg_alpha_2}
\end{figure}
The formal definitions for $(\pilo,\fonelo)$ and $(\pihi,\fonehi)$ are given below (for reproducibility):
\begin{itemize}
	\item[(i)] $s^{(2/3)} = -3.38 + 1.38x_1 + 1.38x_2, ~~~~ g_1^{(2/3)} = \frac{1}{2} N(-2.5, 1) + \frac{1}{2} N(2.5, 1);$
	\item[(ii)] $s^{(2)} = -2.6 + 2.15x_1 + 2.15x_2, ~~~~ g_1^{(2)} = \frac{1}{4} N(-2.5, 1) + \frac{1}{2} \delta_{0} + \frac{1}{4} N(2.5, 1)$.
\end{itemize}
The computed approximation to $\check f_1^{(\alpha)}$, for $\alpha = 2/3, 2$, was supported on $100$ atoms (as prescribed in~\cref{dengauss}) chosen based on the range of $Y_i$'s and thus some manual cleaning was required to reduce to the above mentioned forms. The above expressions are calculated by aggregating $10$ replicates from setting~\eqref{nonidsim} with fixed choices of $\alpha$.
\begin{figure}[H]
	\centering
	\includegraphics[width=\textwidth]{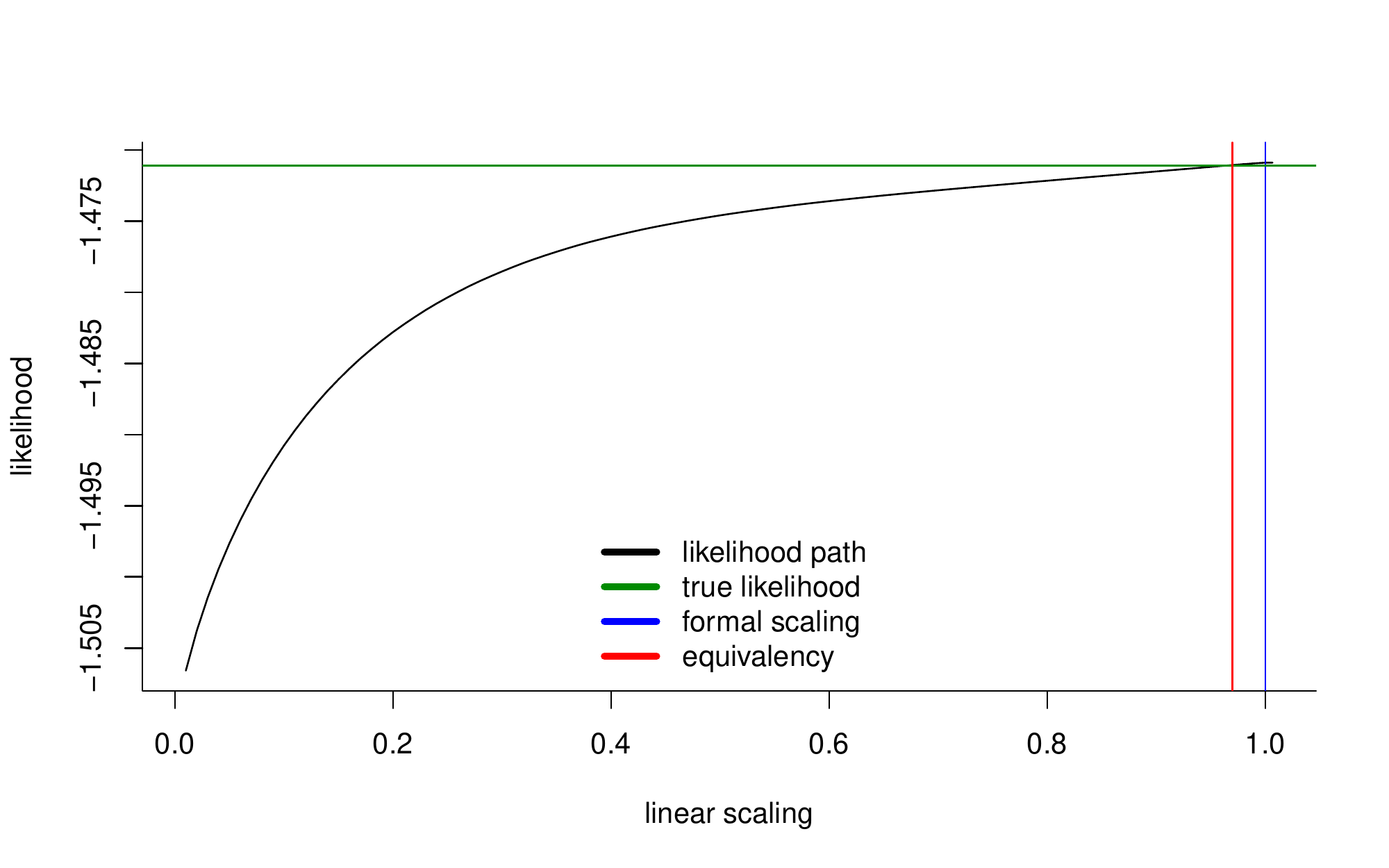}
	\caption{The plot shows the likelihood value as $\alpha$ changes; cf.~Figure~\ref{fig:lik_path}. For this setting, $A = [0,1.007]$.}
	\label{fig:lik_path_strongpi_1}
\end{figure}
Based on our simulations (see Figures~\ref{fig:fdrreg_alpha_2on3} and~\ref{fig:fdrreg_alpha_2}), we believe it is reasonable to conjecture that FDRreg (or in general, any one-step method reliant upon a two-groups estimate of $f_1$) aims for the description indexed by $\underline{\alpha}$. Figure~\ref{fig:fdrreg_alpha_2on3} verifies that for the conservative endpoint, FDRreg provides perfectly accurate estimates of $\pi^*(\cdot)$ and $f_1^*$ and further, is perfectly aligned with the oracle lFDRs. On the other hand, Figure~\ref{fig:fdrreg_alpha_2} demonstrates that for the anti-conservative endpoint $\ahi$, the estimates obtained by FDRreg estimates remain practically unchanged and are consequently very inefficient.

In our previous example, we have seen that FDRreg performs well in estimating $(\pilo, \fonelo)$. We may ask: ``Is this always the case? What would happen if the true parameter was $(\pihi, \fonehi)$?" In the following we illustrate a simulation setting where $\alo \approx \ahi \approx 1$ and yet  FDRreg performs poorly. In our other simulation studies, not reported here, we realized that this phenomenon is more pronounced in settings where there is strong covariate information. Let us consider the following setting:
\begin{equation} \label{model:strong_pi_1}
s(x_1,x_2) = 20(x_1 - 0.75), ~~~~~~~ g_1 = 0.4 \delta_{0} + 0.6 \delta_{1}. 
\end{equation}
where $\mathcal{X} = [0,1]^2$ and $n = 10^4$.
\begin{figure}[h]
	\centering
	\includegraphics[width=\textwidth,height=15cm]{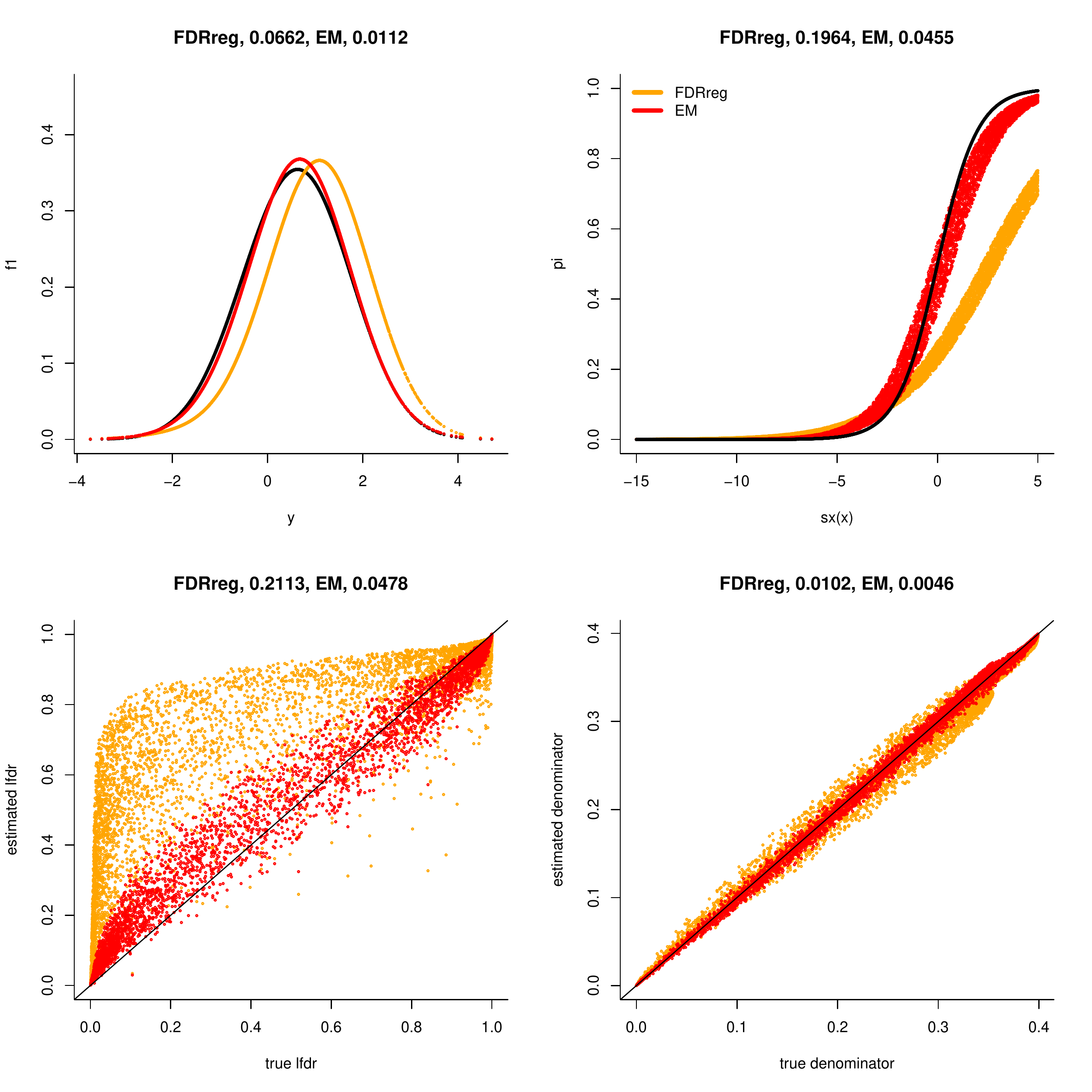}
	\caption{These plots compare the estimates obtained by FDRreg and fMLE for one data set obtained from the setting described in \eqref{model:strong_pi_1}. Clockwise from top left, the plots depict estimates of $f_1^*, \pi^*$, true lFDRs and the denominator of the lFDRs (see~\eqref{lFDRtrue}). The numbers in the title, next to each method, show the RMSE achieved by that method. The black curves represent the truth, while the red and orange dots represent the fMLE and the estimates from FDRreg, respectively.}
	\label{fig:scatter_plot_strongpi_1}
\end{figure}
We begin by drawing a random sample $\{(Y_i,X_i)\}_{i=1}^n$ from setting~\eqref{model:strong_pi_1} and plotting the analogue of Figure \ref{fig:lik_path}, shown in Figure \ref{fig:lik_path_strongpi_1}. The conclusions from Figure~\ref{fig:lik_path_strongpi_1} are strikingly different from those for Figure \ref{fig:lik_path}. Since  $A = [0,1.007]$, $\overline{\alpha}$ is necessarily $\approx 1$. Further $\alo \approx 0.97$. This leads us to conclude that this setting does not suffer from the problem of near non-identifiability, even for small sample sizes. Figure \ref{fig:scatter_plot_strongpi_1} illustrates the performance fMLE and FDRreg and shows a stark difference between their behaviors. From the plots and the RMSE values (depicted in their titles) we conclude that in this setting the fMLE is a far superior estimate of the identifiable parameters $(\pi^*(\cdot), f_1^*)$. 

Although this section does not include any substantive theoretical underpinning, we believe that there is an important take away message: Even under identifiability, estimating parameters in model~\eqref{denmodel} can be difficult for some parameter settings with small to moderate sample sizes. Moreover, given a dataset, it can be hard to know whether the model at hand is nearly non-identifiable.  
\subsection{More tables and plots}\label{tables}
\begin{table}[H]
	\centering
	\begin{tabular}{| l | r | r | r | r |}
		\hline
		Setting & \texttt{FDRreg} & \texttt{Marginal - I} & \texttt{Marginal - II} & \texttt{fMLE} \\
		\hline
		(A)(i) & \textbf{2.6} & 23.9 & 18.4 & 88.7 \\
		(A)(ii) & \textbf{2.9} & 19.7 & 13.9 & 75.8 \\
		(A)(iii) & \textbf{2.8} & 24.7 & 18.6 & 154.1 \\
		(A)(iv) & \textbf{2.9} & 25.3 & 15.4 & 67.7 \\
		\hline
		(B)(i) & \textbf{2.3} & 17.5 & 14.2 & 90.4 \\
		(B)(ii) & \textbf{2.7} & 20.4 & 14.2 & 84.8 \\
		(B)(iii) & \textbf{1.8} & 22.5 & 13.2 & 169.4 \\
		(B)(iv) & \textbf{2.4} & 20.5 & 15.5 & 87.3 \\
		\hline
		(C)(i) & \textbf{2.9} & 24.8 & 15.8 & 83.5 \\
		(C)(ii) & \textbf{3.1} & 22.4 & 16.1 & 65.5 \\
		(C)(iii) & \textbf{1.9} & 20.2 & 13.7 & 85.5 \\
		(C)(iv) & \textbf{2.9} & 24.9 & 17.8 & 88.5 \\
		\hline
		(D)(i) & \textbf{2.5} & 20.9 & 12.9 & 90.1 \\
		(D)(ii) & \textbf{2.7} & 19.4 & 12.7 & 60.2 \\
		(D)(iii) & \textbf{3.6} & 18.5 & 11.6 & 146.3 \\
		(D)(iv) & \textbf{3.1} & 24.1 & 14.6 & 65.5 \\
		\hline
	\end{tabular}
	\caption{Table showing the average time (in seconds) required by each of the algorithms for the different simulation settings. The fastest one in each setting is highlighted in bold. Here $n = 10^4$.}
	\label{table:timecomp}
\end{table}
\begin{table}[H]
	\centering
	\begin{tabular}{| l | r | r | r |}
		\hline
		$n$ & \texttt{FDRreg} & \texttt{Marginal - I} & \texttt{Marginal - II} \\
		\hline
		10000 & \textbf{3.06} & 31.50 & 19.62 \\
		20000 & \textbf{5.83} & 75.06 & 66.21 \\
		30000 & \textbf{8.67} & 111.23 & 69.28 \\
		40000 & \textbf{10.82} & 228.72 & 177.34 \\
		50000 & \textbf{13.60} & 211.94 & 135.04 \\
		60000 & \textbf{13.82} & 180.42 & 91.11 \\
		70000 & \textbf{15.54} & 219.43 & 106.18 \\
		80000 & \textbf{17.22} & 261.91 & 115.44 \\
		90000 & \textbf{19.34} & 323.09 & 137.21 \\
		100000 & \textbf{20.82} & 343.46 & 148.41 \\
		\hline
	\end{tabular}
	\caption{Table showing the time required by each of the methods as $n$ varies from $10000$ to $100000$, in setting of (A)(i). The fastest method for each $n$ is highlighted in bold.}
	\label{table:timereq}
\end{table}

\begin{table}[H]
	\centering
	\begin{tabular}{| l | r | r | r |}
		\hline
		Setting & \texttt{FDRreg} & \texttt{Marginal - I} & \texttt{Marginal - II} \\
		\hline
		(A)(i) & 0.04 & \textbf{0.82} & 0.14 \\
		(A)(ii) & 0.00 & \textbf{0.94} & 0.06 \\
		(A)(iii) & 0.00 & \textbf{1.00} & 0.00 \\
		(A)(iv) & 0.00 & \textbf{0.96} & 0.04 \\
		\hline
		(B)(i) & 0.00 & \textbf{0.98} & 0.02 \\
		(B)(ii) & 0.00 & \textbf{0.94} & 0.06 \\
		(B)(iii) & 0.00 & \textbf{0.96} & 0.04 \\
		(B)(iv) & 0.00 & \textbf{0.94} & 0.06 \\
		\hline
		(C)(i) & 0.08 & \textbf{0.84} & 0.08 \\
		(C)(ii) & 0.00 & \textbf{0.98} & 0.02 \\
		(C)(iii) & 0.00 & \textbf{0.98} & 0.02 \\
		(C)(iv) & 0.02 & \textbf{0.88} & 0.10 \\
		\hline
		(D)(i) & 0.26 & \textbf{0.68} & 0.06 \\
		(D)(ii) & 0.02 & \textbf{0.88} & 0.10 \\
		(D)(iii) & 0.02 & \textbf{0.94} & 0.04 \\
		(D)(iv) & 0.12 & \textbf{0.74} & 0.14 \\
		\hline
	\end{tabular}
	\caption{Table showing the proportion of times each of the methods FDRreg, Marginal - I and Marginal - II were used to initialize fMLE. The most common method in each setting is highlighted in bold.}
	\label{table:init_proportions}
\end{table}

\begin{figure}[H]
	\centering
	\begin{subfigure}[b]{0.47\textwidth}
		\centering
		\includegraphics[width=\textwidth]{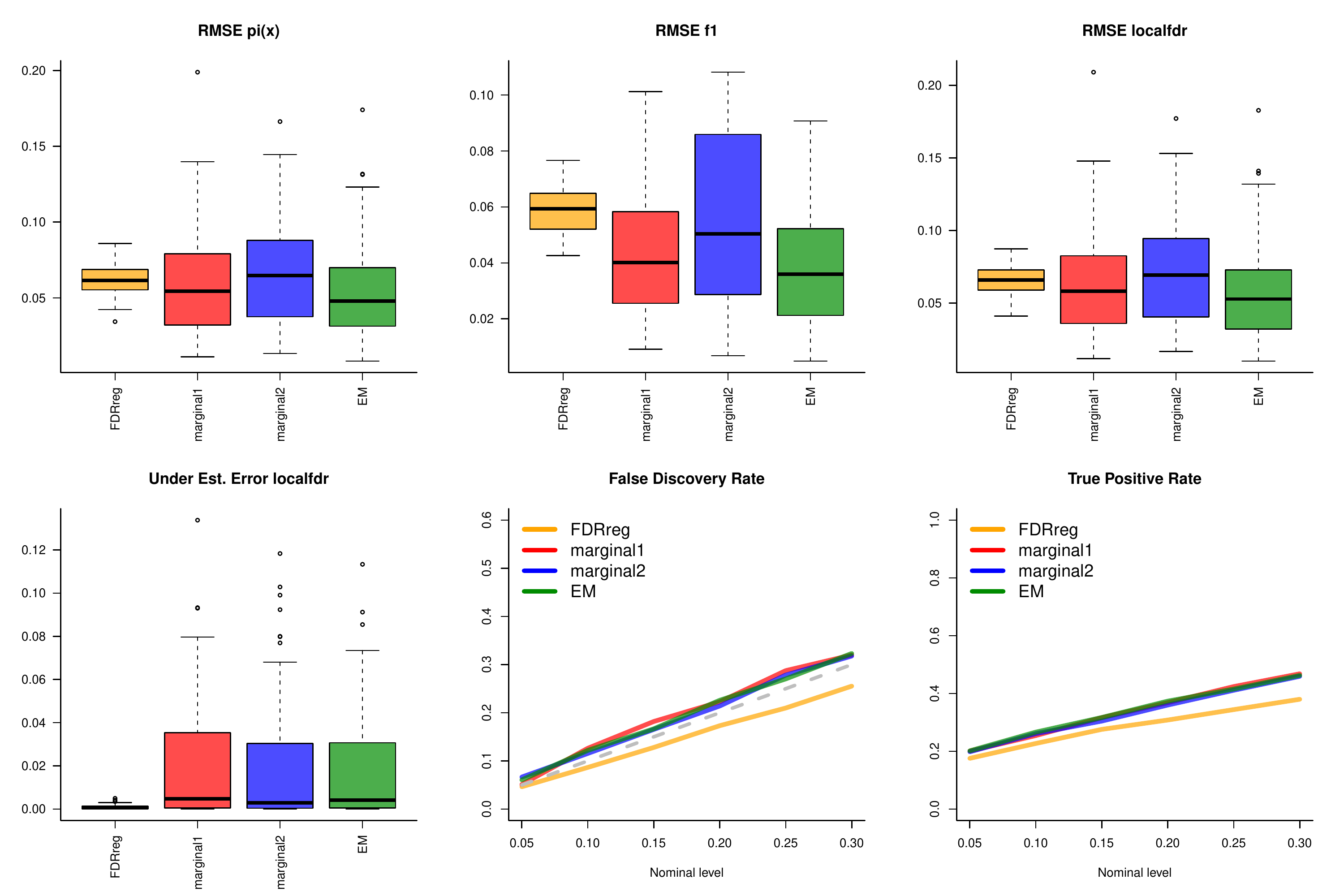}
		\caption{$f_1^*$ as in setting (i). }
		\label{fig:pi_setting_Ci}
	\end{subfigure}
	\hfill
	\begin{subfigure}[b]{0.47\textwidth}
		\centering
		\includegraphics[width=\textwidth]{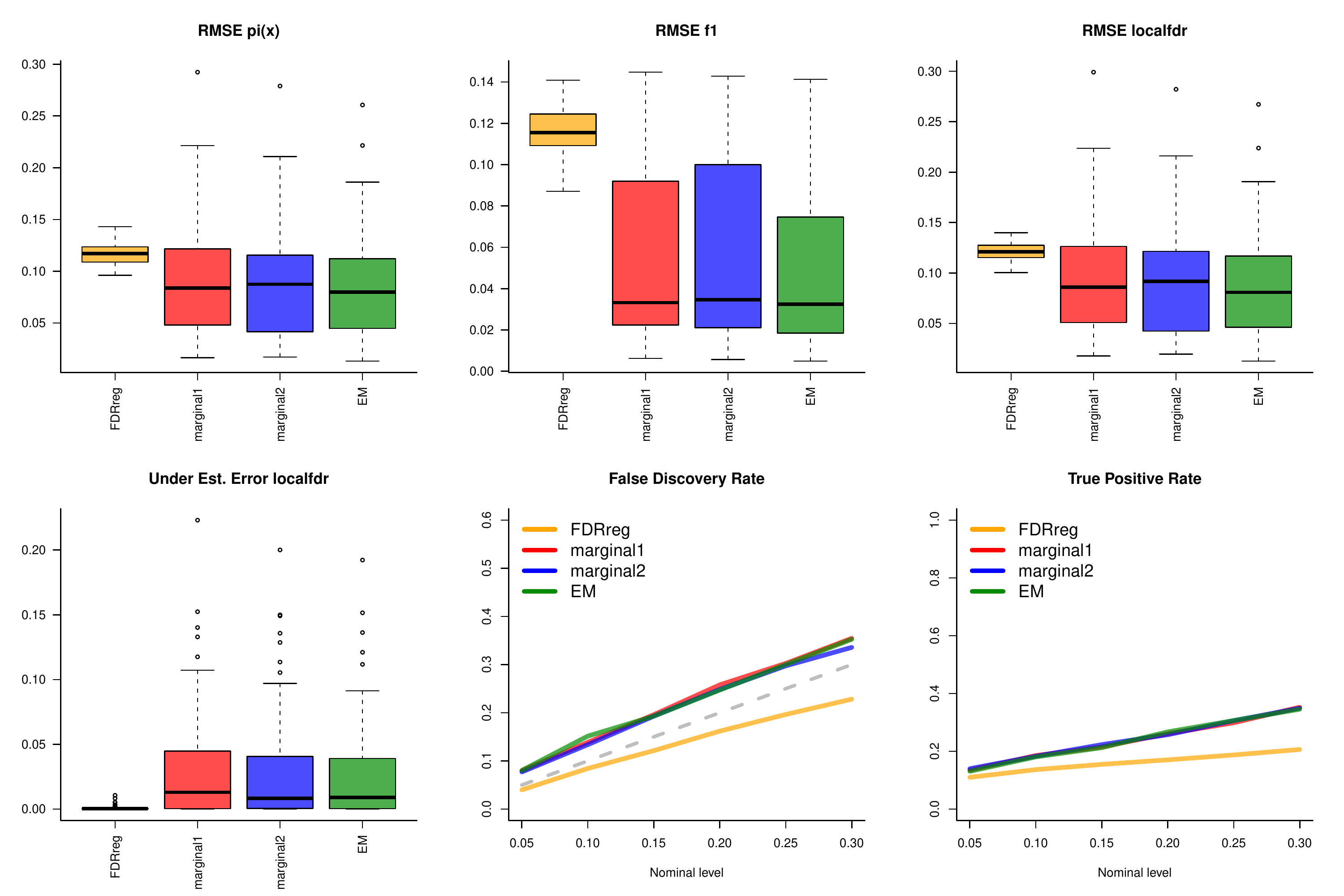}
		\caption{$f_1^*$ as in setting (ii).}
		\label{fig:pi_setting_Cii}
	\end{subfigure}
	\vspace{15mm}
	\begin{subfigure}[b]{0.47\textwidth}
		\centering
		\includegraphics[width=\textwidth]{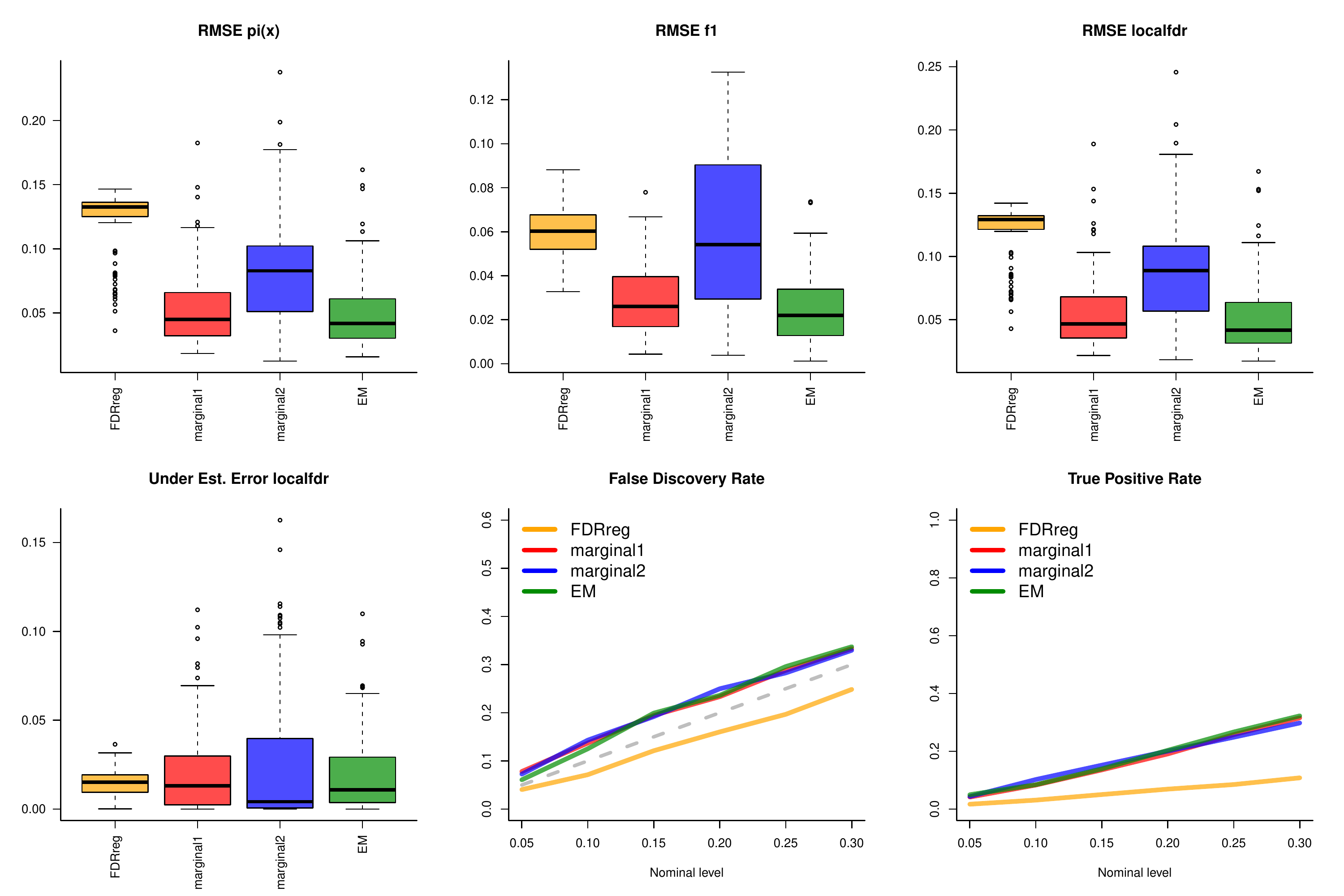}
		\caption{$f_1^*$ as in setting (iii). }
		\label{fig:pi_setting_Ciii}
	\end{subfigure}
	\hfill
	\begin{subfigure}[b]{0.47\textwidth}
		\centering
		\includegraphics[width=\textwidth]{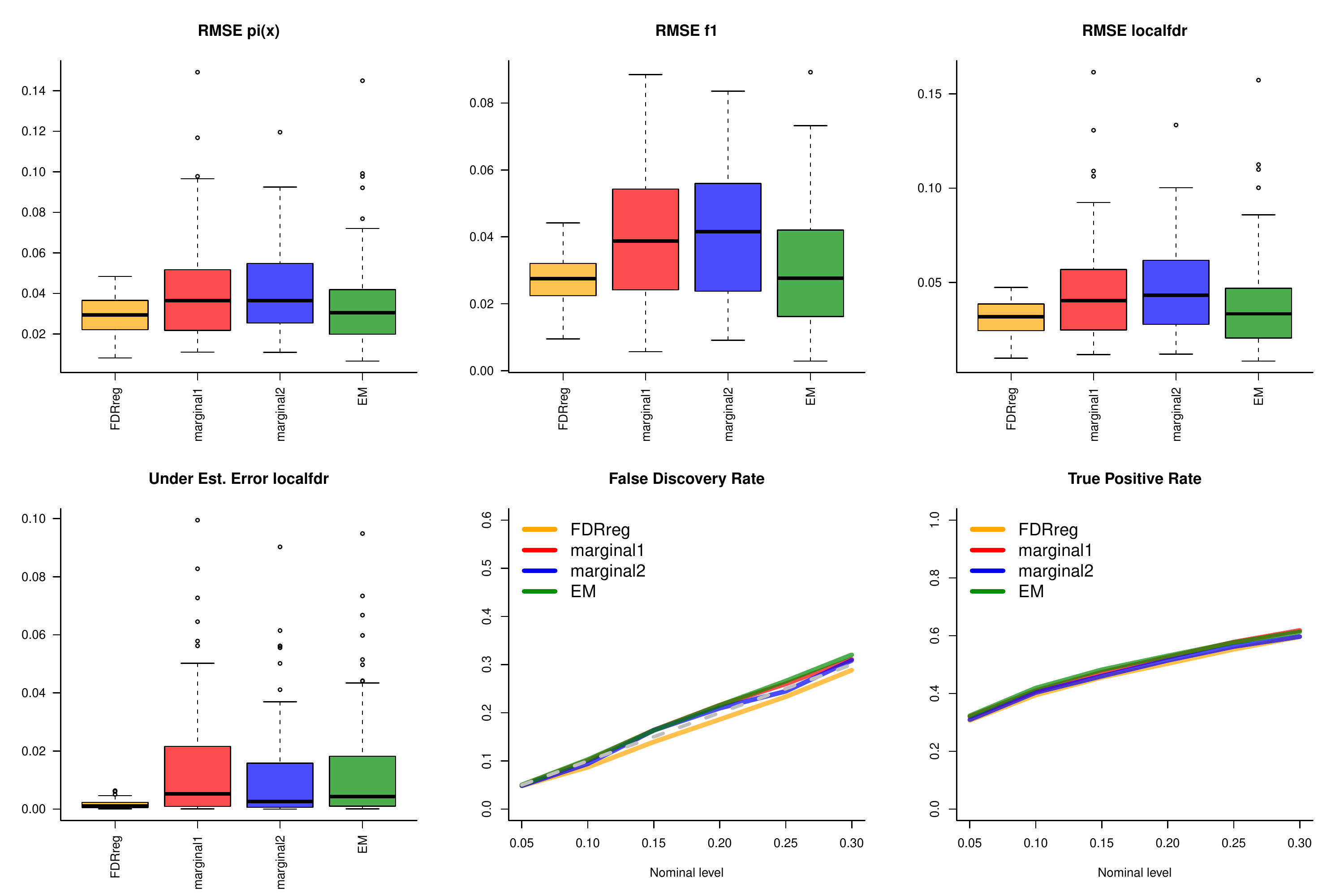}
		\caption{$f_1^*$ as in setting (iv). }
		\label{fig:pi_setting_Civ}
	\end{subfigure}
	\vspace{-1.5cm}
	\caption{Figure depicts the same plots as in Figure~\ref{fig:pi_setting_A} when $s(\cdot)$ is from setting (C).}
	\label{fig:pi_setting_C}
\end{figure}

\section{Proofs of the main results}\label{sec:allproofs}
\subsection{Proof of \cref{lem:bid}}\label{lem:bidpf}

Suppose first that both the conditions \eqref{bid.eq1} and
\eqref{bid.eq2} are true. We shall then argue that $P_{\pi, F_1} =
P_{\pi', F_1'}$. To see this, simply observe that under
\eqref{bid.eq1} and \eqref{bid.eq2}, 
\begin{equation*}
\pi(x) F_1(y) + (1 - \pi(x)) F_0(y) = \pi'(x) F_1'(y) + (1 -
\pi'(x)) F_0(y), \qquad \forall \; m\mbox{-a.e. } x, \forall \; y \in \R. 
\end{equation*}
This therefore implies that under both the probability measures
$P_{\pi, F_1}$ and $P_{\pi', F_1'}$, the conditional distributions of
$Y | X = x$ are equal for $m$-a.e $x$. This further implies that $P_{\pi,
	F_1} = P_{\pi', F_1'}$ which proves the first part of the lemma. 

Let us now turn to the second part where we assume that $P_{\pi, F_1}
= P_{\pi', F_1'}$ and that $F_0 \neq F_1$ and $\pi(x) > 0$ with 
positive probability under $m$. Our aim is to prove that \eqref{bid.eq1} and \eqref{bid.eq2} hold for some $c \neq 1$. Because  $P_{\pi, F_1} = P_{\pi', F_1'}$, the 
conditional distributions of $Y | X = x$ under $P_{\pi, F_1}$ and
$P_{\pi', F_1'}$ are the same for almost all $x$ (under $m$) which
means that  
\begin{equation}\label{st}
\pi(x) F_1(y)  + (1 - \pi(x)) F_0(y) = \pi'(x) F_1'(y) + (1 -
\pi'(x)) F_0(y), \quad\mbox{$\forall \; y$ and $m$-a.e~$x$}. 
\end{equation}
Let $\samp'$ denote the class of all $x$ for which the above equality holds for all $y$ and note that we have $m(\samp') = 1$. Further, let $\samp'' \defeq \{x \in \samp' : \pi(x) > 0\}$. Because $\pi(x) > 0$ with positive probability under $m$, the set $\samp''$ is non-empty.  For every $x   \in \samp''$,  the quantity $\pi'(x)$  cannot be equal to 0 for
otherwise, we would have from \eqref{st} that $F_0(y) = F_1(y)$ for all $y$ which contradicts the assumption that $F_0 \neq F_1$. Therefore~\eqref{st} can be  rewritten for $x \in \samp''$ as   
\begin{equation}\label{st1}
F'_1(y) = \frac{\pi'(x) - \pi(x)}{\pi'(x)} F_0(y) + \frac{\pi(x)}{\pi'(x)} F_1(y) \qt{for every $y$}. 
\end{equation}
Next, we claim that under the assumption that $F_0 \neq F_1$, the above implies that the function $x \mapsto (\pi'(x) - \pi(x))/\pi'(x)$ is constant for $x \in \samp''$. To see this, take $x_1,x_2 \in \X''$ and let $\alpha_i \defeq (\pi'(x_i) - \pi(x_i))/ \pi'(x_i)$, for $i = 1,2$. Then, from the above display, we have, for $i=1,2$,
\begin{eqnarray*}
	F_1'(y) = \alpha_i F_0(y) + (1 - \alpha_i) F_1(y), \qt{for every $y$}. 
\end{eqnarray*}
Thus, by subtraction, $$(\alpha_1 - \alpha_2)( F_0(y) - F_1(y)) = 0, \qt{for every $y$},$$ which shows that  $\alpha_1$ must equal $\alpha_2$ (note that $F_0 \neq F_1$). Thus $x \mapsto (\pi'(x) - \pi(x))/\pi'(x)$ is constant for $x \in \samp''$ and we denote this constant value by $c$ i.e., 
\begin{equation}\label{st9}
\frac{\pi'(x) - \pi(x)}{\pi'(x)} = c \qt{for all $x \in \samp''$}. 
\end{equation}
Note first that this constant $c$ cannot be equal to one because then $\pi(x)$ would be zero for $x \in \samp''$ contradicting the definition of $\samp''$. Therefore $c \neq 1$ and we deduce that  
\begin{equation}\label{kkh}
\pi'(x) = \frac{\pi(x)}{1 - c} 
\end{equation}
for all $x \in \samp'$ with $\pi(x) > 0$.  We shall now argue that \eqref{kkh} is also true for $x \in \samp'$ with $\pi(x) = 0$. Suppose this is not true so that there exists $x \in \samp'$ with $\pi(x) = 0$ and $\pi'(x) > 0$. Then \eqref{st} for this particular $x$ gives 
\begin{equation*}
F_0(y) = \pi'(x) F_1'(y) + (1 - \pi'(x)) F_0(y) \qt{for all $y$} 
\end{equation*}
and cancelling $\pi'(x)$ (note that $\pi'(x) > 0$), we deduce that $F_0 = F_1'$. But $F_0 = F_1'$ along with the equality \eqref{st} implies that 
\begin{equation*}
\pi(z) F_1(y) + (1 - \pi(z)) F_0(y) = F_0(y) \qt{for all $y$ and $z \in \samp'$}
\end{equation*}
which implies (by using the above for $z \in \samp''$) that $F_0 = F_1$ which contradicts the hypothesis that $F_0 \neq F_1$. We have therefore proved that \eqref{kkh} holds for all $x \in \samp'$. This proves \eqref{bid.eq1} (note that $m(\samp') = 1$). The identity \eqref{bid.eq2} is simply a consequence of \eqref{st1} and \eqref{st9}.

\subsection{Proof of \cref{ide}}\label{idepf}
We will prove this lemma in two parts: first with $g(x)=(1+\exp(x))^{-1}$ and second with $g(x)=\Phi(x)$.
\begin{proofpart}
	Suppose that $P_{(\pi^*, F_1^*)} = P_{(\pi, F_1)}$ for some $\pi \in \logi$ and $F_1 \in \mathcal{F}$. Lemma \ref{lem:bid} immediately implies the existence of a real number $c \neq 1$ such that 
	\begin{equation}\label{v0}
	\pi(x) = \frac{\pi^*(x)}{1 - c} \qt{ for $m$-a.e.~$x$} ~~ \text{ and
	} ~~     F_1 = c F_0 + (1 - c) F_1^*. 
	\end{equation}
	If, for $x\in \X$, $\pi(x) = 1/(1 + e^{-\beta_0 - \beta^\top x})$, for some $\beta_0
	\in \R$ and $\beta \in \R^p$, then the first condition above implies that 
	\begin{equation}\label{llnl}
	\beta_0 + \beta^\top x = - \log \left((1 - c) e^{-\beta_0^* -
		(\beta^*)^\top x} - c
	\right) 
	\end{equation}
	for $m$-a.e.~$x$. 
	
	We shall now prove that \eqref{llnl} holds for every $x \in \samp'$. To see this, fix $x \in \samp'$. Because $\samp'$ is open, there exists $u > 0$ such that
	\begin{equation*}
	B(x, u) \defeq \left\{z \in \R^p : \norm{z - x} \leq u \right\}  \subseteq \samp'
	\end{equation*}
	where $\norm{\cdot}$ denotes the usual Euclidean norm on $\R^p$. Because of the assumption on the probability measure $m$, the ball $B(x, u/\ell)$ has strictly positive probability under $m$ for every integer $\ell \geq 1$. This, along with the fact that \eqref{llnl} holds for $m$-a.e.~$x$, implies the existence of a sequence $x_{\ell} \in B(x, u/\ell)$ such that \eqref{llnl} holds for every $x = x_{\ell}$. Letting $\ell \rightarrow \infty$, we deduce that \eqref{llnl} holds for every $x \in \samp'$. As a result,  the Hessians (taken with respect to $x$) of both sides of \eqref{llnl} must agree for all $x  \in \samp'$. This gives  
	\begin{equation*}
	\frac{c(1 - c) \beta_i^* \beta_j^* e^{\beta_0^* + (\beta^*)^\top x} }{\left(1 - c - ce^{\beta_0^* + (\beta^*)^\top x} \right)^2} = 0 \qt{for all $x \in \samp'$ and $1 \leq i, j \leq p$}.   
	\end{equation*}
	Because $c \neq 1$ and $\beta^* \neq 0$, the above can happen only if $c = 0$. The equalities in \eqref{v0}  then imply that $\pi = \pi^*$ and $F_1 = F_1^*$ which proves the identifiability of $(\pi_1^*, F_1^*)$ according to Definition \ref{def:Iden}. 
\end{proofpart}

\begin{proofpart}
	Suppose that $P_{(\pi^*, F_1^*)} = P_{(\pi, F_1)}$ for some $\pi \in \Pi_{probit}$ and $F_1 \in \mathcal{F}$. Lemma \ref{lem:bid} immediately implies the existence of a real number $c \neq 1$ such that 
	\begin{equation*}
	\pi(x) = \frac{\pi^*(x)}{1 - c} \qt{ for $m$-a.e.~$x$} ~~ \text{ and
	} ~~     F_1 = c F_0 + (1 - c) F_1^*. 
	\end{equation*}
	If, for $x\in \X$, $\pi(x) = \Phi\left(\beta_0 + \beta^\top x\right)$, for some $\beta_0
	\in \R$ and $\beta \in \R^p$, then the exact same calculations as in \textbf{Part 1} imply 
	\begin{equation}\label{lln2}
	\Phi\left(\beta_0+\beta^\top x\right)=\frac{\Phi\left(\beta_0^*+(\beta^*)^\top x\right)}{1-c}
	\end{equation}
	for $m$-a.e.~$x$ and in particular for all $x\in \samp'$. As $\beta^*$ is a non-zero vector, without loss of generality, assume that $\beta^*_1\neq 0$ where $\beta_1^*$ denotes the first component of $\beta^*$ and suppose $\beta_1$ denotes the first component of $\beta$. As a result the gradients (taken with respect to $x$) of both sides of \eqref{lln2} must agree for all $x\in \samp'$. Therefore,
	\begin{equation}\label{lln3}
	(1-c)\exp\left(-\frac{1}{2}\left(\beta_0+\beta^\top x\right)^2\right) \beta_1=\exp\left(-\frac{1}{2}\left(\beta_0^*+(\beta^*)^\top x\right)^2\right)\beta_1^*.
	\end{equation}
	Note that for \eqref{lln3} to hold, $\beta_1^*\neq 0$ and $\beta_1$ must have the same sign. Therefore, dividing both sides by $\beta_1^*$ and taking $\log$ on both sides,
	\begin{equation*}
	-2\log(1-c)+\left(\beta_0+\beta^\top x\right)^2-2\log\left(\beta_1/\beta_1^*\right)=\left(\beta_0^*+(\beta^*)^\top x\right)^2.
	\end{equation*}
	As the above holds for all $x\in \samp'$, which is an open set, we can equate coefficients of the quadratic and the linear term to get
	\begin{equation}\label{lln5}
	\beta_0\beta^\top=\beta_0^*(\beta^*)^\top\qquad\mbox{and}\qquad \beta\beta^\top=\beta^*(\beta^*)^\top
	\end{equation}
	Note that, if $\beta=0$, then \eqref{lln5} implies $\beta^*=0$ which is a contradiction. Therefore, assume $\beta\neq 0$. By \eqref{lln5}, if $\beta_0^*=0$, we must have $\beta_0=0$. In this case,~\eqref{lln3} can be rewritten as
	\begin{equation}\label{llndcom}
	(1-c)\exp\left(-\frac{1}{2}x^\top\beta\beta^\top x\right)\beta_1=\exp\left(-\frac{1}{2}x^\top\beta^*(\beta^*)^\top x\right)\beta_1^*.
	\end{equation}
	By~\eqref{lln5}, $\beta\beta^\top=\beta^*(\beta^*)^\top$ and so the exponential terms in~\eqref{llndcom} cancel out and we get $\beta_1^*=(1-c)\beta_1$. As a result, \eqref{lln5} yields $(1-c)^2=1\implies c=0$ which again by \eqref{lln2} gives identifiability. 
	
	So, from here on in, we will take $\beta_0^*\neq 0$. Note that both $\beta_1$ and $\beta_1^*$ are non-zero. In the second condition of~\eqref{lln5}, we can equate the entries in the first row and first column on both sides; to get $\beta_1^2=(\beta_1^*)^2$. Next, in the first condition of~\eqref{lln5}, we can equate the first entries of the vectors on both sides to get $\beta_0\beta_1=\beta_0^*\beta_1^*$. Combining these two conditions implies $(\beta_0/\beta_0^*)^2=1$. This gives rise to two natural cases:
	\begin{description}
		\item[Case 1:] If $\beta_0=\beta_0^*$, \eqref{lln5} implies $\beta=\beta^*$ which gives identifiability.
		\item[Case 2:] If $\beta_0=-\beta_0^*$, \eqref{lln5} implies $\beta=-\beta^*$. Note that this is a contradiction, as we have already argued that $\beta_1^*\neq 0$ and $\beta_1$ have the same sign.
	\end{description}
	This completes the proof.
\end{proofpart}

\subsection{Proof of~\cref{revide}}\label{revidepf}
The proof of this corollary is identical to the proof of~\cref{ide}. Therefore, we will only discuss part $1$, i.e., the case when $g(x)=(1+\exp(x))^{-1}$.

Suppose that $P_{(\pi^*, F_1^*)} = P_{(\pi, F_1)}$ for some $\pi \in \logi$ and $F_1 \in \mathcal{F}$. Lemma \ref{lem:bid} immediately implies the existence of a real number $c \neq 1$ such that 
\begin{equation}\label{ru0}
\pi(x) = \frac{\pi_1^*(x)}{1 - c} \qt{ for $m$-a.e.~$x$} ~~ \text{ and
} ~~     F_1 = c F_0 + (1 - c) F_1^*. 
\end{equation}
If, for $x=(x_1,x_2)\in \X_1\times \X_2$, $\pi(x) = 1/(1 + e^{-\beta_0 - \beta_1^\top x_1-\beta_2^\top x_2})$, for some $\beta_0
\in \R$ and $\beta_1 \in \R^{p_1}$ and $\beta_2\in\R^{p_2}$, then the first condition in~\eqref{ru0} implies that 
\begin{equation}\label{rullnl}
\beta_0 + \beta_1^\top x_1 + \beta_2^\top x_2 = - \log \left((1 - c) e^{-\beta_0^* -
	(\beta_1^*)^\top x_1-(\beta_2^*)^\top x_2} - c
\right) 
\end{equation}
for $m$-a.e.~$x$. For the purpose of this proof let us write $\beta_2^*=(\beta_{21}^*,\beta_{22}^*,\ldots ,\beta_{2p_2}^*)$. 

We shall now prove that \eqref{rullnl} holds for every $x = (\tilde{x}_1,x_2)$ with $x_2\in \samp_2'$. To see this, fix a particular $x_2 \in \samp_2'$. Because $\samp_2'$ is open, there exists $u > 0$ such that
\begin{equation*}
B(x_2, u) \defeq \left\{z \in \R^{p_2} : \norm{z - x_2} \leq u \right\}  \subseteq \samp_2'
\end{equation*}
where $\norm{\cdot}$ denotes the usual Euclidean norm on $\R^{p_2}$. Because of the assumption on the probability measure $m$, the ball $B(x_2, u/\ell)$ has strictly positive probability under $m_{2|1}(\cdot,\tilde{x}_1)$ for every integer $\ell \geq 1$. Note that $$m\left(\{\tilde{x}_1\},B(x_2,u/l)\right)=m_1(\{\tilde{x}_1\})m_{2|1}(B(x_2,u/l),\tilde{x}_1)>0$$ by the conditions stated in the statement of~\cref{revide}. This observation, along with the fact that \eqref{rullnl} holds for $m$-a.e.~$x$, implies the existence of a sequence $x_{2,\ell} \in B(x_2, u/\ell)$ such that \eqref{rullnl} holds for every $x = (\tilde{x}_1,x_{2,\ell})$. Letting $\ell \rightarrow \infty$, we deduce that \eqref{rullnl} holds for every $x=(\tilde{x}_1,x_2)$ with $x_2\in \samp_2'$.
Note that this implies
\begin{equation}\label{rulln2}
\beta_0 + \beta_1^\top \tilde{x}_1 + \beta_2^\top x_2 = - \log \left((1 - c) e^{-\beta_0^* -
	(\beta_1^*)^\top \tilde{x}_1-(\beta_2^*)^\top x_2} - c
\right) 
\end{equation}
holds for every $x_2\in\samp_2'$. Here $\tilde{x}_1$ is a fixed element of $\mathbb{R}^{p_1}$ whose existence is guaranteed by the assumptions in~\cref{revide}. Write $\tilde{\beta}=\beta_0+\beta_1^\top\tilde{x}_1$ and $\tilde{\beta}^*=\beta_0^*+(\beta_1^*)^\top\tilde{x}_1$. By renaming the variable $x_2$ as $x$,~\eqref{rulln2} can be rewritten as
\begin{equation}\label{rulln3}
\tilde{\beta} + \beta_2^\top x = - \log \left((1 - c) e^{-\tilde{\beta}^*-(\beta_2^*)^\top x} - c
\right) 
\end{equation}
where $x\in\samp_2'$, an open subset of $\mathbb{R}^{p_2}$. Since~\eqref{rulln3} holds for all vectors in an open subset of $\mathbb{R}^{p_2}$, we can equate the Hessians calculated on both sides (with respect to $x$). This yields,
\begin{equation}\label{rfin}
\qquad\qquad\frac{c(1 - c) \beta_{2i}^* \beta_{2j}^* e^{\tilde{\beta}^* + (\beta_2^*)^\top x} }{\left(1 - c - ce^{\tilde{\beta}^* + (\beta_2^*)^\top x} \right)^2} = 0\qquad \mbox{ for all }x\in\samp_2'\mbox{ and }1\leq i,j \leq p_2.   
\end{equation}
If $c\neq 0$, then~\eqref{rfin} implies $\beta_2=0$ which contradicts the assumptions in~\cref{revide}. Therefore, $c=0$ and identifiability follows.

\subsection{Proof of Theorem \ref{sr1}}\label{sr1pf}
Fix $t \geq 1$. Because of the assumption on $f_1^*$, we first argue
that 
\begin{small}
	\begin{equation}\label{add}
	\P(A_t|X_1,X_2,\ldots ,X_n) \geq 1 -n^{-t^2} \qt{where $A_t \defeq \bigcap_{i=1}^n \left\{Y_i
		\in [-M - 2 t \sqrt{\log n}, M + 2 t \sqrt{\log n}] \right\}$}.   
	\end{equation}
\end{small}
To see \eqref{add}, note first that 
\begin{equation*}
1 - \P(A_t|X_1,X_2,\ldots ,X_n) \leq \sum_{i=1}^n \P \left\{|Y_i| > M + 2 t \sqrt{ \log n}\big|X_1,X_2,\ldots ,X_n
\right\}. 
\end{equation*}
Because of the assumption on $f_1^*$, it is easy to see that each
$Y_i$ can be written as $\theta_i + Z_i$ where $\theta_i$ and $Z_i$
are independent random variables. In fact, $\theta_i$ is drawn from $(1 - \pi^*(X_i))
\delta_{\{0\}} + \pi^*(X_i) G^*$ where $\delta_{\{0\}}$
is the Dirac probability measure concentrated at 0 and
$Z_i$ is simply standard normal. Note that $\P\{\theta_i \in [-M, M]\} = 1$. As a result,
\begin{small}
	\begin{align*}
	\P \left\{|Y_i| > M + 2 t \sqrt{\log n}\Big|X_1,X_2,\ldots ,X_n
	\right\} &= \P \left\{|\theta_i + Z_i| > M + 2 t \sqrt{ \log n}\big|X_1,X_2,\ldots ,X_n
	\right\}  \\
	&\leq \P \left\{|\theta_i + Z_i| > |\theta_i| + 2 t \sqrt{ \log n} 
	\big|X_1,X_2,\ldots ,X_n\right\}  \\
	&\leq \P \left\{|Z_i| > 2 t \sqrt{ \log n} \right\} = 2 \left(1 -
	\Phi(2 t \sqrt{ \log n}) \right)
	\end{align*}
\end{small}
where $\Phi$ is the standard normal cumulative distribution
function. The standard inequality $1 - \Phi(s) \leq \phi(s)/s$  for $s
> 0$ now gives 
\begin{equation*}
\P \left\{|Y_i| > M +  2 t \sqrt{ \log n}\big|X_1,X_2,\ldots ,X_n
\right\} \leq \frac{1}{t \sqrt{2 \pi \log n}} e^{-2 t^2 
	\log n} = \frac{n^{-2t^2}}{t\sqrt{2 \pi \log n}}. 
\end{equation*}
Therefore, 
$$1 - \P(A_t|X_1,X_2,\ldots ,X_n) \leq\frac{n^{1-2t^2}}{t\sqrt{2\pi \log{n}}}\leq n^{-t^2}. $$
for $t\geq 1,n\geq 2$. The last inequality follows from the observation that $n^{1-2t^2}=n^{1-t^2-t^2}\leq n^{-t^2}$ and $\log{n}\geq\log{2}$ for $t\geq 1$ and $n\geq 2$. This proves \eqref{add}. 

Let $S\defeq [-M - 2 t \sqrt{\log n}, M + 2 t \sqrt{\log n}]$, $\eta \defeq
n^{-2}$ and let $h_1, \dots, h_{N_1}$ denote an $\eta$-covering 
subset of $\Fga$ under the pseudometric given by $\|\cdot\|_{\infty,
	S}$ where   
\begin{equation*}
\|h - \tilde{h}\|_{\infty, S} \defeq \sup_{x \in S} \left|h(x) -
\tilde{h}(x) \right|. 
\end{equation*}
and $N_1\defeq N(\eta, \Fga, \|\cdot\|_{\infty, S})$. It follows as a
consequence of \cite[Lemma 2]{zhang2009generalized} (note that $\eta
\defeq n^{-2}$)  that there exists a universal positive constant $C$ such
that  
\begin{equation}\label{ad2}
\log N_1 \leq C M^* t (\log n)^{3/2} \qt{where $M^* \defeq \max(M,
	\sqrt{\log n})$}. 
\end{equation}
Fix $\gamma > 0$ and recall that $N_2 = N(\gamma, \Pi_n, L^{\infty})$
denotes the  $\gamma$-covering number of $\Pi_n$ under the $L^{\infty}$
metric and that $H(\gamma, \Pi_n, L^{\infty}) = \log N(\gamma, \Pi_n,
L^{\infty})$ where $\Pi_n=\{(\pi(X_1),\ldots ,\pi(X_n):\pi\in\Pi\}$.

We shall now bound 
\begin{equation*}
\P  \left\{\Di \left((\hat{\pi}^A, \hat{f}_1^A), (\pi^*, f_1^*)
\right) \geq t \delta_n  \right\}
\end{equation*}
for $\delta_n > 0$ and $t \geq 1$. For any $\tilde{\pi}\in \Pi$, define $\tilde{\pi}_i=\tilde{\pi}(X_i)$ for $i=1,2,\ldots ,n$. Let $h_1,\ldots ,h_{N_1}$ form an $\eta-$cover of $\dga$ in $\lVert\cdot\rVert_{\infty,S}$ and $\pi^{(1)},\ldots ,\pi^{(N_2)}$ form a $\gamma-$cover for $\Pi_n$ in $L^{\infty}$. Let $J \subseteq \{(j, k) : 1 \leq
j \leq N_1, 1 \leq k \leq 
N_2\}$ be the set of all $(j, k)$ for which there exist $h_{0j} \in
\Fga$ and $\pi^{(0k)} \in \Pi_n$ such that 
\begin{equation}\label{ad1}
\|h_{0j} - h_j\|_{\infty, S} \leq \eta, ~
\max_{1 \leq i \leq n}|\pi^{(k)}_i - \pi^{(0k)}_i | \leq \gamma ~ \text{ and } ~
\Di((\pi^{(0k)}, h_{0j}), (\pi^*, f_1^*))  \geq t \delta_n.  
\end{equation}
Now if $\Di \left((\hat{\pi}^A, \hat{f}_1^A), (\pi^*, f_1^*) \right) \geq
t \delta_n$, then there
exist $1 \leq j  \leq N_1$ and $1 \leq k \leq N_2$ such that  
\begin{equation*}
\|\hat{f}_1^A - h_j\|_{\infty, S} \leq \eta ~~~ \text{ and } ~~~
\max_{1 \leq i \leq n} |\hat{\pi}_i^A - \pi^{(k)}_i| \leq \gamma. 
\end{equation*}
Then clearly $(j, k) \in J$ and 
\begin{equation*}
\|\hat{f}_1 - h_{0j} \|_{\infty, S} \le 2 \eta ~~~ \text{ and } ~~~
\max_{1 \leq i \leq n} |\hat{\pi}_i^A - \pi^{(0k)}_i | \leq 2 \gamma. 
\end{equation*}
Consequently, whenever $Y_i \in S$, we have 
\begin{align*}
\left(1 - \hat{\pi}_i^A \right) f_0(Y_i) + \hat{\pi}_i^A \hat{f}_1(Y_i)
&\leq  \left(1 - \pi_i^{(0k)} + 2 \gamma \right) f_0(Y_i) +
\left(\pi_i^{(0k)} + 2 \gamma \right) \left(h_{0j}(Y_i) + 2 \eta
\right) \\
&= \left(1 - \pi_i^{(0k)} \right) f_0(Y_i) + \pi_i^{(0k)} h_{0j}(Y_i) +
B_{ijk} 
\end{align*}
where
\begin{equation*}
B_{ijk} \defeq 2 \gamma f_0(Y_i) + 2 \gamma h_{0j}(Y_i) + 4 \gamma \eta
+ 2 \eta \pi_i^{(0k)}. 
\end{equation*}
Because every density in $\Fga$ is bounded from above by $(2
\pi)^{-1/2}$  and that $\eta = n^{-2} \leq 1/4$, we see that   
\begin{equation*}
B_{ijk} \leq 4 \gamma (2 \pi)^{-1/2} + 2 \eta + 4 \gamma \eta \leq
1.6 \gamma + 2 \eta + \gamma = 2.6 \gamma + 2 \eta. 
\end{equation*}
In the subsequent discussion, we will write $\mathbb{P}^*(\cdot)$ for $\mathbb{P}(\cdot|X_1,\ldots ,X_n)$ and $\mathbb{E}^*(\cdot)$ for $\mathbb{E}^*(\cdot|X_1,\ldots ,X_N)$.
Define $b \defeq 2.6 \gamma + 2 \eta$ so that
$B_{ijk} \leq b$ for all $i, j, k$. This allows us to deduce that  
\begin{equation*}
\prod_{i=1}^n \frac{\left(1 - \hat{\pi}_i^A \right) f_0(Y_i) +
	\hat{\pi}_i^A \hat{f}_1^A(Y_i)}{(1 - \pi_i^*) f_0(Y_i) + \pi_i^*
	f^*_1(Y_i)}  \leq \max_{(j, k) \in J} \prod_{i=1}^n \frac{\left(1
	- \pi_i^{(0k)} \right) f_0(Y_i) + \pi_i^{(0k)}  h_{0j}(Y_i) +
	b}{(1 - \pi_i^*) f_0(Y_i) + \pi_i^* f^*_1(Y_i)}. 
\end{equation*}
on the event $\{\Di \left((\hat{\pi}^A, \hat{f}_1^A), (\pi^*, f_1^*) \right) \geq
t \delta_n\} \cap A_t$. The left hand side of the above inequality
is at least 1 because $(\hat{\pi}^A, \hat{f}_1^A)$ is an AMLE. We thus have 
\begin{align*}
&\;\;\;\;\;\P^* \left( \left\{\Di \left((\hat{\pi}^A, \hat{f}_1^A), (\pi^*, f_1^*) \right) \geq
t \delta_n  \right\} \cap A_t \right)\\ &\leq \P^* \left\{\max_{(j, k)
	\in J}  \prod_{i=1}^n \frac{\left(1
	- \pi_i^{(0k)} \right) f_0(Y_i) + \pi_i^{(0k)}  h_{0j}(Y_i) +
	b }{(1 - \pi_i^*) f_0(Y_i) + \pi_i^* f^*_1(Y_i)} \geq 1, Y_i
\in S ~\forall i\right\} \\
&\leq \sum_{(j, k) \in J} \P^* \left\{ \prod_{i=1}^n \frac{\left(1
	- \pi_i^{(0k)} \right) f_0(Y_i) + \pi_i^{(0k)}  h_{0j}(Y_i) +
	b}{(1 - \pi_i^*) f_0(Y_i) + \pi_i^* f^*_1(Y_i)} \geq 1, Y_i
\in S ~\forall i\right\} \\
&\leq \sum_{(j, k) \in J} \E^* \prod_{i=1}^n \sqrt{\frac{\left(1 
		- \pi_i^{(0k)} \right) f_0(Y_i) + \pi_i^{(0k)}  h_{0j}(Y_i) +
		b}{(1 - \pi_i^*) f_0(Y_i) + \pi_i^* f^*_1(Y_i)}} I \left\{Y_i
\in S \right\} \\
&= \sum_{(j, k) \in J} \prod_{i=1}^n \E^* \sqrt{\frac{\left(1 
		- \pi_i^{(0k)} \right) f_0(Y_i) + \pi_i^{(0k)}  h_{0j}(Y_i) +
		b}{(1 - \pi_i^*) f_0(Y_i) + \pi_i^* f^*_1(Y_i)}} I \left\{Y_i
\in S \right\}. 
\end{align*}
Now for a fixed $(j, k) \in J$, we can bound 
\begin{align*}
D_{jk} \defeq  \prod_{i=1}^n \E^* \sqrt{\frac{\left(1 
		- \pi_i^{(0k)} \right) f_0(Y_i) + \pi_i^{(0k)}  h_{0j}(Y_i) +
		b}{(1 - \pi_i^*) f_0(Y_i) + \pi_i^* f^*_1(Y_i)}} I \left\{Y_i
\in S \right\} 
\end{align*}
as
\begin{align*}
D_{jk} &= \exp \left(\sum_{i=1}^n \log \E^* \sqrt{\frac{\left(1 
		- \pi_i^{(0k)} \right) f_0(Y_i) + \pi_i^{(0k)}  h_{0j}(Y_i) +
		b}{(1 - \pi_i^*) f_0(Y_i) + \pi_i^* f^*_1(Y_i)}} I \left\{Y_i
\in S \right\} \right) \\
&\leq \exp \left(\sum_{i=1}^n \E^*  \sqrt{\frac{\left(1 
		- \pi_i^{(0k)} \right) f_0(Y_i) + \pi_i^{(0k)}  h_{0j}(Y_i) +
		b}{(1 - \pi_i^*) f_0(Y_i) + \pi_i^* f^*_1(Y_i)}} I \left\{Y_i
\in S \right\} - n  \right)
\end{align*}
where the last inequality follows by using $\log{x}\leq x-1$ for $x>0$. The expectation above is bounded as 
\begin{align*}
E &\defeq \int_S \sqrt{(1 - \pi^{(0k)}_i) f_0 + \pi_i^{(0k)} h_{0j} + b}
\sqrt{(1 - \pi_i^*) f_0 + \pi_i^* f_1^*}  \\ 
&\leq \int_S \sqrt{(1 - \pi^{(0k)}_i) f_0 + \pi_i^{(0k)} h_{0j}}
\sqrt{(1 
	- \pi_i^*) f_0 + \pi_i^* f_1^*} +\sqrt{b} \int_S \sqrt{(1
	- \pi_i^*) f_0 + \pi_i^* f_1^*}  \\ 
&\leq 1 - \frac{1}{2} h^2 \left((1 - \pi^{(0k)}_i) f_0 + \pi_i^{(0k)}
h_{0j}, (1 - \pi^*_i) f_0 + \pi_i^* f_1^* \right) + \sqrt{b}
(2 \pi)^{-1/4} |S| 
\end{align*}
where, in the last inequality above, we have used that the density $(1
- \pi_i^*) f_0 + \pi_i^* f_1^*$ is bounded from above by $(2
\pi)^{-1/2}$ ($|S|$ in the above refers to the length of the interval
$S$). Because $|S| = 2 M + 4 t \sqrt{\log n}$ and $\sqrt{b} \leq
\sqrt{2.6 \gamma} + \sqrt{2\eta} \leq \sqrt{2.6 \gamma} + \sqrt{2}/n$,
we obtain 
\begin{align*}
E &\leq 1 - \frac{1}{2} h^2 \left((1 - \pi^{(0k)}_i) f_0 + \pi_i^{(0k)}
h_{0j}, (1 - \pi^*_i) f_0 + \pi_i^* f_1^* \right) \\ &\;\;\;+2 (2 \pi)^{-1/4}
\left(\sqrt{2.6 \gamma} + \sqrt{2}/n \right)  
\left( M + 2 t \sqrt{\log n} \right) \\
&\leq 1 - \frac{1}{2} h^2 \left((1 - \pi^{(0k)}_i) f_0 + \pi_i^{(0k)}
h_{0j}, (1 - \pi^*_i) f_0 + \pi_i^* f_1^* \right) + 7
\left(\sqrt{\gamma} + n^{-1}\right) M^* t
\end{align*}
where, as before, $M^* \defeq \max(M, \sqrt{\log n})$. This gives 
\begin{align*}
D_{jk} &\leq \exp \left( \frac{-n}{2} \Di^2 \left((\pi^{(0k)},
h_{0j}),(\pi^*, f_1^*) \right) + 7 M^* t + 7 n \sqrt{\gamma} M^*
t\right) \\
&\leq \exp \left( \frac{-n t^2 \delta_n^2}{2} + 7 M^* t + 7 n
\sqrt{\gamma} M^*  t\right)
\end{align*}
where we have also used \eqref{ad1}. We have therefore proved that  
\begin{small} 
	\begin{align*}
	\P^* \left( \left\{\Di \left((\hat{\pi}^A, \hat{f}_1^A), (\pi^*, f_1^*) \right) \geq
	t \delta_n  \right\} \cap A_t \right) &\leq |J| \exp \left( \frac{-n
		t^2 \delta_n^2}{2} + 7 M^* t +
	7 n 
	\sqrt{\gamma} M^*  t\right)  \\  
	&\leq N_1 N_2 \exp \left( \frac{-n
		t^2 \delta_n^2}{2} + 7 M^* t +
	7 n 
	\sqrt{\gamma} M^*  t\right)
	\end{align*}
\end{small}
From \eqref{ad2}  and $\log N_2 = H(\gamma, \Pi_n, L^{\infty})$, we
deduce 
\begin{align*}
&\;\;\;\;\P^* \left( \left\{\Di \left((\hat{\pi}^A, \hat{f}_1^A), (\pi^*, f_1^*) \right) \geq
t \delta_n  \right\} \cap A_t \right) \\ &\leq \exp \left(-\frac{n t^2 
	\delta_n^2}{2} + 7 M^* t + 7 n \sqrt{\gamma} M^* t + C M^* t (\log
n)^{3/2} + H(\gamma) \right) \\
&\leq \exp \left(-\frac{n t^2 
	\delta_n^2}{2} + \tilde{C} M^* t (\log n)^{3/2} + t \left\{7 n \sqrt{\gamma}
M^* + H(\gamma) \right\}  \right).   
\end{align*}
where $H(\gamma)$ is shorthand for $H(\gamma, \Pi_n,
L^{\infty})$ and $\tilde{C}=C+30$ (this is because $7\leq 30(\log{n})^{3/2}$ for all $n\geq 2$. Because $\gamma > 0$ is arbitrary, we have in fact
proved that 
\begin{align*}
&\;\;\;\;\P^* \left( \left\{\Di \left((\hat{\pi}^A, \hat{f}_1^A), (\pi^*, f_1^*) \right) \geq
t \delta_n  \right\} \cap A_t \right) \\ &\leq \exp \left(-\frac{n t^2 
	\delta_n^2}{2} + \tilde{C} M^* t (\log n)^{3/2} + t \inf_{\gamma > 0}\left\{7 n \sqrt{\gamma} M^* +
H(\gamma) \right\}  \right).  
\end{align*}
Suppose now, we choose
\begin{equation}\label{dco}
\delta_n^2 \geq \frac{K}{n} \max \left(M^* (\log n)^{3/2},
\inf_{\gamma > 0} \left\{n \sqrt{\gamma} M^* + H(\gamma)
\right\}\right) \defeq K\epsilon_n^2 
\end{equation}
for some constant $K \geq \max(8 \tilde{C}, 56)$, then  
\begin{equation*}
\P^* \left( \left\{\Di \left((\hat{\pi}^A, \hat{f}_1^A), (\pi^*, f_1^*) \right) \geq
t \delta_n  \right\} \cap A_t \right) \leq \exp \left( \frac{-n t^2
	\delta_n^2}{4} \right). 
\end{equation*}
Further $n \delta_n^2 \geq K (\log n)^2 \geq 4 \log n$ and hence  
\begin{equation*}
\P^* \left( \left\{\Di \left((\tilde{\pi}, \tilde{f}_1), (\pi^*, f_1^*) \right) \geq
t \delta_n  \right\} \cap A_t \right) \leq n^{-t^2}. 
\end{equation*}
Combining with \eqref{add}, we deduce finally that
\begin{equation*}
\P^*  \left\{\Di \left((\hat{\pi}^A, \hat{f}_1^A), (\pi^*, f_1^*) \right) \geq
t \delta_n  \right\} \leq 2n^{-t^2} \qt{for all $t \geq 1$, $n\geq 2$}
\end{equation*}
provided \eqref{dco} holds for some $K \geq \max(8\tilde{C}, 56)$. This
completes the proof of inequality \eqref{sr1.eq}.    

For \eqref{expe.eq}, we split the resulting integral into two integrals over disjoint intervals, namely $t\in [0,1]$ and $t\in (1,\infty)$.  This yields,
\begin{align*}
\frac{1}{K \epsilon_n^2} \mathbb{E}^*\left[\Di^2\left((\hat{\pi}^A, \hat{f}_1^A),
(\pi^*, f_1^*) \right)\right]\;\; &\leq 1+2\int\limits_1^{\infty} t\P^* \left( \left\{\Di \left((\hat{\pi}^A, \hat{f}_1^A), (\pi^*, f_1^*) \right) \geq
Kt \epsilon_n  \right\} \right)\,dt
\\ &=1+ 2\int\limits_1^{\infty} tn^{-t^2}\,dt\\ &=1+\frac{1}{n\log{n}}
\end{align*}
which immediately implies \eqref{expe.eq} with $C^*=2K$.

\subsection{Proof of \cref{sr.li}}\label{sr.lipf}
We shall use Theorem \ref{sr1}. The main step is to bound the metric
entropy $H(\gamma, \Pi_n, L^{\infty})$ under the given assumptions
on $\Pi$. We claim here that 
\begin{equation}\label{bb}
H(\gamma, \Pi_n, L^{\infty}) \leq p \log \left(1 + \frac{2 L T
	R}{\gamma} \right). 
\end{equation}
Assuming \eqref{bb} is true, it is easy to see that 
\begin{align*}
\inf_{\gamma > 0} \left\{n \sqrt{\gamma} M^* + H(\gamma, \Pi_n,
L^{\infty}) \right\} &\leq \inf_{\gamma > 0}
\left\{n\sqrt{\gamma} M^* + p \log \left(1
+ \frac{2 L T R}{\gamma}\right) \right\}
\\
&\le M^* + p \log \left(1 + 2 L T R n^2 \right)
\end{align*}
where, in the last inequality, we have taken $\gamma \defeq n^{-2}$. This,
along with Theorem \ref{sr1}, completes the proof of Corollary
\ref{sr.li}. 

It therefore only remains to prove \eqref{bb}. Towards this direction, note that for
two vectors $(g(x_1^T \beta), \dots, g(x_n^T \beta))$ and 
$(g(x_1^T \tilde{\beta}), \dots, g(x_n^T
\tilde{\beta}))$ in the set $\Pi_n$, we have 
\begin{equation*}
|g(x_i^T \beta) - g(x_i^T \tilde{\beta})| \leq L |x_i^T \beta -
x_i^T \tilde{\beta} | \leq L \|x_i\| \|\beta - \tilde{\beta} \| \leq
L T \|\beta - \tilde{\beta}\|. 
\end{equation*}
This, together with the fact that we assumed $\|\beta\| \leq R$,
implies that $N(\gamma, \Pi_n, L^{\infty})$ is bounded from above by the
$\gamma/(LT)$ covering number of the Euclidean ball $B(0, R) \subseteq
\R^p$ of radius $R$ centered at the origin in the Euclidean
metric. This observation implies \eqref{bb} via a standard volumetric
argument (see e.g.~\cite[Lemma 4.5]{DeGeer00}). 

\subsection{Proof of \cref{marg-1}}\label{marg-1pf}
Recall that for any $\alpha\in (0,1)$ (in particular for $1\geq \alpha\geq \overline{\pi}=\mathbb{E}_m[\pi(X)]$), we defined
\begin{equation}\label{tempref}
\hat{f}^{(\alpha)}=\frac{1}{n}\argmax\limits_{f\in\dga}\sum\limits_{i=1}^n \log{\left(\alpha f(Y_i)+(1-\alpha)f_0(Y_i)\right)}
\end{equation}
Clearly $\hat{f}^{(\alpha)}$ depends on $n$. We drop it here for notational simplicity.

Define $t\defeq \overline{\pi}/\alpha$ and $\tilde{f}\defeq tf_1^*+(1-t)f_0$. From the given conditions, $t\in [0,1]$ and so $\tilde{f}\in\dga$. By \eqref{tempref}, we thus have
\begin{align*}
\frac{1}{n}\sum\limits_{i=1}^n \log{\left(\alpha \hat{f}^{(\alpha)}(Y_i)+(1-\alpha)f_0(Y_i)\right)}&\geq \frac{1}{n}\sum\limits_{i=1}^n \log{\left(\alpha \tilde{f}(Y_i)+(1-\alpha)f_0(Y_i)\right)}\\ &=\frac{1}{n}\sum\limits_{i=1}^n \log{\left(\alpha t f_1^*(Y_i)+(\alpha-\alpha t+1-\alpha)f_0(Y_i)\right)}\\ &=\frac{1}{n}\sum\limits_{i=1}^n \log{\left(\overline{\pi}f_1^*(Y_i)+(1-\overline{\pi})f_0(Y_i)\right)}
\end{align*} 
Hence, $(\alpha,\hat{f}^{(\alpha)})$ is an AMLE for estimating $(\overline{\pi},f_1^*)$ where
$$Y\sim \overline{\pi}f_1^*+(1-\alpha)f_0.$$
Note that the above model is of the same form as~\eqref{denmodel} with $\pi^*=\overline{\pi}\in \Picon$ and $f_1^*\in\dga$. This brings us in the setup of \cref{sr1} and a direct application yields
$$\mathbb{E}\left[h^2\left(\left(\alpha,\hat{f}^{(\alpha)}\right),(\overline{\pi},f_1^*)\right)\right]\leq \tilde{C}\epsilon_n^2.$$
Here $\epsilon_n^2=\frac{1}{n}\max\left(M^*(\log{n})^{3/2},\inf\limits_{\gamma>0}\left(n\sqrt{\gamma}M^*+H(\gamma,\Pi_n,L^{\infty})\right)\right)$, $\tilde{C}$ is a universal constant and $\Pi_n=\{(\pi,\pi,\ldots ,\pi):\pi\in [0,1]\}\subset \mathbb{R}^n$.

A simple volumetric argument implies $H(\gamma,\Pi_n,L^{\infty})\leq \log\left(\frac{1}{\gamma}+1\right)$. By putting \newline $\gamma\defeq n^{-2}$, we get
\begin{equation*}
\frac{1}{n}\inf\limits_{\gamma>0}\left(n\sqrt{\gamma}M^*+H(\gamma,\Pi_n,L^{\infty})\right)\leq \frac{1}{n}\left(M^*+\log{(1+n^2)}\right)\leq \frac{4}{n}\left(M^*+\log{n}\right)
\end{equation*}

The conclusion then follows with $C=4\tilde{C}$.
\subsection{Proof of \cref{cor:Mar-2}}\label{cor:Mar-2pf}

Recall the notations introduced in the statement of~\cref{cor:Mar-2}, in particular, the fact that our parameter $\theta\equiv (\beta,\mu)$ belongs to a set $\Theta$ which is compact in $\mathbb{R}^{p+1}$, the definitions of  $m_{\theta}(x,y)\defeq -(y-\mu g(X^\top \beta))^2$, $\dot m_{\theta}=\nabla_{\theta}m_{\theta}$, $M_{\theta}=\mathbb{E}[m_{\theta}(X,Y)]$ and $V_{\theta}=\nabla_{\theta}^2 M(\theta)$. Throughout this proof, $\lVert\cdot\rVert$ will denote the usual Euclidean norm in $\mathbb{R}^{p+1}$. The proof of~\cref{cor:Mar-2} is based on~\cite[Theorem 5.23]{van2000asymptotic}. Let us first restate what we need to show in order to  apply~\cite[Theorem 5.23]{van2000asymptotic} directly:
\begin{enumerate}
	\item The LSE $\hat{\theta}_n$ is consistent for estimating $\theta^*$ (in the interior of $\Theta$), i.e.,
	\begin{equation}\label{con1}
	\qquad\qquad\hat{\theta_n}\overset{P}{\longrightarrow}\theta^* \qquad \mbox{as }n\to\infty.
	\end{equation} 
	\item There exists a function $\tilde{m}(\cdot,\cdot)$ such that $\mathbb{E}[\tilde{m}(X,Y)^2]< \infty$ and satisfying,
	\begin{equation}\label{con2}
	|m_{\theta_1}(X,Y)-m_{\theta_2}(X,Y)|\leq \tilde{m}(X,Y)\lVert \theta_1-\theta_2\rVert \qquad \mbox{for all }\theta_1,\theta_2\in\Theta.
	\end{equation}
	Note that it is sufficient to prove~\eqref{con2} for $\theta_1,\theta_2$ in a neighborhood of $\theta^*$. However we will prove the stronger version here.
	\item For all $\theta\in\Theta$, $M_{\theta}$ satisfies
	\begin{equation}\label{con3}
	M_{\theta}=M_{\theta^*}+\frac{1}{2}(\theta-\theta^*)^\top V_{\theta^*}(\theta-\theta^*)+o(\lVert \theta-\theta^*\rVert^2).
	\end{equation}
\end{enumerate}
As $\Theta$ is a compact set, we may assume that there exists $K>0$ such that $\lVert \theta \rVert_{\infty}\leq K$ for all $\theta\in\Theta$ and consequently $m_{\theta}(X,Y)\leq 2(Y^2+K^2)$ which is integrable with respect to the joint distribution of $(X,Y)$. Also, $m_{\theta}(X,Y)$ is a continuous function in $\theta$, $X$, and $Y$ and as a result we can use Theorem 2.4.1 in~\cite{van1996weak} and Example 19.8 in~\cite{van2000asymptotic} to get that $\{m_{\theta}\}_{\theta\in\Theta}$ is a Glivenko-Cantelli class of functions. Further, observe that the identifiability assumption on $\theta^*$ from the model implies that the unique maximizer of $M_{\theta}$ (as a function in $\theta$) is at the point $\theta=\theta^*$. Then,~\cite[Theorem 3.2.3]{van1996weak} implies $\hat{\theta}_n\overset{P}{\longrightarrow}\theta^*$ as $n\to\infty$. This proves~\eqref{con1}.

By the smoothness assumption on $g(\cdot$), for any fixed $(x,y)$, $m_{\theta}(x,y)$ is at least thrice differentiable in $\theta$ for all $(x,y)$. Let us denote by $\nabla^j m_{\theta}(x,y)$ the $j$-th derivative of $m_{\theta}(x,y)$ with respect to $\theta$ at $(x,y)$, for $j=1,2,3$.

For any $\delta > 0$, let $B(\theta^*, \delta)$ denote the Euclidean ball of radius $\delta>0$ centered at $\theta^*$. For any $\theta_1, \theta_2 \in B(\theta^*, \delta)$ and any fixed $(x,y) \in \X \times \R$,
\begin{equation*}
|m_{\theta_1}(x,y) - m_{\theta_2}(x,y)| \leq \sup_{\theta \in B(\theta^*, \delta)} \| \nabla m_{\theta}(x,y)\| \|\theta_1 - \theta_2\|.
\end{equation*}
Note that
\begin{equation*}
\nabla m_{\theta}(x,y) = \Big( 2(y - \mu g(x^\top \beta))g(x^\top \beta), \;2\mu(y - \mu g(x^\top\beta))g^{(1)}(x^\top\beta)x \Big),
\end{equation*}
where we have used the fact that $\nabla_{\beta}\; g(x^\top\beta) = g^{(1)}(x^\top\beta)x$, $x \in \X$. As $0 \leq g(x^\top\beta) \leq 1$ and $|g^{(1)}(x^\top\beta)|\leq c_1$, this gives
\begin{equation*}
\| \nabla m_{\theta}(x,y) \| \leq 2 | y - \mu g(x^\top\beta) | \left\{ 1 + |\mu| c_1\|x\| \right\}\leq 2(|y|+K)(1+Kc_1\lVert x\rVert).
\end{equation*}
Define $\tilde{m}(x,y)\defeq 2(|y|+K)(1+Kc_1\lVert x\rVert)$ and observe that $$\mathbb{E}[\tilde{m}(X,Y)^2]\leq 16\mathbb{E}[Y^2+K^2c_1^2Y^2\lVert X\rVert^2+K^2+K^4c_1^2\lVert X\rVert^2]<\infty$$ by the moment assumptions on $X$ and $Y$.
This completes the proof of~\eqref{con2}.

We shall now verify~\eqref{con3}. By smoothness of $m_{\theta}$ (in particular, the continuity of its hessian at all $\theta\in\Theta$, for all $(X,Y)$), the following second order Taylor expansion holds for any fixed $(x,y)$,
\begin{equation}\label{maineq}
m_{\theta} = m_{\theta^*} + (\theta - \theta^*)^{\top} \nabla m_{\theta^*} + \frac{1}{2} (\theta - \theta^*)^{\top} \nabla^2 m_{\theta^*} (\theta - \theta^*) + \frac{1}{6} R_3(\nabla^3 m_{\xi^*}, \theta - \theta^* )
\end{equation}
for some $\xi^* \in \X \times \R$ satisfying $\| \xi^* - \theta^* \| \le \|\theta - \theta^*\|$ and
\begin{equation*}
R_3(\nabla^3 m_{\xi^*}, \theta - \theta^*) \defeq \sum_{i,j,k} [\nabla^3 m_{\xi^*}]_{i,j,k} (\theta_i - \theta^*_i) (\theta_j - \theta^*_j) (\theta_k - \theta^*_k)
\end{equation*}
so that
\begin{align*}
\big|\E [R_3(\nabla^3 m_{\xi^*}, \theta - \theta^*)]\big| &= \Big|\sum_{i,j,k} \E[\nabla^3 m_{\xi^*}]_{i,j,k} (\theta_i - \theta^*_i) (\theta_j - \theta^*_j) (\theta_k - \theta^*_k)\Big| \\
& \leq (p+1)^3 \|\theta - \theta^*\|^3 \max_{i,j,k} \E\left( |[\nabla^3 m_{\xi^*}]_{i,j,k}| \right),
\end{align*}
where $c>0$ is a constant.
Using the fact that $g^{(i)}(x^\top\beta)\leq c_i$ for all $\beta$ and m-a.e.~x, we observe that for $j = 1,2,3$, the absolute value of every entry in $\nabla^j m_{\xi^*}(x,y)$ is bounded by the sum of finitely many terms, each of the form, $\tilde{K}|Y|^i|X_r|^l$, $i\in\{0,1\}$, $1\leq l\leq j$, $1\leq r\leq p$ and some big positive constant $\tilde{K}$. To prove that every term of the form $|Y|^i|X_r|^l$ as mentioned above, is integrable, it suffices to show that $|Y||X_r|^3$ is integrable. This follows from our moment assumptions and by applying H\H{o}lder's inequality with weights $4$ and $4/3$ for $|Y|$ and $|X_r|^3$ respectively. This observation leads to the following consequences:
\begin{enumerate}
	\item $\big|\E [R_3(\nabla^3 m_{\xi^*}, \theta - \theta^*)]\big|=o(\lVert \theta-\theta^*\rVert^2).$
	\item An application of the dominated convergence theorem (DCT) implies $\mathbb{E}[\nabla_{\theta}m_{\theta^*}]=\nabla_{\theta}M_{\theta^*}$. Also, note that $M_{\theta}$ is a smooth function in $\theta$ which is maximized at $\theta=\theta^*$. Therefore, $\nabla_{\theta} M_{\theta^*}=0$.
	\item Once again, by an application of DCT, $\mathbb{E}[\nabla_{\theta}^2 m_{\theta^*}]=\nabla_{\theta}^2 M_{\theta^*}=V_{\theta^*}$.
\end{enumerate}
Combining all the above observations, and taking expectations on both sides of~\eqref{maineq}, we get:
\begin{equation*}
M(\theta) = M(\theta^*) + \frac{1}{2} (\theta - \theta^*)^{\top} V_{\theta^*} (\theta - \theta^*) + o(\| \theta - \theta^* \|^2)
\end{equation*}
which completes the proof of~\eqref{con3} and hence proves~\cref{cor:Mar-2}. \qed	

\end{document}